# Coarse-Grained Molecular Dynamics Modeling of Defective Erythrocyte Membrane and Sickle Hemoglobin Fibers

He Li

University of Connecticut, 2014

Understanding the complex behavior of the normal and defective red blood cell (RBC) membrane requires the development of detailed computational models. In this work, we show that implementation of coarse-grained molecular dynamics (CGMD) methods can answer fundamental questions related to protein diffusion and membrane loss in erythrocytes. In particular, we first developed a two-component CGMD erythrocyte membrane model comprising the lipid bilayer and an implicitly-represented cytoskeleton. The model reproduced the mechanical properties of the RBC membrane illustrating both the fluidic behavior of the lipid bilayer and the elastic properties of the erythrocyte cytoskeleton. By applying shear deformation, we found that the shear stress was mainly due to the cytoskeleton at low shear strain rates while the viscosity of the lipid bilayer contributed to the resulting shear stress at higher strain rates. Building up on the experience acquired from the model above, we then created a RBC membrane model with explicit representation of the cytoskeleton. In addition to the normal RBC membrane, the model can describe the membrane of defective RBCs from patients with hereditary spherocytosis (HS) and hereditary elliptocytosis (HE) by introducing defects that reduce the connectivity between the lipid bilayer and the cytoskeleton (vertical connectivity) or the connectivity of the cytoskeleton (horizontal connectivity), respectively. The model was then applied in the study of band-3 protein diffusion in the normal RBCs and in the RBCs with membrane protein defects. We showed that the diffusion of band-3 protein is more pronounced in the defective RBCs than normal RBC, particularly in HE RBCs, which is in agreement with experimental observations. Further, the explicit two-component CGMD RBC membrane model was used to simulate vesiculation in the normal and defective RBC membrane induced by the spontaneous curvature of the membrane domain or by compression on the lipid bilayer. We found that the vesicle size depended on the membrane connectivity. Lower vertical or horizontal connectivity caused the generation of larger vesicles than the high connectivity. Finally, we introduced a CGMD model to simulate single hemoglobin S (HbS) fibers as a chain of CG particles. We showed that the proposed model was able to efficiently simulate the mechanical behavior of single and interacting HbS fibers. Simulations of the zippering process between two HbS fibers illustrated that the depletion forces induced by HbS monomers were comparable to direct fiber-fiber interaction via the Van der Waals forces.

# Coarse-Grained Molecular Dynamics Modeling of Defective Erythrocyte Membrane and Sickle Hemoglobin Fibers

He Li

B.S., Beijing University of Technology, 2005

M.S., University of Saskatchewan, 2008

A Dissertation

Submitted in Partial Fulfillment of the

Requirements for the Degree of

Doctor of Philosophy

at the

University of Connecticut

2014

APPROVAL PAGE

Doctor of Philosophy Dissertation

Coarse-Grained Molecular Dynamics Modeling of Defective Erythrocyte Membrane and Sickle Hemoglobin Fibers

Presented by

He Li, B.S., M.S.

Major Advisor ___________________________________________________
George Lykotrafitis

Associate Advisor ___________________________________________________
Tai-Hsi Fan

Associate Advisor ___________________________________________________
Greg Huber

Associate Advisor ___________________________________________________
Horea Ilies

Associate Advisor ___________________________________________________
David Pierce

University of Connecticut

2014

# Acknowledgment

I am very honored to have Professor George Lykotrafitis as my research advisor. During my Ph.D. study, Professor George Lykotrafitis introduced me to the interesting and exciting field of cellular mechanics. Meanwhile, Professor George Lykotrafitis gave me unlimited guidance, encouragement and patience. He helped me to improve my presentation skills and supported me to attend research conferences where I can exchange and discuss ideas with other researchers. He also gave me important suggestions on the developing my future career.

Second, I would like to thank my committee members Professor Tai-Hsi Fan, Professor Greg Huber, Professor Horea Ilies and Professor David Pierce for taking time to review my research work and offering their valuable advices.

Third, I am thankful for my colleagues working in the cellular mechanics lab who create a enjoyable research environment.

Last but not least, I would like to send my most sincere and deepest appreciation to my parents and my wife who are being supportive and taking care of me in my personal life so that I can concentrate on my research work. In addition, I would like to thank ME Entertainment Frontier Group, especially all the "ME Big4" members and their families, QiangGe and Xinxuan for giving me so much joy and happiness during the past six years.

# Table of Contents

**Chapter 5. Modeling of Vesiculation in Healthy and Defective Human Erythrocyte Membrane**

# List of Figures

# Abstract of the Dissertation

In this work, we develop a two-component coarse-grained molecular dynamics (CGMD) model for simulating the erythrocyte membrane. This proposed model possesses the key feature of combing the lipid bilayer and the erythrocyte cytoskeleton, thus showing both the fluidic behavior of the lipid bilayer and elastic properties of the erythrocyte cytoskeleton. The proposed model facilitates simulations that span large length-scales (~ µm) and time-scales (~ ms). By tuning the interaction potential parameters, diffusivity and bending rigidity of the membrane model can be controlled. In the study of the membrane under shearing, we find that in low shear strain rate, the developed shear stress is mainly due to the spectrin network, while the viscosity of the lipid bilayer contributes to the resulting shear stress at higher strain rates. In addition, the effects of the reduction of the spectrin network connectivity on the shear modulus of the membrane are investigated.

We extend the previously developed two-component red blood cell (RBC) membrane model and develop a mesoscale implicit-solvent CGMD model of the erythrocyte membrane which explicitly describes the phospholipid bilayer and the cytoskeleton in order to study the cytoskeleton and lipid bilayer interactions, protein diffusion in the lipid bilayer and the membrane vesiculation in the normal and defective RBC membrane. We show that the proposed model represents RBC membrane with the appropriate bending stiffness and shear modulus. The timescale and self-consistency of the model are established by comparing our results to experimentally measured viscosity and thermal fluctuations of the RBC membrane. Furthermore, we measure the pressure exerted by the cytoskeleton on the lipid bilayer. We find that defects at the anchoring points of the cytoskeleton to the lipid bilayer (as in spherocytes) cause a reduction in the pressure compared to an intact membrane, while defects in the dimer-dimer association of a spectrin filament (as in elliptocytes) cause an even larger decrease in the pressure. We conjecture that this finding may explain why the experimentally measured diffusion coefficients of band-3 proteins are higher in elliptocytes than in spherocytes, and higher than in normal RBCs.

We employ the RBC membrane model which explicitly describes the phospholipid bilayer and the cytoskeleton, to study the band-3 protein diffusion in blood disorders such as hereditary spherocytosis (HS) and hereditary elliptocytosis (HE). Furthermore, we investigate the effect of the attraction between the spectrin filaments and lipid bilayer on the band-3 diffusion. First, we measure the band-3 protein diffusion coefficients from healthy RBC membrane and from the membrane after the cytoskeleton is removed. Comparison of these two coefficients clearly illustrates the steric hindrance of the cytoskeleton on the band-3 mobility. Second, we measure the band-3 diffusion from defective RBC membrane and quantify the relation between the band-3 diffusion coefficients and degree of protein defects, via which we interpret the experimentally measured band-3 diffusion coefficients in HS and HE. The diffusion coefficients measured from our simulation are consistent with the experimentally measured diffusion coefficients as well as the diffusion coefficients calculated based on the percolation analysis. By comparing the diffusion coefficients measured from different types of protein defects, we find that the band-3 mobility is primarily controlled by the connectivity of the spectrin network, in agreement with previous experimental results. Meanwhile, we study the dependence of band-3 anomalous diffusion exponent on protein defects in the cytoskeleton and we find that the effect of cytoskeleton on the anomalous diffusion exponent is small. At last, we applied the attractive forces between spectrin filaments and lipid bilayer, and find that the attractive forces have the functions of decelerating the band-3 diffusion. The band-3 diffusion measurements conducted when the spectrin filaments are fully attached to the lipid bilayer generally agree with the calculation based on the percolation theory. The simulation results and the comparisons with the existing experimental measurements are also used to predict the scale of the undetermined attractive forces between the spectrin filament and lipid bilayer.

We employ a two-component RBC membrane model to simulate the diffusion of band-3 proteins in the normal RBC and in the RBCs with protein defects. We introduce protein defects which reduce the connectivity between the lipid bilayer and the cytoskeleton or reduce the connectivity of the cytoskeleton and these defects are associated with the blood disorders of HS and HE, respectively. We first measure the band-3 diffusion coefficients in the normal RBC membrane and in the RBC membrane without cytoskeleton. Comparison of these two coefficients demonstrates that the cytoskeleton limits the band-3 lateral mobility. Second, we study band-3 diffusion in defective RBC membranes and quantify the relation between the band-3 diffusion coefficients and the percentage of protein defects in HS and HE RBCs. By comparing the diffusion coefficients measured in these two cases, we conclude that the band-3 mobility is primarily controlled by the cytoskeleton connectivity. Third, we study how the band-3 anomalous diffusion exponent depends on the percentage of protein defects in the membrane cytoskeleton. Our measurements show that the effect of the cytoskeleton connectivity on the anomalous diffusion exponent is small. Finally, we show that introduction of attraction between the lipid bilayer and the spectrin network can reduce band-3 diffusion. By comparing our measurements to experimental data, we predict that the attractive force between the spectrin filament and the lipid bilayer is at least 20 times smaller than the binding forces at the two membrane major binding sites at band-3 and glycophorin.

.

Normal and defective RBCs release microvesicles of different compositions and sizes. Recent advances have revealed that RBC-derived microvesicles play essential roles in coagulation and in immune response. In blood disorders induced by membrane protein defects, microvesiculation is a result of deteriorated stability of the lipid bilayer. Despite extended experimental and theoretical work conducted on the vesiculation, the effects of the RBC cytoskeleton and its connectivity on vesicle formation are not clear. In this work, we employ a recently developed RBC membrane model, which explicitly describes the lipid bilayer and the spectrin network, to study mechanisms of vesicle formation in the normal RBC membrane and in membranes with defects related to HS and HE. We specifically correlate the size of the vesicle to microvesiculation procedure and to membrane defects. We also determine conditions under which the released microvesicles contain cytoskeleton fragments. Our simulations show that lateral compression on the membrane can cause formation of vesicles with size similar to the size of the basic cytoskeleton corral. When we consider a lipid bilayer model with different areas of spontaneous curvature, corresponding to different phospholipid composition, then the produced vesicles have a homogeneous composition. If lateral compression is applied on the previous system, then the formation of vesicles originated from the curved membrane domain is facilitated. In HS, where the vertical connectivity between the lipid bilayer and the spectrin network is reduced, and in HE, where the lateral network connectivity is reduced, we find that the sizes of microvesicles is diverse compared to the sizes of the microvesicles released from the healthy RBC. This is due to the reduced confinement of the lipid bilayer by the RBC cortex. An increased vertical connectivity between the lipid bilayer and the cytoskeleton causes the generation of multiple vesicles with sizes similar to the cytoskeleton corral dimension. In the case of a low vertical connectivity, the membrane tends to release larger vesicles under the same compression ratio as above. It is noted that vesicles released from the HE RBCs may contain cytoskeleton components due to the fragmentation of the cytoskeleton while vesicles released from the HS RBCs are depleted of cytoskeleton components.

We develop a solvent-free CGMD HbS fiber model which represents a single hemoglobin fiber as four tightly bonded chains. Because the intracellular polymerization of deoxy-sickle cell hemoglobin has been identified as the main cause of sickle cell disease, the material properties and biomechanical behavior of polymerized HbS fibers becomes important in the studies of the sickle cell disease. In this developed fiber model, a finitely extensible nonlinear elastic (FENE) potential, a bending potential, a torsional potential, a truncated Lennard-Jones potential and a Lennard-Jones potential are implemented along with the Langevin thermostat to simulate the behavior of a polymerized HbS fiber in the cytoplasm. The parameters of the potentials are identified via comparison of the simulation results to the experimentally measured values of bending and torsional rigidity of single HbS fibers. After it is shown that the developed model is able to very efficiently simulate the mechanical behavior of single HbS fibers, it is employed in the study of the interaction between HbS fibers. It is illustrated that frustrated fibers and fibers under compression require a much larger interaction force to zipper than free fibers resulting to partial unzippering of these fibers. Continuous polymerization of the unzippered fibers via heterogeneous nucleation and additional unzippering under compression can explain the formation of HbS fiber networks and consequently the wide variety of shapes of deoxygenated sickle cells.

Finally, we simplified the four-chain fiber model and simulate single HbS fibers as a chain of particles. The overall strategy for this one-chain model is similar to the one applied in the four-chain fiber model. However, the representation of the cross-section by one CG particle discards the need for calculating the interactions between neighboring chains. More importantly, the one-chain fiber model can be easily used to simulate the formation of fiber bundles and fiber domains. For example, the zippering process between two HbS fibers is studied and the effect of depletion forces is investigated. Simulation results illustrate that depletion forces play a role comparable to direct fiber-fiber interaction via Van der Waals forces.

# Chapter 1.

# Introduction

In this chapter, I present a brief review on the RBC and the structure of healthy RBC membrane. Following that, HS and HE, which are caused by the proteins defects in the RBC membrane, will be introduced. Afterwards, structure of the HbS fibers and sickle cell disease (SCD) are introduced and described. This section ends with a review on the existing models for RBC Membrane.

**1.1 Human Erythrocytes**

A healthy quiescent RBC is biconcave in shape. Its diameter is approximately 8 μm and its thickness is approximately 2.5 μm. Its main function is to deliver oxygen to the body tissues via the blood flow through the blood circulations. An erythrocyte completes almost 250,000 cycles during its 100 to 120 days life span and often passes through areas that induce high shear stress such as the aortic valve, and endures multiple cycles of swelling and shrinkage at lungs and kidneys. A RBC not only has to be robust but it must be very compliant too, as it undergoes very large deformations when it passes through narrow capillaries of diameters as low as 3 - 4 μm in the brain (1) and when it squeezes through spleen slits, which are roughly 1 μm ×2 μm openings and about 2 μm deep formed by adjacent endothelial cells (2).

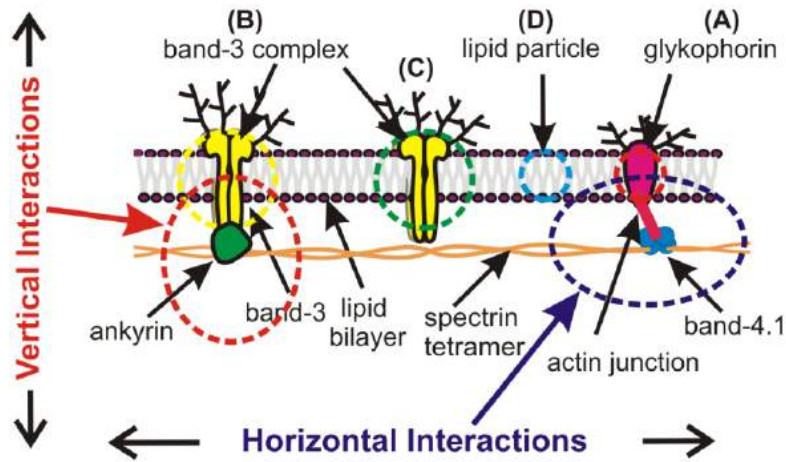

Figure 1.1. Schematic model of the RBC membrane. The vertical and horizontal interactions between its components are indicated. Disruption of vertical interactions causes spherocytosis via loss of the unsupported PB, while disruption of actin-spectrin junctions causes elliptocytosis because the network cannot recover its initial shape after undergoing large deformations in the circulation (3)

The erythrocyte's remarkable mechanical properties originate from the unique architecture of its cell membrane (4), which is the main load bearing component, as there are no stress fibers inside a normal RBC. A schematic representation of a part of the membrane is shown in Fig. 1.1(5, 6). The cell membrane consists of spectrin tetramers that form a 2D six-fold network tethered to a phospholipid bilayer (PB) (width: approximately 5 nm) comprising various types of phospholipids, sphingolipids, cholesterol, and intramembrane proteins. The spectrin tetramers are formed of spectrin heterodimers consisting of intertwined α-spectrin and β-spectrin filaments running antiparallel to each other (Fig. 1.1). The intramembrane proteins band 3 and glycophorin tether the spectrin network to the bilayer via additional binding proteins (e.g., ankyrin and 4.1).

The spectrin tetramers are connected via actin filaments and additional proteins forming junctional complexes (5-8). The PB behaves as a 2D fluid and it resists bending but it cannot resist shear loading that is sustained by the spectrin network. The mechanical properties and shape of the RBC rely on the competition between the elastic energy of the lipid membrane, including membrane tension and bending stiffness, and the shear and dilation elasticity of the cytoskeleton. Within the framework of continuum mechanics, the free energy of a lipid membrane can be described by the classical Canham-Helfrich theory (9), as

$$E_{\text{Lipid Bilayer}} = \int \left[ \gamma + \frac{\kappa}{2}(C_1 + C_2 - C_0)^2 + \bar{\kappa} C_1 C_2 \right] dA, \qquad (1.1)$$

where $C_1$ and $C_2$ are the principal curvatures, $C_0$ is the spontaneous curvature, $\gamma$ is the surface tension, $\kappa$ and $\bar{\kappa}$ are the bending and Gaussian rigidities, respectively. The first term on the right-hand side of Eq. (1.1) is the area-expansion energy, while the second and third terms represent the normal and Gaussian bending energies, respectively. Since the topology of the membrane remains unchanged the Gaussian bending energy results in a constant contribution, and this energy term can be dropped. Surface tension is defined as $\gamma = -(\sigma_{11} + \sigma_{22})L_z/2$, where $\sigma_{11}$ and $\sigma_{22}$ are the components of the in-plane Virial stress (10) in a 3D periodic supercell calculation, and $L_z$ is height of the supercell, so that $\gamma$ has the unit of N/m. Note that the so-called nonlocal bending energy (11-13) which is induced by the lipid inner and outer layer area difference is not included. This term appears in the area-difference elasticity model of bending energy, but not in the classical spontaneous curvature model (9).

The cytoskeleton shrinks to a 3 to 5-fold smaller area after the entire membrane of the RBC is removed (14), indicating that the cytoskeleton is stretched by its attachment to the lipid bilayer (13, 15), and thus exerts a compression force on the lipid bilayer. The compression force is balanced by the membrane bending stress. The elastic energy of the cytoskeleton is expressed as

$$E_{Cytoskeleton} = \frac{K_\alpha}{2}\int \alpha^2 dA + \mu \int \beta dA \qquad (1.2)$$

where $\alpha = \lambda_1\lambda_2 - 1$ and $\beta = (\lambda_1 - \lambda_2)^2/2\lambda_1\lambda_2$ are the local area and shear strain invariants, respectively. $\lambda_1$ and $\lambda_2$ are the local principal stretches. $K_\alpha$ and $\mu$ are the linear elastic moduli for stretching and shear, respectively. We add the elastic energy of the cytoskeleton to Eq. (1.1) and therefore obtain the total free energy for the RBC membrane

$$E_{Total} = E_{Lipid\ Bilayer} + E_{Cytoskeleton} = \int \left[\gamma + \frac{\kappa}{2}(C_1 + C_2 - C_0)^2 + \bar{\kappa}C_1C_2\right]dA +$$

$$\frac{K_\alpha}{2}\int \alpha^2 dA + \mu \int \beta dA \qquad (1.3)$$

In addition, the RBC membrane is composed of a host of different lipids and membrane proteins, forming a heterogeneous bilayer. The steric interactions mismatch between different types of lipids or between lipids and proteins results in a line tension which induces lateral phase separation. The phase boundary line tension prefers to reduce the phase domain boundary length. The line tension could promote the membrane vesiculation and the energy induced by the line tension is described by

$$E_{line} = \oint \sigma\, dl \qquad (1.4)$$

where $l$ is boundary of the phase separation and $\sigma$ is the energy per unit length at the phase boundary.

## 1.2 Hereditary Spherocytosis and Elliptocytosis

Because of the importance of the membrane skeleton for the stability of a RBC, defects in any of its components lead to blood disorders (3, 5), such as HS and HE, which will be briefly introduced as follows.

The hemolytic disorders of hereditary spherocytosis, and hereditary elliptocytosis affect the lives of millions of individuals worldwide. HS is by far the most common congenital hemolytic anemia in northern European descendants with a frequency of at least 1:5,000 (16). HE has a worldwide distribution but is more common in people with African and Mediterranean ancestry (17). Erythrocytes in patients with HS are characterized by a spherical shape of a smaller diameter than healthy RBCs (Fig. 1.2A). In HE and its variants erythrocytes are elongated into a cigar or oval shape (Fig. 1.2B).

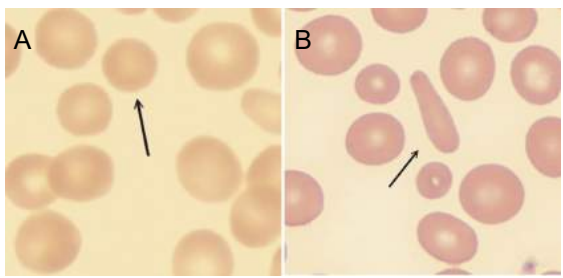

Figure 1.2. (A)Spherocytes.(B) Elliptocytes. (18)

In HS and HE, connections between components of the RBC membrane break resulting to loss of structural and functional integrity of the membrane. Two types of interactions are identified as: vertical interactions between the spectrin network and the PB and horizontal interactions at the actin-spectrin junctions of the cytoskeleton (Fig. 1.1) (3). The prevailing hypothesis is that defects in vertical interactions lead to spherocytosis via loss of unsupported membrane, while defects in horizontal interactions lead to elliptocytosis because the network loses the ability to recover its initial shape after undergoing large deformations in circulation (3, 16, 19-24).

The typical clinical picture of HS and HE combines evidences of hemolysis with spherocytosis. Patients with severe HS, by definition, have life-threatening hemolytic anemia that requires blood transfusion and they show an incomplete response to splenectomy. In addition to the risks of recurrent transfusions, these patients are particularly prone to aplastic crises (bone marrow failure), which occasionally develop retardation and delayed sexual maturation. Most cases of HS are heterozygous since homozygosity is lethal (25, 26).

**1.3 Sickle Cell Disease and Sickle Hemoglobin Fiber**

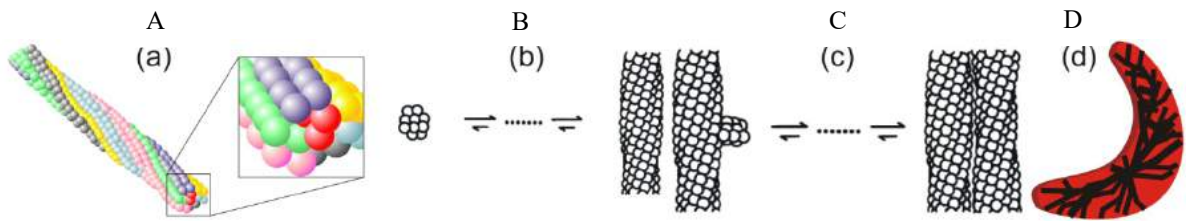

Figure 1.3. (A) The HbS fiber is composed of 7 double strands, shown as like-colored pairs (after (27)). (B) Homogeneous nucleation. (C) Heterogeneous nucleation (27). (D) 2D fiber domain, after (28).

SCD affects approximately 1:400 African-American individuals in the United States (29-37). It is an inherited blood disorder caused by a single point mutation in one of the genes encoding hemoglobin which results in polymerization of deoxygenated abnormal sickle hemoglobin (HbS) and the formation of long stiff rodlike fibers causing sickling of RBCs (27, 32, 38-40). Electron microscopy has revealed that the HbS fiber consists of 7 double strands in a hexagonally shaped cross-section twisted about a common axis in a rope-like fashion retaining their atomic contacts (Fig. 1.3A) (41-43). The fibers have an average radius of approximately 11 nm and a mean helical path of approximately 270 nm, defined as the length over which the fibers twist through 180° (41, 44). In vivo, HbS fibers are formed in an unusual fashion by two types of nucleation processes. Homogeneous nucleation entails the nucleation of a new fiber from an aggregate of monomers (see Fig. 1.3B) (27, 45). In heterogeneous nucleation, the surface of a formed polymer assists the nucleation of a new polymer (46, 47) (see Fig. 1.3C). Since the homogeneous nucleation is very slow, there are only a few homogeneous nuclei in the deoxygenated state leading to a very low number of polymer domains via the dominant process of heterogeneous nucleation, which generates aligned polymers. The process of fiber domain formation is shown in Fig.1.4. First, the fiber polymerization initiates from in the HbS solution and form few nucleuses, which develop to HbS fibers through homogeneous nucleation. Then, the HbS fibers elongate with new fibers growing on the surface of the existing fibers, which eventually generates fiber arrays and form fiber domains (See Fig.1.4B-F).

In fact, most RBCs gel with only a single polymer domain (48, 49). Another important observation is that the sickle RBC membrane considerably enhances the nucleation by a factor of 6 (50) and that HbS fibers are directly attached to the cytoplasmic tail of the band 3 protein(51). However, because the bulk majority of band 3 proteins are not attached to the cytoskeleton, HbS fibers can create long protrusions consisting of PB without support from the spectrin network (6, 35). Due to the presence of the fiber domains inside the cell, sickled RBC shows increased cell rigidity and adhesion, along with morphologies, such as sickle, holly leaf, granular, which result from the interactions between the RBC membrane and intracellular HbS fiber domains (52). The increased cell rigidity and adhesion raise the blood flow resistance and potentially trigger vaso-occlusion in the microcirculation.

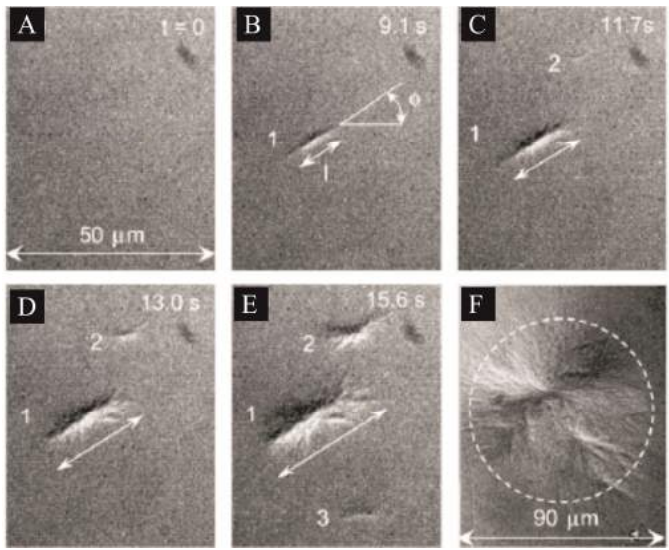

Figure 1.4. (A–F) Evolution of HbS fiber polymerization (53). Image in f corresponds to equilibrium between polymerized fiber and solution after 1 min of polymerization. Elongated spherulites in (B–E) evolve into isometric spherulites in (F).

Pathology in SS-SCD begins with the increased rigidity and cell adherence of the sickled RBCs leading to vasoocclusion (36, 50, 54-57), which often causes significant tissue damage all over the body and severe complications including acute and chronic pulmonary dysfunction, aseptic necrosis of the hip, sickle cell retinopathy, and severe and chronic pain (31, 58-60). Overt stroke caused by the occlusion of large blood vessels is the most deadly complication of SCD (61-64). Silent stroke due to cerebral microinfarcts is the most common neurological complication in children with the SCD resulting to both motor and neurocognitive impairment and to progressive cerebral injury (65-68).

## 1.4 Existing RBC Membrane Models

*Continuum membrane models* based on membrane elasticity (9, 13, 69-73) have been adopted to elucidate vesicle shape transitions and estimate thermal fluctuations of fluid membranes at length scales much larger than the bilayer thickness (74, 75). At the opposite end of the length scale spectrum, *atomistic simulations* have been extensively used to rationalize the molecular mechanisms of various functions of bilayer membranes (76-79). However, owing to the prohibitive computational cost, atomistic models are inadequate for direct comparisons with the length and time scales of typical laboratory experiments.

The limitations of atomistic simulations and continuum approaches have motivated a continual search for *CGMD methods* that bridge atomistic and continuum models (80-101). The complete picture of CGMD and relevant references can be found in recent reviews (82, 96, 100). CGMD models can be generally categorized into explicit solvent and implicit solvent (solvent-free) schemes. Explicit solvent schemes employ the hydrophobic interactions between membrane and solvent particles (a water molecule or a group of several water molecules) to stabilize the 2D membrane (84, 87, 94). Explicit solvent models frequently employ dissipative particle dynamics (DPD), a very efficient method that represents a large volume of the solvent with a soft bead, thus significantly accelerating the computations (102-105). The technique has been extended to lipid bilayers by introducing spring forces between representative particles in the polymer chains (87, 106). In the case of implicit solvent schemes, the solvent particles are not directly represented in the simulation and their effect is taken into account by employing effective multibody interaction potentials, based on the local particle density (83, 86, 95), or by implementing different pair-potentials between particles representing the hydrophobic tail and those representing the hydrophilic head of the lipids (88, 91, 92).

In order to largely extend the accessible length and time scale of membrane model, an early CGMD lipid membrane model featuring orientation-dependent interactions was developed by Drouffe et al. (83), where an anisotropic attractive interaction between spherical particles, along with a hard core repulsive interaction and a density-dependent "hydrophobic" multibody potential, was used to describe the lipid interactions. Branningan and Brown (107) developed a solvent-free model in which lipids are represented as rigid, asymmetric spherocylinders interacting through orientation-dependent attractions. Kohyama (101) extended the Drouffe's model by introducing an extra degree of freedom that corresponds to the effective curvature caused by thermal fluctuations. Yuan et al. (108) simulated the biological fluid membranes by introducing a one-particle-thick, solvent-free, coarse-grained model, in which the interparticle interaction is described by a soft-core pairwise potential. The model essentially combines Drouffe's (83) and Cooke's approaches (92). The interaction strength is also dependent on the relative orientations of the particles, but there is no membrane cytoskeleton involved. In all the models, the orientation-dependent interactions are utilized to describe the hydrophobicity of the tail groups of the lipids, and are essential to the self-assembly of lipid bilayer in aqueous environment.

**1.5 Research Motivation**

Previously developed CGMD models of the RBC membrane either only simulated the RBC cytoskeleton with application of a bending potential to represent the bending stiffness induced by the lipid bilayer (80, 81, 91, 109, 110) or only modeled the lipid bilayer (84, 86, 91, 92, 95, 108) with implicit spectrin network (111). Therefore, we develop a two-component CGMD model for simulating the erythrocyte membrane. The developed model possesses the key feature of combing the lipid bilayer and the erythrocyte cytoskeleton, thus showing both the fluidic behavior of the lipid bilayer and elastic properties of the erythrocyte cytoskeleton. Then, by extending this two-component RBC membrane model, we developed a mesoscale implicit-solvent CGMD model of the erythrocyte membrane which explicitly describes the phospholipid bilayer and the cytoskeleton.

We employed the developed model to study:

**(1)** D*iffusion of the mobile band 3 particles in the normal RBC membrane and in the membrane with defective membrane proteins.* Recent experimental studies have demonstrated that the properties of band 3 change significantly in HS and HE (112). In HS cells, the fast population of band 3 (free band 3 proteins) takes the prominence while the slow population (fixed band 3) dominates the behavior in healthy cells (73% of the total band 3). Similarly, a substantial fraction of band 3 was found to redistribute from the slower to faster diffusing population in HE samples. In addition, the diffusion coefficient for the HE samples were approximately 10 times faster than in normal cells (112). With the proposed model, we can quantitatively study the dependence of the band 3 diffusion coefficient on the proteins defects. First, all the linkages between the spectrin proteins and the immobile band 3 proteins will be broken (vertical interactions), which commonly happens in the HS where defective band 3, protein 4.2 and ankyrin proteins are found in the RBC membrane. Then, in the next case, the spectrin chains are dissociated in the middle or from the actin junctions to represent the proteins defects in horizontal interactions, which happen in HE. At last, the comparison of the band 3 diffusion coefficients measured from the reduced vertical connections and horizontal connections is performed to find out which is the major determinant of the lateral diffusion rate of band 3. Moreover, comparison is made between band 3 diffusion of the proposed model and the previous two-component WLC RBC membrane model (111) and demonstrate the hindering effect of the spectrin filaments on the band 3 mobility.

**(2)** *Vesicluation in the healthy and defective erythrocyte membrane*. The aggregation of membrane proteins of specific types that have mutual affinity, is the dominate process driving the formation of the nanovesicles (diameter of 60 ~ 100 nm)(113), while the compression forces induced by cytoskeleton stiffening lead to buckling and formation of the microvesicles (diameter of 60 ~ 300 nm (15, 113). The proposed model will be used to simulate the membrane vesiculation process and explore the dependence of vesiculation on the compression ratio and cytoskeleton connectivity. Moreover, different types of the lipids or proteins can be introduced into the model. The aggregation of the same types of the particles will form membrane domains. The domain boundary line tension prefers to reduce the domain boundary length and is expected to contribute to the membrane vesiculation. In addition, we will use our model to test the two predominant mechanisms that have been hypothesized to explain how defects in vertical interactions result to HS (25). One suggests that membrane loss is induced by defective interactions between the lipid bilayer and the cytoskeleton. The unsupported lipid bilayer in the cytoskeleton deficient area will bud off and form vesicles. The other hypothesis assumes that the membrane is stabilized by the interactions between band 3 proteins and their neighboring lipids. Defects in membrane proteins enhance the lateral mobility of band 3 proteins (112, 114-118), and the band 3, therefore, may aggregate and form clusters in the defective RBC membrane, which lead to loss of the unsupported lipids in band 3 deficient areas (25).

## 1.6 Dissertation Outline

The rest of this dissertation is organized as follows. A two-component CGMD model for the erythrocyte membrane is introduced in Chapter 2, which possesses the key feature of combing the lipid bilayer and the erythrocyte cytoskeleton, thus showing both the fluidic behavior of the lipid bilayer and elastic properties of the erythrocyte cytoskeleton. In Chapter 3, we extend the two-component RBC membrane model introduced in the Chapter 2 and develop a mesoscale implicit-solvent CGMD model of the erythrocyte membrane which explicitly describes the phospholipid bilayer and the cytoskeleton in order to study the cytoskeleton and lipid bilayer interactions, protein diffusion in the lipid bilayer and the membrane vesiculation in the normal and defective RBC membrane. In Chapter 4, we applied the RBC membrane model which explicitly describes the phospholipid bilayer and the cytoskeleton, to study the band-3 protein diffusion in blood disorders such as hereditary spherocytosis (HS) and hereditary elliptocytosis (HE). In Chapter 5, we employ the RBC membrane model which explicitly describes the phospholipid bilayer and the cytoskeleton, to study vesiculation in the healthy and Defective Human Erythrocyte Membrane. In Chapter 6, we develop a solvent-free CGMD HbS fiber model which represents a single hemoglobin fiber as four tightly bonded chains to simulate the material properties and biomechanical behavior of polymerized HbS fibers. In Chapter 7, we simplified the four-chain fiber model introduced in the Chapter 6 and simulate single HbS fibers as a chain of particles, which can be easily used to simulate the formation of fiber bundles and fiber domains. Epilogue is provided in Chapter 8.

# Chapter 2.

# Two-Component Coarse-Grain Molecular Dynamics Model for the Human Erythrocyte Membrane

**Abstract**

We present a two-component coarse-grain molecular dynamics model for simulating the erythrocyte membrane. The proposed model possesses the key feature of combing the lipid bilayer and the erythrocyte cytoskeleton, thus showing both the fluidic behavior of the lipid bilayer and elastic properties of the erythrocyte cytoskeleton. In this model, three types of coarse-grained particles are introduced to represent clusters of lipid molecules, actin junctions and band-3 complexes, respectively. The proposed model facilitates simulations that span large length-scales (~ μm) and time-scales (~ ms). By tuning the interaction potential parameters, diffusivity and bending rigidity of the membrane model can be controlled. In the study of the membrane under shearing, we find that in low shear strain rate, the developed shear stress is mainly due to the spectrin network, while the viscosity of the lipid bilayer contributes to the resulting shear stress at higher strain rates. In addition, the effects of the reduction of the spectrin network connectivity on the shear modulus of the membrane are investigated.

## 2.1 Introduction

Mature human erythrocytes do not have an internal structure, meaning that their mechanical properties originate from the cell membrane which is the only structural element of the cell (6).

The red blood cell (RBC) membrane is composed of a two-dimensional (2D) six-fold spectrin network tethered to a lipid bilayer comprising various types of phospholipids, sphingolipids, cholesterol and integral membrane proteins (see Fig. 2.1). The spectrin network consists of spectrin tetramers that are connected via actin filaments and additional proteins forming junctional complexes. Each spectrin tetramer comprises two heterodimers, which consist of intertwined and antiparallel α-spectrin and β-spectrin filaments (6). The integral membrane proteins band-3 and glycophorin tether the cytoskeleton network to the phospholipid bilayer via additional peripheral proteins (e.g., Ankyrin and 4.1) (3). The lipid bilayer is essentially a 2D fluid-like structure embedded in a three-dimensional (3D) space. It resists bending but cannot sustain in-plane shear stress because the lipids and most of the proteins can diffuse freely within the membrane to relax the shear stress. The stiffness of RBCs arises primarily from the spectrin network. It is also noted that the material properties of the lipid bilayers play important roles in the function and distribution of the membrane proteins (119).

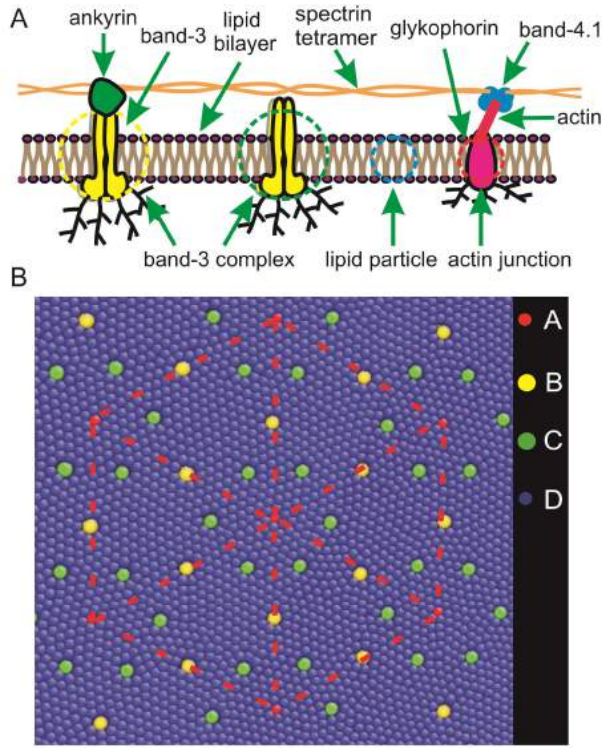

Figure 2.1. (A) Schematic of the membrane of human RBC. The blue circle represents the lipid particle and the red circle signifies the actin junction. The yellow and green circles correspond to a band-3 complex connected to the spectrin network and a free band-3 complex respectively. (B) Two-component human RBC membrane model. "A" type particles represent actin junctions. "B" type particles represent band-3 complex that are connected to the spectrin network. "C" type particles represent band-3 complex that are not connected to the network. "D" type particles represent lipid particles. Red dash line highlights the structure of 2D six-fold spectrin network.

Continuum models based on membrane elasticity (9, 13, 69-73) have been adopted to elucidate vesicle shape transitions and estimate thermal fluctuations of fluid membranes at length scales much larger than the bilayer thickness (74, 75). At the opposite end of the length scale spectrum, atomistic simulations have been extensively used to rationalize the molecular mechanisms of various functions of bilayer membranes (76-79). However, owing to the prohibitive computational cost, atomistic models are inadequate for direct comparisons with the length and time scales of typical laboratory experiments.

The limitations of atomistic simulations and continuum approaches, along with the practical need to treat the heterogeneous nature of RBC membrane have motivated a continual search for Coarse-Grained Molecular Dynamics (CGMD) methods that bridge atomistic and continuum models (80-101, 120, 121). The complete picture of CGMD and relevant references can be found in recent reviews (82, 96, 100). CGMD models can be generally categorized into explicit solvent and implicit solvent (solvent-free) schemes. Explicit solvent schemes employ the hydrophobic interactions between membrane and solvent particles (a water molecule or a group of several water molecules) to stabilize the 2D membrane (84, 87, 94). Explicit solvent models frequently employ dissipative particle dynamics (DPD), a very efficient method that represents a large volume of the solvent with a soft bead, thus significantly accelerating the computations (102-105). The technique has been extended to lipid bilayers by introducing spring forces between representative particles in the polymer chains (87, 106). In the case of implicit solvent schemes, the solvent particles are not directly represented in the simulation and their effect is taken into account by employing effective multibody interaction potentials, based on the local particle density (83, 86, 95), or by implementing different pair-potentials between particles representing the hydrophobic tail and those representing the hydrophilic head of the lipids (88, 91, 92).

In order to largely extend the accessible length and time scale of membrane model, an early CGMD lipid membrane model featuring orientation-dependent interactions was developed by Drouffe et al. (83), where an anisotropic attractive interaction between spherical particles, along with a hard core repulsive interaction and a density-dependent "hydrophobic" multibody potential, was used to describe the lipid interactions. Branningan and Brown (107) developed a solvent-free model in which lipids are represented as rigid, asymmetric spherocylinders interacting through orientation-dependent attractions. Kohyama (101) extended the Drouffe's model by introducing an extra degree of freedom that corresponds to the effective curvature caused by thermal fluctuations. Yuan et al. (108) simulated the biological fluid membranes by introducing a one-particle-thick, solvent-free, coarse-grained model, in which the interparticle interaction is described by a soft-core pairwise potential. The model essentially combines Drouffe's (83) and Cooke's approaches (92). The interaction strength is also dependent on the relative orientations of the particles, but there is no membrane cytoskeleton involved. In all the models, the orientation-dependent interactions are utilized to describe the hydrophobicity of the tail groups of the lipids, and are essential to the self-assembly of lipid bilayer in aqueous environment.

Here, we extend the coarse-grained solvent-free approach followed in the simulation of lipid bilayers (108), to model the entire RBC membrane by introducing the two-component membrane model. Three types of particles are used to represent lipid molecules, actin junctions and band-3 complexes. The actin junction particles along with a number of band-3 particles are connected via the worm-like chain (WLC) potential to form a hexagonal network linked to the lipid bilayer. Additional band-3 particles that do not sense the WLC potential are employed to simulate the band-3 protein complexes that can diffuse freely in the lipid bilayer. An orientation-dependent potential, similar to the potentials introduced in (108), is employed between all particles to simulate the two-dimensional fluidic nature of the membrane.

## 2.2 Simulation method

*2.2.1 Model*

The model describes the RBC membrane as a two-component system, including the cytoskeleton and the lipid bilayer. Three types of particles are introduced (see Fig. 2.1B). The blue color particles represent a cluster of lipid molecules, which have a diameter of 5 nm, a similar value to the thickness of the lipid bilayer. The red particles signify actin junctions and they form a canonical hexagonal network representing the spectrin network. An actin junction has a diameter of approximately 35 nm and resides in the cytoplasm (122, 123). It is connected to the lipid bilayer via glycophorin which has a size comparable to thickness of the lipid bilayer. Thus, the fluidic behavior of the membrane is not affected by the size of the actin junctions but only by the size of the glycophorin. Therefore, the particles that represent the actin junctions in the model have a diameter of 5 nm, similar to the size of glycophorin. The yellow particles denote band-3 complexes with diameter of 7.5 nm (6) and they are connected to the spectrin network, while the green particles represent the free band-3 complexes that are not connected to the spectrin network. The masses of the particles are given in the section 2.3.

In the model, each particles carry both translational and rotational degrees of freedom $(\mathbf{x}_i, \mathbf{n}_i)$, where $\mathbf{x}_i$ and $\mathbf{n}_i$ are the position and the orientation (director vector) of particle $i$, respectively. The rotational degrees of freedom obey the normality condition $|\mathbf{n}_i|=1$. Thus, each particle effectively carries 5 degrees of freedom. We define $\mathbf{x}_{ij} = \mathbf{x}_j - \mathbf{x}_i$ as the distance vector between particles $i$ and $j$. Correspondingly, $r_{ij} \equiv |\mathbf{x}_{ij}|$ is the distance, and $\hat{\mathbf{x}}_{ij} = \mathbf{x}_{ij}/r_{ij}$ is a unit vector. The particles interact with one another via a pair-wise additive potential

$$u_{ij}(\mathbf{n}_i,\mathbf{n}_j,\mathbf{x}_{ij}) = u_R(r_{ij}) + A(\alpha, a(\mathbf{n}_i,\mathbf{n}_j,\mathbf{x}_{ij})) u_A(r_{ij}), \tag{2.1}$$

with

$$A(\alpha, a(\mathbf{n}_i,\mathbf{n}_j,\mathbf{x}_{ij})) = 1 + \alpha(a(\mathbf{n}_i,\mathbf{n}_j,\mathbf{x}_{ij}) - 1), \tag{2.2}$$

where $u_R(r_{ij})$ and $u_A(r_{ij})$ are the repulsive and attractive components of the pair potential, respectively. $\alpha$ is a tunable linear amplification factor. The energy well of this potential is essential to regulate the fluid-like behavior of the membrane, therefore we define a function $A(\alpha, a)$ to enable tuning the energy well. The effects of function $A(\alpha, a)$ in the potential will be discussed in the section 2.2. The interaction between two particles depends not only on their distance but also on their relative orientation via the function $a(\mathbf{n}_i,\mathbf{n}_j,\mathbf{x}_{ij})$ which varies from -1 to +1 and adjusts the attractive part of the potential. We specify that $a=1$ corresponds to the case when $\mathbf{n}_i$ is parallel to $\mathbf{n}_j$ and both are normal to vector $\mathbf{x}_{ij}$ $((\mathbf{n}_i \uparrow\uparrow \mathbf{n}_j) \perp \hat{\mathbf{x}}_{ij})$, and the value $a=-1$ to the case when $\mathbf{n}_i$ is anti-parallel to $\mathbf{n}_j$ and both are perpendicular to vector $\mathbf{x}_{ij}$ $((\mathbf{n}_i \uparrow\downarrow \mathbf{n}_j) \perp \hat{\mathbf{x}}_{ij})$. The former instance is energetically favored due to the maximum attractive interaction between particles $i$ and $j$, while the latter is energetically disfavored due to the maximum repulsive interaction. In essence, the energy difference between these two cases acts as the thermodynamic driving force responsible for the self-assembly of lipid bilayer. One simple form of $a(\mathbf{n}_i, \mathbf{n}_j, \mathbf{x}_{ij})$ that captures these characteristics is

$$a(\mathbf{n}_i, \mathbf{n}_j, \hat{\mathbf{x}}_{ij}) = (\mathbf{n}_i \times \hat{\mathbf{x}}_{ij}) \cdot (\mathbf{n}_j \times \hat{\mathbf{x}}_{ij}) = \mathbf{n}_i \cdot \mathbf{n}_j - (\mathbf{n}_i \cdot \hat{\mathbf{x}}_{ij})(\mathbf{n}_j \cdot \hat{\mathbf{x}}_{ij}). \tag{2.3}$$

The equations of motion include the translational and rotational components

$$m_i \ddot{\mathbf{x}}_i = -\frac{\partial(V)}{\partial \mathbf{x}_i}, \tag{2.4}$$

$$\tilde{m}_i \ddot{\mathbf{n}}_i = -\frac{\partial(V)}{\partial \mathbf{n}_i} + \left(\frac{\partial(V)}{\partial \mathbf{n}_i} \cdot \mathbf{n}_i\right)\mathbf{n}_i - \tilde{m}_i (\dot{\mathbf{n}}_i \cdot \dot{\mathbf{n}}_i)\mathbf{n}_i, \tag{2.5}$$

where $V = \sum_{j=1}^{N} u_{ij}$, $\tilde{m}_i$ is a pseudo-mass with units of energy $\times$ time$^2$, and the right-hand side (RHS) of Eq. (2.5) conforms to the normality constraint $|\mathbf{n}_i| = 1$.

*2.2.2 Pair potential*

To stabilize a fluid membrane, the lipid particles should be allowed to move past each other with relatively low resistance, while the overall cohesive integrity is preserved. Cooke et al. (92) discovered that the classical LJ6-12 potential, $u(r) = 4\varepsilon\left((d/r)^{12} - (d/r)^6\right)$, where $\varepsilon$ has the unit of energy and $d$ has the unit of length, possesses very strong repulsive core below the equilibrium distance $r < 2^{1/6}d$. This suppresses position exchange of particles and makes it difficult to form a liquid phase. As temperature $T$ is raised from absolute zero, the initial crystal phase sublimes to the vapor phase without going through the liquid phase. Compared to traditional particle systems, the vapor phase in the model with orientation disorder is thermodynamically more favorable because of the entropic contributions from its rotational degrees of freedom. To describe a fluid membrane by self-assembled moving particles, we employ the following repulsive and attractive potentials:

$$\begin{cases} u_R(r_{ij}) = \varepsilon\left((R_{cut}-r_{ij})/(R_{cut}-r_{min})\right)^8 & \text{for } r_{ij}<R_{cut} \\ u_A(r_{ij}) = -2\varepsilon\left((R_{cut}-r_{ij})/(R_{cut}-r_{min})\right)^4 & \text{for } r_{ij}<R_{cut} \\ u_R(r_{ij}) = u_A(r_{ij}) = 0, & \text{for } r_{ij} \geq R_{cut} \end{cases} \quad (2.6)$$

We chose $r_{min}^{l-l} = 2^{1/6}d$ for the pair potentials between lipid and lipid particles, so that the potential described in (2.6) has the same minimum energy and equilibrium distance as the LJ6-12 potential. The association energy between lipid and lipid particles is $\varepsilon$. While $r_{min}^{l-a}$ and $r_{min}^{a-a}$ in the pair potentials between lipid and actin particles, and between actin and actin particles are also $2^{1/6}d$, $r_{min}^{l-b}$, $r_{min}^{a-b}$ and $r_{min}^{b-b}$ in the pair potentials between lipid and band-3 particles, between actin and band-3 particles, between band-3 and band-3 particles are $1.25 \cdot 2^{1/6}d$, $1.25 \cdot 2^{1/6}d$ and $1.5 \cdot 2^{1/6}d$, respectively, as diameter of band-3 particles is 1.5 of diameter for the lipid and actin particles. The association energy between lipid and band-3 particles, between lipid and actin particles, as well as between actin and band-3 particles is chosen to be $1.4\varepsilon$. The Figs. 2.2A and 2.2B only show the potential between lipid particles. Another appealing characteristic of the above potential (hereon denoted by "Poly4-8"), is that the 1st, 2nd and 3rd derivatives of $u(r_{ij})$ with respect to $r_{ij}$ are all continuous at $r_{ij} = R_{cut}$.

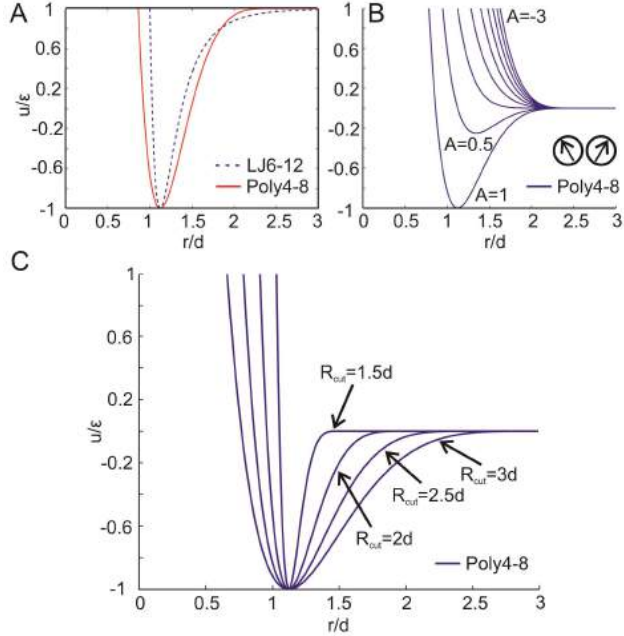

Figure 2.2 The pair-wise interaction potential between lipid and lipid particles. (A) Comparison with LJ6-12 potential. (B) The interaction potential profile as a function of $A(\alpha,a)$. (C) Increasing the parameter $R_{cut}$ widens the potential well, giving rise to a softer core of the inter-particle interaction $(A(\alpha,a)=+1)$. The soft core is essential for the stabilization of membrane at the fluid phase.

Fig. 2.2A shows that the effective potential in the case of $A(\alpha,a)=+1$ has a wider energy well than the LJ6-12 potential, which gives rise to a larger "maneuverability" near the equilibrium distance than that of the classical LJ6-12 potential. This characteristic facilitates particles squeezing past each other and stabilizes the liquid phase of the particle ensemble against the gas phase. The effect of the function $A(\alpha,a)$ on the potential is illustrated in Fig. 2.2B. As the function $A(\alpha,a)$ changes from +1 to negative values, the total potential shifts from attractive to repulsive, penalizing the non-parallel arrangement of neighboring particles. This property of the potential ensures the formation of a 2D membrane sheet rather than a dense 3D structure with higher coordination numbers. We emphasize that because the parameter $\alpha$ regulates the energy penalty paid by membrane particles when their relative orientation changes, it plays an important role in adjusting the bending rigidity of the membrane.

The restoring force near the equilibrium distance between same types or different types of CGMD particles can be tuned by adjusting the parameter $R_{cut}$. Fig. 2.2C shows that a decreasing $R_{cut}$ narrows the potential well between the lipid and lipid particles, which results in a rapid increase of the repulsive force. This increase in repulsive force, in turn, prevents the particles from passing each other and stabilizes them into a solid-like crystal phase. The influence of $R_{cut}$ on the stabilization of fluid membranes is a general feature of soft-core potentials independent of the precise functional form of the potentials (92).

A WLC potential (124, 125) is employed to simulate the spectrin network. The WLC forces between the connected actin and band-3 particles is described by

$$f_{WLC}(L) = -\frac{k_B T}{p}\left[\frac{1}{4(1-x)^2} - \frac{1}{4} + x\right], \tag{2.7}$$

where $x = L/L_{max} \in [0,1)$, $L_{max}$ = 100 nm is maximum or the contour length of the spectrin chain between an actin particle and a band-3 particle. $L$ is the instantaneous chain length or the instantaneous distance between an actin and the corresponding band-3 particles. $p$ = 10 nm is the persistent length of the spectrin filaments (126). $k_B$ is the Boltzmann constant and $T$ is the temperature. We used the WLC model to represent the RBC cytoskeleton without introducing bending rigidity to the spectrin network because the RBC membrane cytoskeleton is a highly flexible structure with bending rigidity $\kappa \sim 0.024 - 0.24 k_B T$ (127, 128), which is about two orders of magnitude smaller than the bending rigidity of the lipid bilayer $\kappa \sim 10 - 20\ k_B T$ (126, 129).

*2.2.3 Simulation details*

We perform (CGMD) simulations of a planar membrane to extract the elastic properties of the membrane. Periodic boundary conditions (PBCs) were imposed in the plane of the membrane (*x*- and *y*-directions), while the *z*-direction (perpendicular to the membrane surface) is completely free without any constraints. The system consists of $N = 35500$ particles, including 34600 lipid particles, 90 actin junctions particles and 810 band-3 complex particles out of which, 540 band-3 particles are not connected to the spectrin network. The dimension of the membrane is approximately 1 μm × 1 μm. The numerical integration of the equations of motion (Eqs. (2.4) and (2.5)) was performed using the Beeman algorithm (10, 130). The Nose-Hoover thermostat was employed to control the temperature. The model was implemented in the NAT ensemble (95, 131, 132). Since the model is solvent free and the membrane is a two-dimensional structure, we controlled the projected area instead of the volume. The projected area was adjusted to result in zero tension for both the lipid bilayer and the spectrin network at the equilibrium state. The time scale that guides the choice of the time step in the CGMD simulations is $t_s = (m_i d^2 / \varepsilon)^{1/2}$. The chosen time step is $0.01 t_s$. In addition, to match the time steps used for translational and rotational degrees of freedom, the pseudomass $\tilde{m}_i$ is chosen to be $\tilde{m}_i \sim m_i^l d^2$, where $m_i^l$ is the mass of the lipid particles. In section 3.1, we determined an appropriate timescale by comparing the in-plane diffusion coefficient obtained from the simulations with experimental values (133).

We use $\varepsilon$ as the energy unit, and $d$ as the unit length to quantify the geometrical and mechanical properties of lipid bilayer and membrane proteins. Given the diameter of the lipid particles $r_{min}^{l-l} = 2^{1/6} \cdot d = 5\text{nm}$, one finds $d \cong 4.45\text{nm}$. Each coarse-grained lipid particle carries a mass of $m^l = 95.6\text{kDa}$, which was obtained by averaging over those of various lipids that compose an area of approximately 20 nm$^2$ of the erythrocytic membrane (6). The band-3 complex consists of band-3 protein and glycophorin A, with molecular weights of 102 kDa and 14 kDa, respectively. It also includes Ankyrin, which connects the band-3 protein with the spectrin cytoskeleton, and protein 4.2 which binds with the band-3 and Ankyrin. Ankyrin and 4.2 proteins have molecular weights of 210 kDa and 72 kDa respectively (3). Therefore, the total mass of the band-3 complex is approximately 398 kDa, corresponding to $4m^l$. As mentioned above, the band-3 complex is approximated as a sphere with diameter of 7.5 nm, equivalent to $1.5 r_{min}^{l-l}$. An actin junction is composed of 14 kDa glycophorin C, short actin filaments (with a weight of 546 kDa), and 80 kDa protein 4.1, 200 kDa adducin, 41 kDa tropomodulin and a 28 kDa tropomyosin (3). Thus, the actin junction is approximated as a sphere with total weight of 1000 kDa and a diameter of 5 nm, corresponding to $10m^l$ and $r_{min}^{l-l}$. The maximum or the contour length of the spectrin chain between actin particles and band-3 particles is equivalent to $L_{max} = 20 r_{min}^{l-l}$. As it will be shown below, the system is found to be a fluid membrane with bending rigidity well within the experimentally established range at $k_B T/\varepsilon = 0.22$ and $R_{cut} = 2.6d$. The amplification factor $\alpha$ in Eq. (2.2) was chosen to be 1.55.

### 2.3. Results

*2.3.1. Self-Diffusion*

In this section, we show that the proposed model simulates a membrane with the appropriate mechanical and physical properties. Simulations are initiated from a flat triangular crystalline lattice that occupies the entire mid-plane of the super-cell. At zero temperature, the particles form a 2D solid phase with a long-range order. As the temperature increases, the membrane starts to fluctuate and is gradually equilibrated at a fluid phase at the temperature of $k_\mathrm{B}T/\varepsilon = 0.22$ (see Fig. 2.3A). To ensure that the model represents a fluid membrane, it is necessary to examine whether the lipid and band-3 particles can diffuse in a fluidic manner such that the particle's positions are exchangeable. Fig. 2.3A and 2.3B show the positions of the particles of a thermally equilibrated fluid membrane at two separate times. The tiles with different colors are used to differentiate the positions of the particles at the initial moment. After $5\times10^5$ time steps the colored particles are mixed due to diffusion, demonstrating the fluidic nature of the membrane.

We further verify the fluidic characteristics of the membrane with two other methods described below. First, we show in Fig. 2.3C that the mean square displacement (MSD) of lipid particles, defined by $\frac{1}{N}\left\langle \sum_{j=1}^{j=N}\left[\mathbf{x}_j(t) - \mathbf{x}_j(0)\right]^2 \right\rangle$ increases linearly with time, indicating that self-diffusion and the changes of the particle connectivity take place easily and that the membrane is essentially a fluid within the simulation time. Using the expression of the diffusion coefficient $D = \lim_{t \to \infty}(1/4t\, MSD)$, we obtained from Fig. 2.3C that $D = 2.7\times10^{-2}\, d^2/t_\mathrm{s}$, where $t_\mathrm{s}$ is the time scale. A typical value of the diffusion coefficient for the lipid of the phospholipid bilayer membranes is $D \approx 10^{-7}\mathrm{cm}^2/\mathrm{s}$ (133). By comparing with the real value, we can determine the time scale $t_s \approx 0.07\ \mu\mathrm{s}$. Second, we found by examining the radial distribution function at $R_\mathrm{cut} = 2.6d$ and $k_B T/\varepsilon = 0.22$ (see Fig. 2.3D) that the membrane does not possess a crystalline order at a distance larger than the equilibrium distance, further demonstrating the characteristic of the fluid membrane.

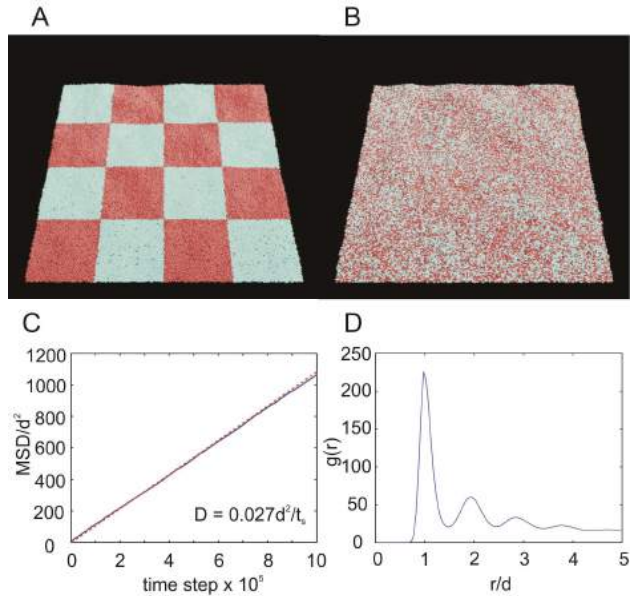

Figure 2.3. (A) Thermally equilibrated fluid membrane of N = 35500 particles at a reference time, representing a membrane with dimension of approximately 1 μm × 1 μm The tiles with different colors tiles are used to differentiate the positions of the particles at a reference time. (B) After $5\times10^5$ time steps, the particles are seen to be mixed due to diffusion, demonstrating the fluidic behavior of the model membrane. (C) Linear time dependence of MSD of the two-component membrane model. (D) Radial distribution function in the 2D fluid membrane embedded in 3D

*2.3.2. Membrane elasticity*

Within the framework of continuum mechanics, the free energy of a fluid membrane can be described by the classical Canham-Helfrich theory (69), as

$$F = \int \left[ \gamma + \frac{\kappa}{2}(C_1 + C_2 - C_0)^2 + \bar{\kappa} C_1 C_2 \right] dA, \tag{2.8}$$

where $C_1$ and $C_2$ are the principal curvatures, $C_0$ is the spontaneous curvature, $\gamma$ is the surface tension, $\kappa$ and $\bar{\kappa}$ are the bending and Gaussian rigidities (134), respectively. The first term on the right-hand side of Eq. (2.8) is the area-expansion energy, while the second and third terms represent the normal and Gaussian bending energies, respectively. Since the topology of the membrane remains unchanged, the Gaussian bending energy results in a constant contribution, and thus this energy term can be dropped. Surface tension is defined as $\gamma = -(\sigma_{11} + \sigma_{22}) L_z / 2$, where $\sigma_{11}$ and $\sigma_{22}$ are the components of the in-plane Virial stress (10) in a 3D periodic supercell calculation, and $L_z$ is height of the supercell, so that $\gamma$ has the unit of N/m. Since the spectrin network is described by the WLC model, which does not directly introduce resistance to bending, we expect that the Canham-Helfrich theory for lipid bilayers is still valid in the equilibrium. In what follows, we extract the bending rigidity from the power spectrum of the height thermal fluctuations.

Applying the equipartition theorem on the Helfrich free energy in the Monge representation (9, 69, 74, 126), one can express the power spectrum as

$$\left\langle |\tilde{h}(q)|^2 \right\rangle = \frac{K_B T}{l^2 (\gamma q^2 + \kappa q^4)}, \tag{2.9}$$

where $\tilde{h}(q)$ is the discrete Fourier transform of the out-of-plane displacement $h(\mathbf{r})$ of the membrane, defined as

$$\tilde{h}(q) = \frac{l}{L}\sum_n h(\mathbf{r})e^{i\mathbf{q}\cdot\mathbf{r}}, \tag{2.10}$$

with $L$ being the lateral size of the supercell, $l = r_{min}^{l-l} = 2^{1/6}d$ sets the mesh size equal to size of smallest particles, and $q$ is the norm of the wave vector $\mathbf{q} = (q_x, q_y)$, i.e., $(q_x, q_y) = 2\pi(n_x, n_y)/L$.

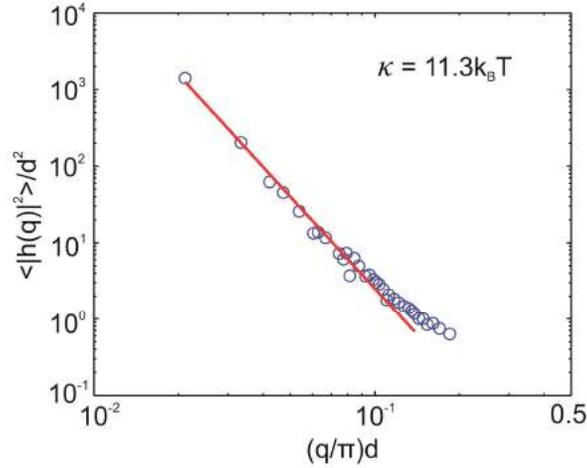

Figure 2.4. Vertical displacement fluctuation spectrum of two-component membrane as a function of the wave number $q$.

To compute the bending rigidity, we first constructed an equilibrated membrane at $k_B T/\varepsilon = 0.22$ and with zero tension. This effectively excludes the membrane tension effect, and consequently, one expects that the power spectrum exhibits only a $q^{-4}$ dependence on the wave vector. Then, we calculated $|\tilde{h}(q)|^2$ from the raw data of each recorded configuration and averaged them over all the available configurations to evaluate $\langle |\tilde{h}(q)|^2 \rangle$. Finally, by fitting the numerical data to Eq.(2.9) for $\gamma = 0$, via a minimization procedure based on a nonlinear regression scheme, one obtains the bending rigidity $\kappa = 11.3 k_\text{B} T$ of the membrane model (see Fig. 2.4). The numerical result lies within the experimental range of $(10 k_B T - 20 k_B T)$ for lipid bilayer (126, 129).

*2.3.3 Phase diagram and the effects of the interaction potential parameters on membrane properties*

The phase diagram of the membrane model based on the potential parameters $R_\text{cut}$ and $T$ is plotted. The dependence of the diffusion coefficient $D$ on the temperature $T$ and $R_\text{cut}$ as well as the dependence of the bending rigidity on the parameter $\alpha$ are investigated to study the effects of the parameters of the interaction potential on the behavior of the membrane model.

First, we determined the conditions under which the model behaves as a stable fluid membrane. The main parameters which govern the behavior of the model are $R_\text{cut}$ and $T$. We constructed a two-dimensional phase-diagram, as shown in Fig. 2.5A, where simulations are performed to determine the diffusivity of the membrane model. A similar analysis has been conducted by Cooke et al.(135). The phase diagram is based on a system consisted of N = 35500 coarse-grained particles. At low temperature and for small $R_\text{cut}$, the model behaves as a gel with very low diffusivity. At large values of $R_\text{cut}$, the model is unstable independently of the temperature. As the temperature increases, the range of $R_\text{cut}/d$ within which the membrane behaves as stable fluid widens. The gel-fluid boundary is identified by a sudden increase in the diffusion coefficient and the disappearance of the crystalline order at a distance larger than the equilibrium distance. The transition from fluid to unstable state (gas phase) is characterized by the observation that the tensionless membrane starts to break apart (135).

Second, the dependence of the diffusion coefficient $D$ on the temperature $T$ for $R_{cut}$= 2.6$d$ and 2.8$d$, and $α$ = 1.55 is studied. Measurements of $D$ are limited in the fluid phase of the membrane. As shown in Fig.2.5B, $D$ increases monotonically with $T$ for both cases of $R_{cut}$= 2.6$d$ and 2.8$d$, as higher temperature leads to larger kinetic energy. On the other hand, an increase in $R_{cut}$ from 2.6$d$ to 2.8$d$ broadens the interaction potential well between the particles, as shown in Fig. 2.2C, resulting in a decrease in the repulsive force. This decrease in repulsive force makes the particles pass each other with lower resistance and thus contribute to an increase in the diffusion coefficient. In addition, we also studied the dependence of the bending rigidity on the parameter $α$ at three temperatures. $R_{cut}$ is chosen to be 2.6$d$. The main role of $α$ in the interaction potential is to regulate the energy well when the two interacting particles are disoriented from their favorable relative orientation (see Fig. 2.2B) and therefore is closely related to the bending rigidity of the membrane model. As shown in the Fig. 2.5C, the bending rigidities $κ$ increases monotonically with $α$ for three different temperatures. We note that $κ$ increases faster when the temperature is lower. As $α$ increases, the function $A(α,a)$ decreases and the total potential shifts from attractive to repulsive, imposing greater penalty for the non-parallel arrangement of neighboring particles and forcing the membrane to become a flat 2D sheet. It is also illustrated in the Fig.2.5C that the temperature affects the bending rigidity of the membrane. The bending rigidity drops significantly when the temperature increases because a higher temperature increases the equilibrium distance between the membrane particles.

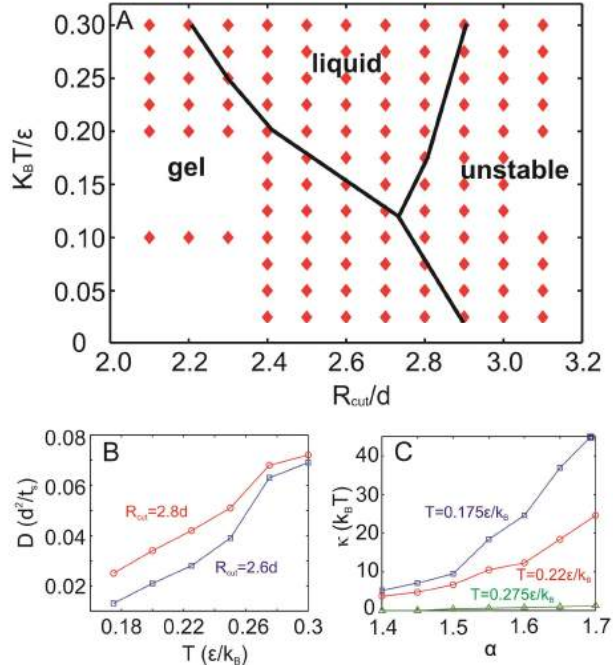

Figure 2.5. (A) Phase diagram. The hand-drawn lines show the phase boundaries. (B) Dependence of the diffusion coefficient $D$ on the temperature $T$ for $R_{cut}$= 2.6$d$ and 2.8$d$, respectively. (C) Dependence of the bending rigidity $\kappa$ on the parameter $\alpha$ at temperature $T$ = 0.175, 0.22 and 0.275, respectively.

## 2.4. Application

In this section, the developed two-component membrane model is applied in shearing experiments where the shear moduli of the membrane are measured. The numerical results are compared to experimentally obtained values. We also examine the effect of the reduction of the spectrin network connectivity on the shear modulus.

As discussed above, phospholipid bilayers behave as 2D fluid-like structures and thus they resist bending but cannot sustain static in-plane shear stress because the lipids and the proteins can diffuse freely within the membrane. The resistance of the membrane under low shearing strain rate is due only to the cytoskeleton. At the high shear strain rate, however, the viscosity of the lipid bilayer also contributes to the total resistance to the deformation. The membrane is sheared up to a shear strain of 1 at a temperature of $k_B T/\varepsilon = 0.22$ (see Fig. 2.6 A and B) at the two strain rates of $0.001 d/t_s$ and $0.01 d/t_s$, respectively. The response of the membrane to shearing at the low strain rate of $0.001 d/t_s$ is illustrated by the blue line in Fig. 2.6C. At this strain rate, the shear viscosity of the lipid bilayer is not detectable and the total resistance is due only to the spectrin network. This becomes apparent from the fact that the red line in Fig. 2.6C, which describes the shear stress – shear strain response of the corresponding WLC network, follows the pattern of the blue line, which describes the shear stress-shear strain response of the entire membrane. It is worth noting that the WLC model captures the experimentally identified stiffening behavior of the RBC membrane (136) and the behavior of a detailed spring-based model of cytoskeleton network (109). When the strain rate is increased to $0.01 d/t_s$, the viscosity of the lipid bilayer contributes to the total shear resistance (black line in Fig. 2.6C). As a result the initial value of the shear stress is not zero but 8 µN/m, which is equal to the shear stress measured during the shear deformation of a one-component lipid bilayer model.

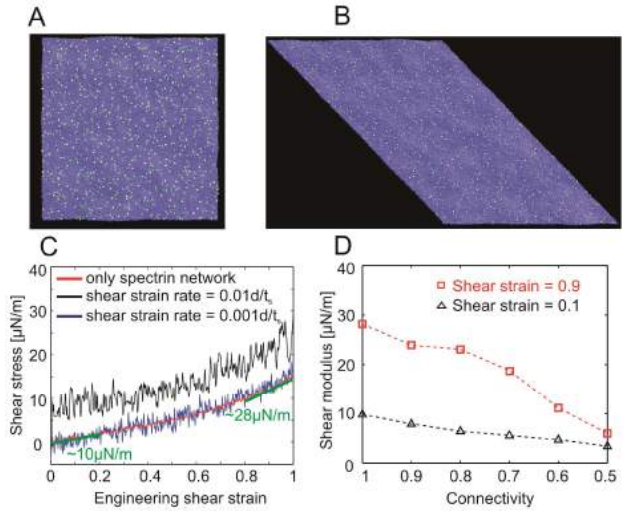

Figure 2.6. (A) Thermally equilibrated fluid membrane at a temperature of $k_BT/\varepsilon=0.22$. (B) Sheared membrane at a temperature of $k_BT/\varepsilon=0.22$ at engineering shear strain of 1. (C) Shear stress-strain response of the membrane at two strain rates. Red curve represents the response of the "bare" spectrin network. The blue and black curves signify the shear stress obtained for the strain rates of $0.001d/t_s$ and $0.01d/t_s$, respectively. (D) Shear moduli of the membrane versus spectrin network connectivity at the engineering shear strains of 0.1 and 0.9, respectively.

Up to this point, the connectivity of the spectrin network is maintained at 100%, meaning that each actin junction is connected with its 6 neighboring junctions through the WLC potential forming a perfect 2D six-fold structure. However, there are both experimental evidence (137-140) and sound theoretical studies (141, 142) showing that the connectivity of the spectrin network in the RBC membrane is not 100% and that the network is not a perfect 2D six-fold structure. Electron microscopy images demonstrate that while most of actin junctions are connected with other 6 actin junctions, a small amount of actins junctions are connected with 5 or 7 actin junctions (137, 138). Subsequent atomic force microscopy (AFM) results showed a lower connectivity of cytoskeleton is about 3 and 4 links per actin junction, leading to square, pentagon and hexagon-like structures in the network (139, 143, 144). A CGMD model of the RBC cytoskeleton predicted that the biconcave shape of RBCs can be attained only if the network undergoes constant remodeling to relax the in-plane shear elastic energy to zero during the deformation meaning that the connectivity of the spectrin network is changeable (142). Recent experimental results presented the evidence that the remodeling of the spectrin network plays a vital role in determining the cell deformation and biconcave shape of the RBCs (140).

As shown in Fig.2.6C, when the spectrin network connectivity is 100%, the shear modulus of the membrane at small deformations is approximately 10 μN/m while it is 28 μN/m at 90% engineering shear strain. The linear elastic shear modulus of the model is larger than the experimentally measured values of 7.3 μN/m (136) and 4 - 9 μN/m (126) as well as the simulation results of 8.3 μN/m (142). This is probably because the connectivity of the employed spectrin network is 100% while, as it was mentioned above, the actual spectrin network is not perfect. Since the membrane resistance to the shearing results mainly from the spectrin network, it is expected that the shear modulus of the membrane will decrease as the connectivity of the spectrin network is reduced. To investigate the effect of the spectrin network connectivity on the mechanical properties of the membrane model, the shear modulus of the membrane is measured when the connectivity of the spectrin network is reduced from 100% to 50%. The resulting shear moduli at small and large deformations are shown in Fig. 2.6D, which clearly demonstrated that their values drop dramatically with decreased spectrin network connectivity especially at large deformations. The shear moduli of the membrane fall within the experimentally measured value range when the network connectivity is between 60% and 90%. RBC with lower spectrin network connectivity is softer and it is easier to deform when passing through narrow blood vessels. However, the low connectivity could lead to insufficient surface shear resistance to maintain the cell's biconcave shape and integrity during the blood circulation. Our simulation results show that when the connectivity drops below 50%, the membrane model does not resist to shear deformation indicating that the RBCs completely lose the ability of recovering its biconcave shape after deformation (109).

## 2.5. Summary

We have developed a two-component CGMD model for RBC membranes. In this model, the lipid bilayer and the cytoskeleton are considered as an ensemble of discrete particles that interact through a direction-dependent pair potential. Three types of coarse-grained particles are introduced, representing a cluster of lipid molecules, actin junctions and band-3 complexes. The large grain size extends the accessible length and time scales of the simulations to ~ μm and ~ ms. By tailoring only a few parameters of the inter-grain interaction potential, the model can be stabilized into a fluid phase manifested by the free diffusion of the particles and the reproduction of the essential thermodynamic properties of the RBC membrane. An important feature of the proposed model is the combination of the spectrin network with the lipid bilayer, which represents a more compete representation of the RBC membrane compared to one-component CGMD model and to most of the continuum models. This model allows us to study the behavior of the membrane under shearing. The behavior of the model under shearing at different strain rates illustrates that at low strain rates up to $0.001 d/t_s$, the developed shear stress is mainly due to the spectrin network and it shows the characteristic non-linear behavior of entropic networks, while the viscosity of the fluid-like lipid bilayer contributes to the resulted shear stress at higher strain rates. Decrease of the spectrin network connectivity results to significant decrease of the shear modulus of the membrane, which demonstrates that the cytoskeleton carries most of the load applied on the cell while the lipid bilayer only functions as 2D fluid. The values of the shear moduli measured from the membrane with reduced spectrin network connectivity are in a good agreement with previous experimental, theoretical, and numerical results.

# Chapter 3.

# Erythrocyte Membrane Model with Explicit Description of the Lipid Bilayer and the Spectrin Network

**Abstract**

The membrane of the red blood cell (RBC) consists of spectrin tetramers connected at actin junctional complexes, forming a 2D six-fold triangular network anchored to the lipid bilayer. Better understanding of the erythrocyte mechanics in hereditary blood disorders such as spherocytosis, elliptocytosis, and especially, in sickle cell disease requires the development of a detailed membrane model. Here, we introduce a mesoscale implicit-solvent coarse-grained molecular dynamics (CGMD) model of the erythrocyte membrane which explicitly describes the phospholipid bilayer and the cytoskeleton, by extending a previously developed two-component RBC membrane model. We show that the proposed model represents RBC membrane with the appropriate bending stiffness and shear modulus. The timescale and self-consistency of the model are established by comparing our results to experimentally measured viscosity and thermal fluctuations of the RBC membrane. Furthermore, we measure the pressure exerted by the cytoskeleton on the lipid bilayer. We find that defects at the anchoring points of the cytoskeleton to the lipid bilayer (as in spherocytes) cause a reduction in the pressure compared to an intact membrane, while defects in the dimer-dimer association of a spectrin filament (as in elliptocytes) cause an even larger decrease in the pressure. We conjecture that this finding may explain why the experimentally measured diffusion coefficients of band-3 proteins are higher in elliptocytes than in spherocytes, and higher than in normal RBCs. Finally, we study the effects that possible attractive forces between the spectrin filaments and the lipid bilayer on the pressure applied on the lipid bilayer by the filaments. We discover that the attractive forces cause an increase in the pressure while they diminish the effect of membrane protein defects. As this finding contradicts with experimental results, we conclude that the attractive forces are moderate and do not impose a complete attachment of the filaments to the lipid bilayer.

## 3.1. Introduction

The human red blood cell (RBC) repeatedly undergoes large elastic deformations when passing through narrow blood vessels. The large flexibility of the RBCs is primarily due to the cell membrane, as there are no organelles and filaments inside the cell. The RBC membrane is essentially a two-dimensional (2D) structure, comprised of a cytoskeleton and a lipid bilayer, tethered together. The lipid bilayer includes various types of phospholipids, sphingolipids, cholesterol, and integral membrane proteins, such as band-3 and glycophorin (see Fig. 3.1A). It resists bending but cannot sustain in-plane static shear stress as the lipids and the proteins diffuse within the lipid bilayer at equilibrium. The RBC membrane cytoskeleton is a 2D six-fold structure consisting of spectrin tetramers, which are connected at the actin junctional complexes. The cytoskeleton is tethered to the lipid bilayer via "immobile" band-3 proteins at the spectrin-ankyrin binding sites and via glycophorin at the actin junctional complexes (see Fig. 3.1A). Although the mechanical properties and biological functions of the RBC membrane have been well studied in the past decades, the interactions between the lipid bilayer and cytoskeleton as well as the interactions between the cytoskeleton and transmembrane proteins are not yet fully understood. The cytoskeleton plays a major role in the integrity of the RBC membrane, as is evident in blood disorders where defects in membrane proteins lead to membrane loss and reduced mechanical robustness of the RBC (3, 25, 145). In hereditary spherocytosis (HS), the tethering of the cytoskeleton to the lipid bilayer (vertical interaction in Fig. 3.1A) is partially disrupted resulting in membrane loss and subsequently in the spherical shape of the RBCs. In hereditary elliptocytosis (HE), the cytoskeleton is disrupted at α-β spectrin linkages or at spectrin-actin-4.1R junctional complexes (horizontal interactions in Fig. 3.1A) (3, 25, 145, 146). This partial disruption of the cortex diminishes the ability of the RBC to recover its biconcave shape after undergoing large deformations. In addition, because the spectrin filaments act as barriers, restricting the lateral diffusion of the "mobile" band-3 proteins, defects in the vertical and horizontal interactions between the cytoskeleton and the lipid bilayer modify the regular diffusion of band-3 proteins (115, 147-150).

Several approaches have been followed for the mathematical description and modeling of the RBC membrane. At one end of the spectrum, there are the continuum membrane models based on elasticity theory (9, 69-74, 151, 152). At the other end, atomistic simulations mainly study the behavior of lipids in the lipid bilayer (76-79). However, it is very challenging for continuum models to account for the detailed structure and defects in the RBC cytoskeleton, while it is not feasible for atomistic methods to simulate a representative sample of the RBC membrane including lipids, membrane proteins, and the cytoskeleton. Because of these limitations, particle-based mesoscale models were introduced to study the biomechanical behavior of the RBC membrane. These models fall into two main groups. In one group, the membrane is modeled as a 2D canonical hexagonal network of particles where the immediate neighbors are connected via a worm-like chain (WLC) potential that represents an actin filament. The bending rigidity induced by the lipid bilayer is represented by a bending potential applied between two triangles with a common side (80, 81, 110, 141, 142). In the other category, the lipid bilayer is simulated by coarse-grained methods where each lipid molecule is coarse-grained into several connected beads (84, 86, 88, 91, 92, 95). The lipids are forced to assemble the bilayer by additional solvent particles, in the case of explicit solvent models, or by an additional potential, in the case of solvent-free models. In the explicit solvent models, hydrophobic interactions are employed between the lipids and solvent particles to represent effects of the water molecules (84, 87, 94, 102-105). In the implicit solvent models, the effect of the solvent is taken into account by employing orientation-dependent interaction potentials between the lipid particles (83, 91, 92, 96, 101, 107). At a higher level of coarse-graining, a group of lipid molecules are coarse-grained into one bead (108, 111). Drouffe et al. (83) simulated biological membranes by introducing a one-particle-thick, solvent-free, coarse-grained model, in which the inter-particle interaction is described by a Lennard-Jones (LJ) type pair potential depending not only on the distance between the particles but also on their directionality. Noguchi and Gompper (98) developed a one-particle thick, solvent-free, lipid bilayer model by introducing a multibody potential that eliminated the need of the rotational degree of freedom. Yuan et al. (108) introduced a similar approach, but instead of the LJ potential, a soft-core potential was used to better represent the particle self-diffusion. An overview of particle-based models for the RBC membrane can be found in recent reviews (82, 96, 100, 153).

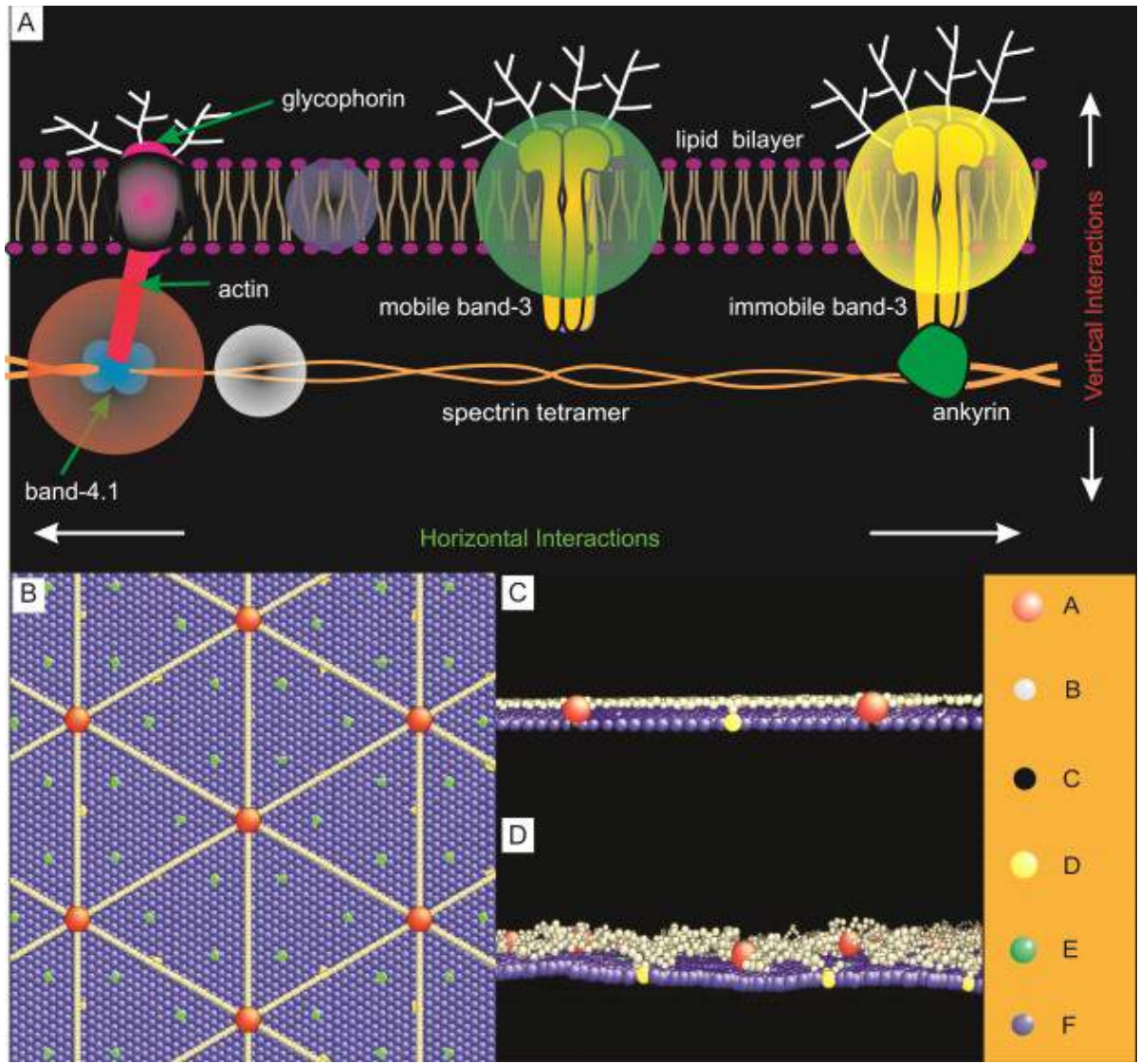

Figure 3.1 (A) Schematic of the human RBC membrane. The blue sphere represents a lipid particle and the red sphere signifies an actin junctional complex. The grey sphere represents a spectrin particle and the black sphere represents a glycophorin particle. The yellow and green circles correspond to a band-3 complex connected to the spectrin network and a mobile band-3 complex respectively. A mesoscale detailed membrane model. (B) top view of the initial configuration, (C) side view of the initial configuration, and (D) a side view of the equilibrium configuration. "A" type particles represent actin junctional complexes, "B" type particles

represent spectrin proteins, "C" type particles represent glycophorin proteins, "D" type particles represent a band-3 complex that are connected to the spectrin network ("immobile" band-3), "E" type particles represent band-3 complex that are not connected to the network ("mobile" band-3), and "F" type particles represent lipid particles.

Absence of explicit representation of either the lipid bilayer or the cytoskeleton in the aforementioned particle-based models limits their applications in the study of the interactions between the cytoskeleton and the lipid bilayer, and between the cytoskeleton and diffusing membrane proteins. Recently, a model consisting of two layers of 2D triangulated networks, where one layer represents the cytoskeleton while the other one represents the lipid bilayer, was introduced (154). This approach is computationally very efficient and it can be used in simulations of blood flow. However, it is not capable of modeling the interactions between the spectrin filaments and the lipid bilayer, which is frequently essential, as exemplified in the study of the diffusion of membrane proteins such as the band-3 (115, 147-150). Li and Lykotrafitis (111) introduced a two-component RBC membrane model where the lipid bilayer consists of CG particles of 5 nm size while the cytoskeleton is comprised of particles that represent actin junctional complexes and form a canonical hexagonal network. These "actin junction" particles are connected by the WLC potential representing a spectrin filament. Due to the implicit representation of the spectrin filaments, this approach does not consider the interactions between the filaments and the lipid bilayer as well as between the filaments and membrane protein during diffusion. Auth et. al (155) developed an analytical model to study the cytoskeleton and lipid bilayer interactions. This analytical model is able to estimate the pressure exerted on the lipid bilayer by the cytoskeleton because of the thermal fluctuations of the spectrin filaments. The computed pressure is then used to obtain the deformation of the lipid bilayer induced by the spectrin filaments, which was calculated to be 0.1 nm ~ 1 nm. This deformation is small compared to the local deformation (~15 nm) of the lipid bilayer caused by the lateral compression applied by the spectrin cytoskeleton (151, 156).

In this chapter, we introduce a two-component implicit-solvent CGMD RBC membrane model comprised of the lipid bilayer and the RBC cytoskeleton by extending a previously developed RBC membrane model (111). The key feature of this new model is the explicit representation of the cytoskeleton and the lipid bilayer by CG particles. We determine the parameters of the model by matching the bending stiffness and shear modulus of the membrane model with existing experimental results (129, 136, 157). Then, we extract the timescale of our simulations by measuring the viscosity of the membrane model and comparing it with the experimentally obtained values. The timescale is later confirmed by measuring the thermal fluctuation frequency of the lipid bilayer and the spectrin filaments. In addition, we measure the pressure exerted by the spectrin filaments on the lipid bilayer. These results are compared with analytically estimated pressure values (118, 155). Finally, we investigate the effects of the attraction between spectrin filaments and the lipid bilayer on the pressure applied by the cytoskeleton on the lipid bilayer in the case of the normal RBC membrane and for membranes with defective proteins.

## 3.2. Model and simulation method

The proposed model describes the RBC membrane as a two-component system, comprised of the cytoskeleton and the lipid bilayer. We first introduce the cytoskeleton, which consists of spectrin filaments connected at the actin junctional complexes forming a hexagonal network. The actin junctional complexes are represented by red particles (see Fig. 3.1B-D) that have a diameter of approximately 15 nm and are connected to the lipid bilayer via glycophorin. Spectrin is a protein tetramer formed by head-to-head association of two identical heterodimers. Each heterodimer consists of an α-chain with 22 triple-helical segments and a β-chain with 17 triple-helical segments (6). In the proposed model, the spectrin is represented by 39 spectrin particles (grey particles in Fig. 3.1B-D) connected by unbreakable springs. The spring potential, $u_{cy}^{s-s}(r) = k_0(r - r_{eq}^{s-s})^2/2$, is plotted as the green curve in Fig. 3.2, with equilibrium distance between the spectrin particles $r_{eq}^{s-s} = L_{max}/39$, where $L_{max}$ is the contour length of the spectrin (~200 nm) and thus $r_{eq}^{s-s} \cong 5$ nm. The spectrin chain is linked to the band-3 particles (yellow particles) at the area where the α-chain and the β-chain are connected. The two ends of the spectrin chains are connected to the actin junctional complexes via the spring potential $u_{cy}^{a-s}(r) = k_0(r - r_{eq}^{a-s})^2/2$, where the equilibrium distance between an actin and a spectrin particle is $r_{eq}^{a-s} = 10$ nm. The spring constant $k_0$ will be determined below. Spectrin particles that are not connected by the spring potential interact with each other via the repulsive part of the L-J potential

$$u_{rep}(r_{ij}) = \begin{cases} 4\varepsilon\left[\left(\dfrac{\sigma}{r_{ij}}\right)^{12} - \left(\dfrac{\sigma}{r_{ij}}\right)^{6}\right] + \varepsilon & r_{ij} < R_{cut,LJ} = r_{eq}^{s-s} \\ 0 & r_{ij} > R_{cut,LJ} = r_{eq}^{s-s} \end{cases} \quad (3.1)$$

where $\varepsilon$ is the energy unit and $\sigma$ is the length unit. $r_{ij}$ is the distance between spectrin particles. The cutoff distance of the potential $R_{cut,LJ}$ is chosen to be the equilibrium distance $r_{eq}^{s-s}$ between two spectrin particles. The potential is plotted as the black curve in Fig. 3.2. The spring constant $k_0 = 57\ \varepsilon/\sigma^2$ is chosen to be identical to the curvature of $u_{LJ}(r_{ij}) = 4\varepsilon\left[\left(\sigma/r_{ij}\right)^{12} - \left(\sigma/r_{ij}\right)^{6}\right] + \varepsilon$ at the energy well bottom to reduce the number of free parameters. The cytoskeleton introduced above is directly connected to the lipid bilayer via the band-3 particles D and the glycophorin particles C (see Fig. 3.1C). In addition, we have investigated the effect of attractive interactions of various strengths between spectrin filaments and the lipid bilayer on the diffusion of band-3 particles. Previous studies on lipid–spectrin filaments interactions have suggested that spectrin binds to the negatively charged lipid surfaces with association constants of 2-10×10$^6$ M$^{-1}$ (158-163), while the association constants of spectrin-ankyrin, ankyrin-band-3, spectrin-protein 4.1-actin are 2×10$^7$ M$^{-1}$, 2×10$^8$ M$^{-1}$ and 2×10$^{12}$ M$^{-2}$, respectively (164, 165). For simplicity, we applied the attractive part LJ potential between spectrin and lipid particles

$$u_{att}(r_{ij}) = \begin{cases} 4n\varepsilon\left[\left(\dfrac{\sigma}{r_{ij}}\right)^{12} - \left(\dfrac{\sigma}{r_{ij}}\right)^{6}\right] + n\varepsilon & r_{ij} > r_{eq}^{l-s} \\ 0 & r_{ij} < r_{eq}^{l-s} \end{cases}, \qquad (3.2)$$

where $n$ is a parameter used to tune the attractive energy between the spectrin filaments and lipid bilayer. $r_{eq}^{l-s} = 5$ nm is the equilibrium distance between spectrin particles and lipid particles. Our simulation results show that the cytoskeleton is completely attached to the lipid bilayer when n ≥ 0.2. Because the cytoskeleton of the RBC membrane is a highly flexible structure with a bending rigidity of $\kappa \sim 0.024 - 0.24 k_B T$ (127, 128), which is about two orders of magnitude smaller than the bending rigidity of the lipid bilayer $\kappa \sim 10 - 20\ k_B T$ (129, 157) as well as the bending rigidity of the RBC membrane κ ~ 10 - 50 $k_BT$ (157), we did not consider bending rigidity for the spectrin network in this model.

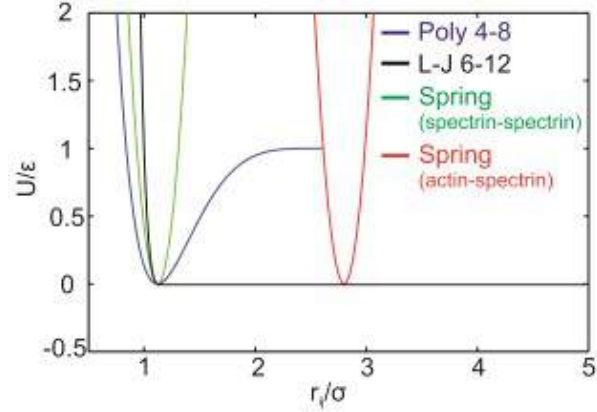

Figure 3.2 The interaction potentials employed in the membrane model. The blue curve represents the pair-wise potential between lipid particles. The green curve represents the spring potential between spectrin particles. The red curve represents the spring potential between actin and spectrin particles. The black curve represents the repulsive L-J potential between the lipid and spectrin particles.

Three types of CG particles are introduced to represent the lipid bilayer and band-3 proteins (see Fig. 3.1B-D). The blue color particles denote a cluster of lipid molecules. Their diameter of 5 nm is approximately equal to the thickness of the lipid bilayer. The black particles represent glycophorin proteins with the same diameter as the lipid particles. The band-3 protein consists of two domains: (i) the cytoplasmic domain of band-3 with a dimension of 7.5×5.5×4.5nm that contains the binding sites for the cytoskeletal proteins, and (ii) the membrane domain, with a dimension of 6×11×8 nm, whose main function is to mediate anion transport (166, 167). We represent the membrane domain of band-3 by a spherical CG particle with a radius of 5 nm. The volume of the particle is similar to the excluded volume of the membrane domain of a band-3. However, when band-3 proteins interact with the cytoskeleton, the effect of the cytoplasmic domain has to be taken into account and thus the effective radius is approximately 12.5 nm. One third of band-3 particles, which are connected to the spectrin network, are depicted as yellow particles (see Fig. 3.1B). The rest of the band-3 particles, which are free to diffuse in the lipid bilayer, simulate the mobile band-3 proteins and they are shown as green particles (see Fig. 3.1B). The CG particles, which form the lipid bilayer and transmembrane proteins, carry both translational and rotational degrees of freedom ($\mathbf{x}_i$, $\mathbf{n}_i$), where $\mathbf{x}_i$ and $\mathbf{n}_i$ are the position and the orientation (direction vector) of particle $i$, respectively. The rotational degrees of freedom obey the normality condition $|\mathbf{n}_i| = 1$. Thus, each particle effectively carries 5 degrees of freedom. $\mathbf{x}_{ij} = \mathbf{x}_j - \mathbf{x}_i$ is defined as the distance vector between particles $i$ and $j$. $r_{ij} \equiv |\mathbf{x}_{ij}|$ and $\hat{\mathbf{x}}_{ij} = \mathbf{x}_{ij}/r_{ij}$ are the distance and the unit vector respectively. The particles, forming the lipid membrane and membrane proteins, interact with one another via the pair-wise potential

$$u_{mem}(\mathbf{n}_i, \mathbf{n}_j, \mathbf{x}_{ij}) = u_R(r_{ij}) + A(\alpha, a(\mathbf{n}_i, \mathbf{n}_j, \mathbf{x}_{ij}))u_A(r_{ij}), \tag{3.3}$$

with

$$u_{4-8}(r_{ij}) = \begin{cases} u_R(r_{ij}) = k\varepsilon\left((R_{cut,mem} - r_{ij})/(R_{cut,mem} - r_{eq})\right)^8 - k\varepsilon & \text{for } r_{ij} < R_{cut,mem} \\ u_A(r_{ij}) = -2k\varepsilon\left((R_{cut,mem} - r_{ij})/(R_{cut,mem} - r_{eq})\right)^4 - k\varepsilon & \text{for } r_{ij} < R_{cut,mem} \\ u_R(r_{ij}) = u_A(r_{ij}) = 0, & \text{for } r_{ij} \geq R_{cut,mem} \end{cases} \tag{3.4}$$

where $u_R(r_{ij})$ and $u_A(r_{ij})$ are the repulsive and attractive components of the pair potential, respectively. $\alpha$ is a tunable linear amplification factor. The function $A(\alpha,a(\mathbf{n}_i,\mathbf{n}_j,\mathbf{x}_{ij}))= 1+\alpha(a(\mathbf{n}_i,\mathbf{n}_j,\mathbf{x}_{ij})-1)$ tunes the energy well of the potential, through which the fluid-like behavior of the membrane is regulated. In the simulations, $\alpha$ is chosen to be 1.55 and the cutoff distance of the potential $R_{cut,mem}$ is chosen to be $2.6\sigma$. The parameters $\alpha$ and $R_{cut,mem}$ are selected to maintain the fluid phase of the lipid bilayer. Detailed information about the selection of the potential parameters can be found from author's previous work (111). $k$ is selected to be 1.2 for the interactions among the lipid particles and $k = 2.8$ for interactions between the lipid and the protein particles, such as glycophorin and band-3. Fig. 3.2 shows only the potential between lipid particles (blue curve). The interactions between the cytoskeleton and the lipid bilayer are represented by the repulsive part of the L-J potential as shown in Eq. (3.1). The cutoff distance of the potential $R_{cut,LJ}$ is adjusted to be the equilibrium distances between different pairs of CG particles. The equilibrium distance $r_{eq}^{a-l}$ between the actin particles and the lipid particles is 10 nm while $r_{eq}^{a-b}$ between the actin particles and the band-3 particles is 20 nm. The equilibrium distance $r_{eq}^{l-s}$ between the spectrin particles and the lipid particles is 5 nm, while the equilibrium distance $r_{eq}^{b-s}$ between the spectrin particles and the band-3 particles is 15 nm.

The equation of translational motion for all the CG particles is

$$m_i \ddot{\mathbf{x}}_i = -\frac{\partial(V)}{\partial \mathbf{x}_i}, \qquad (3.5)$$

where for the CG particles forming the lipid membrane and the transmembrane proteins $V = \sum_{j=1}^{N}(u_{\text{mem},ij} + u_{\text{LJ},ij})$, and for the CG particles comprising the cytoskeleton, $V = \sum_{j=1}^{N}(u_{\text{LJ},ij}) + u_{\text{cy},i}$.

The equation of rotational motion for the CG particles forming the lipid bilayer and proteins in the lipid bilayer is

$$\tilde{m}_i \ddot{\mathbf{n}}_i = -\frac{\partial(\sum_{j=1}^{N} u_{\text{mem},ij})}{\partial \mathbf{n}_i} + \left(\frac{\partial(\sum_{j=1}^{N} u_{\text{mem},ij})}{\partial \mathbf{n}_i} \cdot \mathbf{n}_i\right)\mathbf{n}_i - \tilde{m}_i(\dot{\mathbf{n}}_i \cdot \dot{\mathbf{n}}_i)\mathbf{n}_i, \qquad (3.6)$$

where $\tilde{m}_i$ is a pseudo-mass with units of energy $\times$ time$^2$, and the right-hand side of Eq. (3.6) obeys the normality constraint $|\mathbf{n}_i| = 1$. The pseudomass $\tilde{m}_i$ is chosen to be $\tilde{m}_i \sim m_i^l \sigma^2$, where $m_i^l$ is the mass of the lipid particles.

The system consists of $N$ = 29567 CG particles. The dimension of the membrane is approximately 0.8 μm × 0.8 μm. The numerical integrations of the equations of motion (Eqs. (3.5) and (3.6)) are performed using the Beeman algorithm (10, 130). The temperature of the system is maintained at $k_B T/\varepsilon$ = 0.22 by employing the Nose-Hoover thermostat. The model is implemented in the NAT ensemble (95, 131, 132). Periodic boundary conditions are applied in all three directions. Since the model is solvent-free and the membrane is a two-dimensional structure, we controlled the projected area instead of the volume. The projected area is adjusted to result in zero tension for the entire system at the equilibrium state. Given the diameter of the lipid particles $r_{\text{eq}}^{l-l} = 2^{1/6} \cdot \sigma = 5\text{nm}$, the length unit $\sigma$ is calculated to be $\sigma$ = 4.45 nm. The timescale that guides the choice of the timestep in the MD simulations is $t_s = (m_i d^2/\varepsilon)^{1/2}$. The timestep of the simulation is selected to be $\Delta t = 0.01t_s$. Since the CG particles used in the simulations do not correspond to real molecules, the employed timescale does not have an immediate correlation with the real system. The timescale in our simulation is established by measuring the viscosity of the membrane model and comparing it with the experimentally measured value for the RBC membrane. For an independent confirmation, we also obtain the timescale by measuring the thermal fluctuation frequencies of the spectrin filaments and the lipid membrane. The details of the timescale will be discussed in the section 3.4. For simplicity, we assume that the cytoskeleton possesses a perfect two-dimensional (2D) six-fold triangular structure with a fixed connectivity and that each spectrin filament is anchored to the band-3 proteins at its mid-point. In reality, the RBC cytoskeleton contains numerous defects while the band-3-ankyrin connections and the cytoskeleton undergo dynamic remodeling (109, 140, 168).

## 3.3. Membrane properties

### 3.3.1 Membrane fluidity

In this section, we show that the proposed model of the RBC membrane reproduces the appropriate mechanical properties. To ensure that our system represents the lipid bilayer, it is necessary to examine whether the CG particles can diffuse in a fluidic manner. Figs. 3.3A and B show the positions of the CG particles of a thermally equilibrated fluid membrane at two separate times. The tiles with different colors are used to differentiate the positions of the particles at the initial moment. After $1 \times 10^6$ time steps, the colored particles are mixed due to diffusion, demonstrating the fluidic nature of the membrane. We further verify the fluidic characteristics of the membrane by first showing that the mean squared displacement (MSD), defined by $\frac{1}{N}\left\langle \sum_{j=1}^{j=N} [\mathbf{x}_j(t) - \mathbf{x}_j(0)]^2 \right\rangle$, increases linearly with time (see Fig. 3.3C), and second by showing that that the correlations between the CG lipid particles are lost beyond a few particle diameters, suggesting a typical fluidic behavior of the membrane model at $k_BT/\varepsilon = 0.22$ and $R_{cut,mem}=2.6\sigma$ (see Fig. 3.3D).

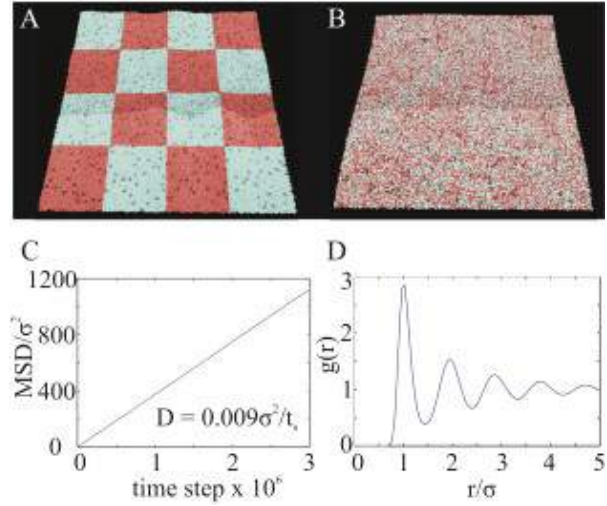

Figure 3.3 (A) Thermally equilibrated fluid membrane of N = 29567 particles at a reference time, representing a membrane with dimension of approximately 0.8 μm × 0.8 μm The tiles with different colors are used to differentiate the positions of the particles at a reference time. (B) After $1\times10^6$ time steps, the particles are mixed due to diffusion, demonstrating the fluidic behavior of the membrane model. (C) Linear time dependence of the mean square displacement (MSD) of the two-component membrane model. (D) Radial distribution function of the 2D fluid membrane embedded in 3D.

*3.3.2 Measurements of membrane bending rigidity, shear modulus and viscosity*

Here, we compute the bending rigidity, shear modulus, and viscosity of the membrane model. We obtain the bending rigidity of the membrane by measuring the fluctuation spectrum of the membrane at zero tension and then fitting the data to the expression of the Helfrich free energy in the Monge representation (9, 69, 74, 157),

$$\left\langle \left|\tilde{h}(q)\right|^2 \right\rangle = \frac{K_B T}{l^2\left(\gamma q^2 + \kappa q^4\right)} \tag{3.7}$$

where $\tilde{h}(q)$ is the discrete Fourier transform of the out-of-plane displacement $h(\mathbf{r})$ of the membrane, defined as

$$\tilde{h}(q) = \frac{l}{L}\sum_n h(\mathbf{r})e^{i\mathbf{q}\cdot\mathbf{r}}, \tag{3.8}$$

where $L$ is the lateral size of the supercell, $l = r_{eq}^{l-l} = 2^{1/6}\sigma$ sets the mesh size equal to size of lipid particles, and $q$ is the wave vector $\mathbf{q} = (q_x, q_y)$, i.e., $(q_x, q_y) = 2\pi(n_x, n_y)/L$. The power spectrum in the Fig. 3.4 exhibits a $q^{-4}$ dependence on the wave vector. The deviation at large wave vectors is due to limitation defined by the size of the CG particles, as at wave lengths smaller than the particle size, the continuum approximation breaks down. The bending rigidity of the membrane is found to be $\kappa = 11.3 k_B T$ (see Fig. 3.4), which lies within the experimental range of ($10 k_B T - 20 k_B T$) for lipid bilayer (129, 157).

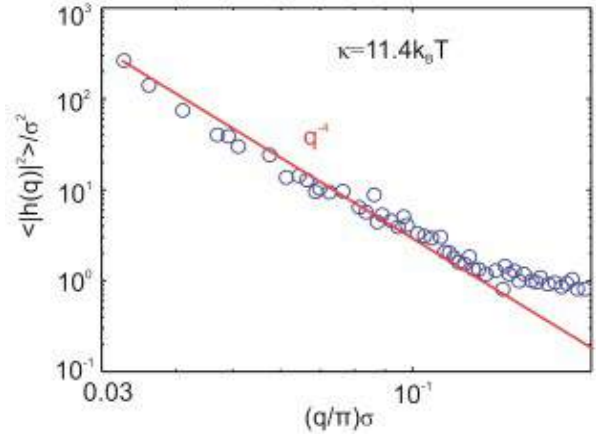

Figure 3.4 Vertical displacement fluctuation spectrum of membrane model as a function of the dimensionless quantity $q\sigma/\pi$, where $q$ is the wave number and $\sigma$ is the unit length corresponding to ~4.45 nm.

For the measurements of the shear moduli, the membrane is sheared up to a shear strain of 1 with the strain rate $\dot{\gamma}=0.001\sigma/t_s$, as shown in Figs. 3.5A and B. The response of the membrane to shearing is illustrated by the blue line in Fig. 3.5C. The proposed model captures the experimentally identified stiffening behavior of the RBC membrane, which is due to the spectrin network. The shear modulus of the membrane at small deformations is approximately 12 µN/m while it is increased to 27 µN/m at engineering shear strain of 0.9. The initial elastic shear modulus of the model is larger than the experimentally measured values of 4 - 9 µN/m (136, 157). This is most likely due to the implementation of a perfect network in the present model, while the spectrin network of the RBC membrane is not perfect (137, 138, 141, 142) probably because of ATP-induced dissociations (151, 155). Experimental measurements showed that ATP-induced dissociations played a crucial role in forming defects in the cytoskeleton. In addition, a larger shear modulus was measured in ATP-depleted RBCs compared to normal RBCs (169). At the strain rate of $\dot{\gamma}=0.001\sigma/t_s$, the shear viscosity of the lipid bilayer is not detectable and the total resistance to the shearing is caused only by the cytoskeleton. This is confirmed by the fact that the shear stress-strain response of the cytoskeleton (red curve in Fig. 3.5C) follows the pattern of the shear stress-strain response of the entire membrane (blue curve in Fig. 3.5C). This result is in agreement with our previous two-component membrane model where the spectrin network is represented implicitly (111). At the higher strain rates of $0.005\sigma/t_s$ and to $0.01\sigma/t_s$, the viscosity of the lipid bilayer contributes to the total resistance during shearing. For example, the total value of the measured shear stress corresponding to $0.01\sigma/t_s$ strain rate (green curve in Fig. 3.5C) is the sum of the shear stress due to the cytoskeleton (purple curve in Fig. 3.5C) and of the shear stress due to the viscosity of the lipid bilayer. By subtracting the shear stress due to the cytoskeleton from the total shear stress, we obtain the viscous shear stress to be $\tau = 0.008 \ \varepsilon/\sigma^2$. At the strain rate of $\dot{\gamma} = 0.005\sigma/t_s$, we computed the viscous shear stress to be $\tau = 0.004 \ \varepsilon/\sigma^2$. We assume that during the shearing process, the homogeneous lipid bilayer behaves as a simple two-dimensional viscous Newtonian fluid. Then, by applying the shear stress - strain rate relation $\tau = \mu\dot{\gamma}$, we calculate the shear viscosity of the lipid bilayer to be $\mu = 0.8\varepsilon t_s L/\sigma^3$, where $L$ is the length of the membrane model. The shear viscosity of the lipid bilayer can also be measured from shearing the membrane without the cytoskeleton. At the same shear strain rate, the measured shear stress remains constant value with respect to different shear strains. At the strain rate of $\dot{\gamma} = 0.001\sigma/t_s$, no shear stress is measured from the lipid bilayer. At the strain rate of $\dot{\gamma} = 0.005\sigma/t_s$, the shear stress is measured to be $\tau = 0.004 \ \varepsilon/\sigma^2$. When the strain rate is further increased to $\dot{\gamma} = 0.01\sigma/t_s$, the shear stress is measured to be $\tau = 0.008 \ \varepsilon/\sigma^2$. Therefore, the obtained shear viscosity of the lipid bilayer is consistent with the value we measured from shearing the lipid bilayer with the cytoskeleton. The comparison between the numerical value predicted by the model and the experimental value, is performed in the section 3.4 where we determine the timescale $t_s$ for our simulations.

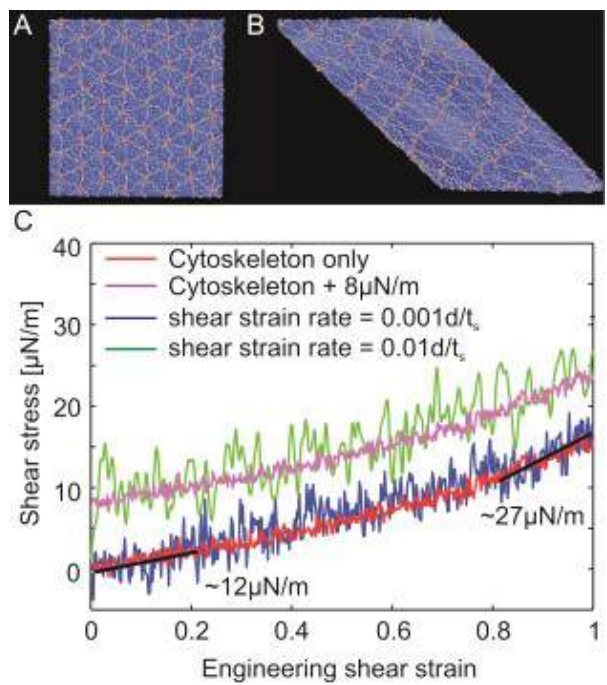

Figure 3.5 (A) Thermally equilibrated fluid membrane at a temperature of $k_BT/\varepsilon = 0.22$. (B) Sheared thermally equilibrated membrane at engineering shear strain of 1. (C) Shear stress-strain response of the membrane at two different strain rates. The red curve represents the response of the "bare" spectrin network. The blue and green curves signify the shear stress obtained at the strain rates of $0.001\sigma/t_s$ and $0.01\sigma/t_s$, respectively. The purple curve represents the shear stress measured from the "bare" spectrin network plus the 8μN/m attributed to viscosity.

*3.3.3 Measurements of fluctuation frequencies of the spectrin filaments and the lipid bilayer*

The fluctuation frequency of the spectrin filaments in the cytoskeleton is measured by tracking the normal displacement with respect to the lipid bilayer of a single spectrin particle in the middle of the two binding sites. Then, we apply Fast Fourier Transform (FFT) on the displacements and compute the power spectral density (PSD) of the vibrations. The corresponding results in Fig. 3.6A show that the dominant fluctuation frequency of the filaments is measured to be approximately $f_c = 0.009/t_s$. Regarding the fluctuation of the lipid bilayer, we measure the average fluctuation displacement of the lipid bilayer patch in one triangular compartment formed by the spectrin filaments. The frequency is then obtained by using the PSD plot and it is found to be $f_l = 0.001/t_s$, which is approximately one order of magnitude smaller than the fluctuation frequency of the filaments. Next, in a different simulation, we measure the fluctuation frequencies of the filaments in the "bare" cytoskeleton, meaning that the lipid bilayer is completely removed from the RBC membrane leaving only the cytoskeleton. Since the cytoskeleton applies compression on lipid bilayer, the lipid bilayer is crumpled at equilibrium and it exerts extension force to the cytoskeleton. Therefore, when the lipid bilayer is removed, an effective repulsive potential is applied between two neighboring actin junctions in the cytoskeleton to represent the effect of the lipid bilayer. The effective repulsive potential $u_{A-A}(r)$ is described by

$$u_{A-A}(r) = \alpha \frac{4\pi \kappa_{lipid}(R_{flat} - r)}{3 R_{flat}} H(R_{flat} - r) \tag{3.9}$$

where $R_{flat}$ is the distance between the two neighboring action junctions when the membrane becomes flat. We take $R_{flat} = 1.2\ R_{eqA\text{-}A}$, where $R_{eqA\text{-}A}$ is the equilibrium distance between the action junctions. $R_{eqA\text{-}A}$ is approximately 90 nm in our simulations. $\kappa_{lipid}$ is the bending stiffness of the lipid bilayer. $H(x)$ is the Heaviside step function. $\alpha$ is chosen to be 0.36 so that the pressure of the "bare" cytoskeleton is nearly zero at equilibrium temperature. More details about the "bare" cytoskeleton can be found in a previous work by one of the authors (109). We found that the frequency of the "bare" spectrin network is $f_c = 0.001/t_s$ (see Fig. 3.6B), which is one order of magnitude smaller than the fluctuation frequency of a spectrin filament in the cytoskeleton that is connected to the lipid bilayer. The reason for this difference is that in the "bare" spectrin network, the spectrin filaments are connected only to the actin junctional complexes, while in the case of the cytoskeleton attached to the lipid bilayer, the spectrin filaments are also anchored to the band-3 proteins. This means that in the "bare" network each filament is connected only at its two ends while in the proposed membrane model each filament has an additional binding site in the middle causing faster vibrations. The result is in agreement with the theoretical prediction that the conformation time $t_c$ of a spectrin filament depends on the number of parts per filament divided by the band-3 binding sites, $N_b$, $\left(t_c \sim 1/N_b^3\right)$ (170). In the proposed membrane model, each filament comprises two parts ($N_b = 2$), as there is one band-3 binding site per filament. Therefore, the conformation time for the filaments in the RBC membrane is about 8 times smaller than the conformation time for the "bare" network in which each filament has only one part ($N_b = 1$).

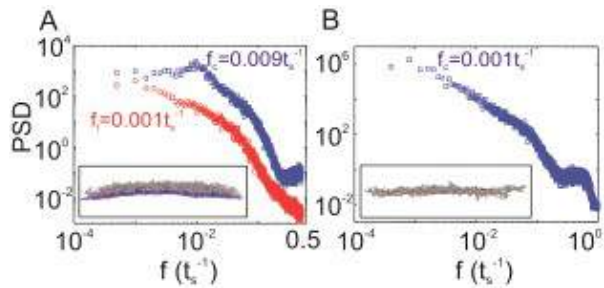

Figure 3.6 (A) Power spectrum density (PSD) corresponding to thermal fluctuations of the membrane model. The data marked as red circles are generated from the average displacement of the particles belonging to a triangular compartment of the lipid bilayer. The data marked as blue squares are generated from a particle positioned in the middle of a spectrin filament. $f_c$ represents the thermal fluctuation frequency of the cytoskeleton and $f_l$ represents the thermal fluctuation

frequency of the lipid bilayer. Inset: equilibrium state of the proposed RBC membrane model. (B) PSD corresponding to thermal fluctuations of a particle positioned in the middle of a spectrin filament in the "bare" cytoskeleton. Inset: equilibrium state of "bare" cytoskeleton.

### 3.3.4 Timescales of the simulations

As we described previously, since the particles introduced in CGMD simulations do not correspond to real atoms or molecules, the timescale $t_s = (m_i \sigma^2/\varepsilon)^{1/2}$ defined in the MD simulation is not directly related to the evolution of a real physical system. Only through comparison with a physical process can such a correspondence be established. Here, we compare the shear viscosity obtained in the section 3.2, with the experimentally measured value of $3.6 \times 10^{-7}$ N·s/m (171), via which we conclude that the timescale of our simulation is approximately $t_s \sim 3 \times 10^{-6}$s. The corresponding timestep of the simulations is then $\Delta t = 0.01 t_s \approx 3 \times 10^{-8} s$. After determining the timescale by employing the viscosity of the RBC membrane, we need to test the self-consistency of the model by confirming that its predictions for the fluctuation frequencies of the spectrin filaments and of the lipid bilayer are in agreement with the expected physical values. The thermal fluctuation frequency of the lipid membrane obtained from our simulation is $f_l = 0.001/t_s \sim 333$Hz, which is comparable with the theoretical and experimentally measured frequency ~1000Hz (151, 169). Analytical estimation shows that the conformation time $\tau_{conform}$ for the spectrin filaments in the cytoskeleton is $\tau_{conform} = \eta_{cytoplasm} d_{ee}^3 / (N_b^3 k_B T) \sim 100 \mu s$, where $d_{ee}$ is the end to end distance of the spectrin filaments, $N_b$ is the number of parts per filament divided by the band-3 binding sites and $\eta_{cytoplasm}$ is the viscosity of the cytoplasm (170). In our model, the thermal fluctuation frequency of the spectrin filament is measured to be $f_c = 0.009/t_s$. By using the timescale obtained above, we determine that the model predicts the conformation time for the spectrin filaments to be $\tau_{conform} \sim 330 \mu s$, which is comparable with the analytically estimated conformation time $\sim 100 \mu s$ (170).

### 3.4. Mechanical interaction between the spectrin network and the lipid bilayer.

The lipid bilayer and the spectrin network along with their interactions play an essential role in the structure and biological functions of the RBC. Experimental measurement of the pressure applied on the lipid bilayer by spectrin filaments is challenging, as the length scale of the interactions between the cytoskeleton and lipid is not easily accessed by dynamic cell mechanics experiments. In this section, we compute the pressure exerted on the lipid bilayer by the spectrin filaments and investigate how it is affected by the attraction between the spectrin filament and the lipid bilayer. The pressure obtained from the simulations is compared with analytically estimated values for the case of flexible linear polymer attached to the lipid bilayer at its two ends (118, 155). Lastly, we examine how defects in the anchoring of the spectrin network to the lipid bilayer and defects in the structure of the spectrin network influence the pressure applied on the lipid bilayer.

In the proposed model, the interaction between the spectrin filaments and the lipid bilayer is described by the repulsive part of the LJ-12 potential, as shown in Eq. (3.1). The filament-induced local pressure is calculated by using the expression proposed by Cheung and Yip (172), who consider the pressure as summation of the momentum flux and force across a planar unit area in a time interval. Because in our model the spectrin filaments do not penetrate the lipid bilayer and the local deformation of the lipid bilayer caused by the spectrin filaments is ~1 nm (155), the pressure is measured by considering only the forces between the lipid bilayer and the spectrin filaments. In the RBC membrane without defects, the two ends of the spectrin filaments are connected at the actin junctional complexes, which are anchored to the lipid bilayer by glycophorin proteins. The spectrin filaments are additionally anchored to the lipid bilayer via band-3 proteins at the spectrin-ankyrin binding sites. The blue curve in Fig. 3.7A shows that the pressure increases at the area close to band-3 binding site, where the spectrin particles are connected to the lipid bilayer. The pressure is large there because of the continuous interactions between the spectrin particles and the lipid bilayer. When moving away from the band-3 binding site, the pressure drops rapidly as the frequency of interactions between the spectrin particles and the lipid bilayer reduces. The pressure measured from the simulation is close to the analytical pressure distribution introduced in (155),

$$P(\rho_1,\rho_2,\rho) = \frac{k_B T}{4\pi R_g^2} e^{(\rho_1-\rho_2)^2/(4R_g^2)} e^{-(|\rho_1-\rho|+|\rho_2-\rho|)^2/(4R_g^2)} \frac{|\rho_1-\rho|+|\rho_2-\rho|}{|\rho_1-\rho|^3|\rho_2-\rho|^3}$$
$$\times [|\rho_1-\rho||\rho_2-\rho|((|\rho_1-\rho|+|\rho_2-\rho|)^2 - 6R_g^2) + 2R_g^2(|\rho_1-\rho|+|\rho_2-\rho|)^2]$$ (3.10)

where $R_g$ is the radius of gyration of the spectrin filaments, $\rho$ is the pressure measurement location, $\rho_1$ and $\rho_2$ are the anchoring locations of the filament to the lipid bilayer. Assuming that the measurement points always remain on the line connecting the two anchoring locations, the pressure profile based on Eq. (3.10) is plotted as the purple curve in Fig. 3.7A. It is noted that the pressures in the analytical estimation are significantly higher than the simulation measurements at the two ends of the filaments. The reason is that in the analytical calculations, the filaments are assumed to be directly attached to the lipid bilayer, while in our numerical simulation and in the actual RBC membrane, they are connected to the actin junctional complexes which then bind to the lipid bilayer through the glycophorin proteins.

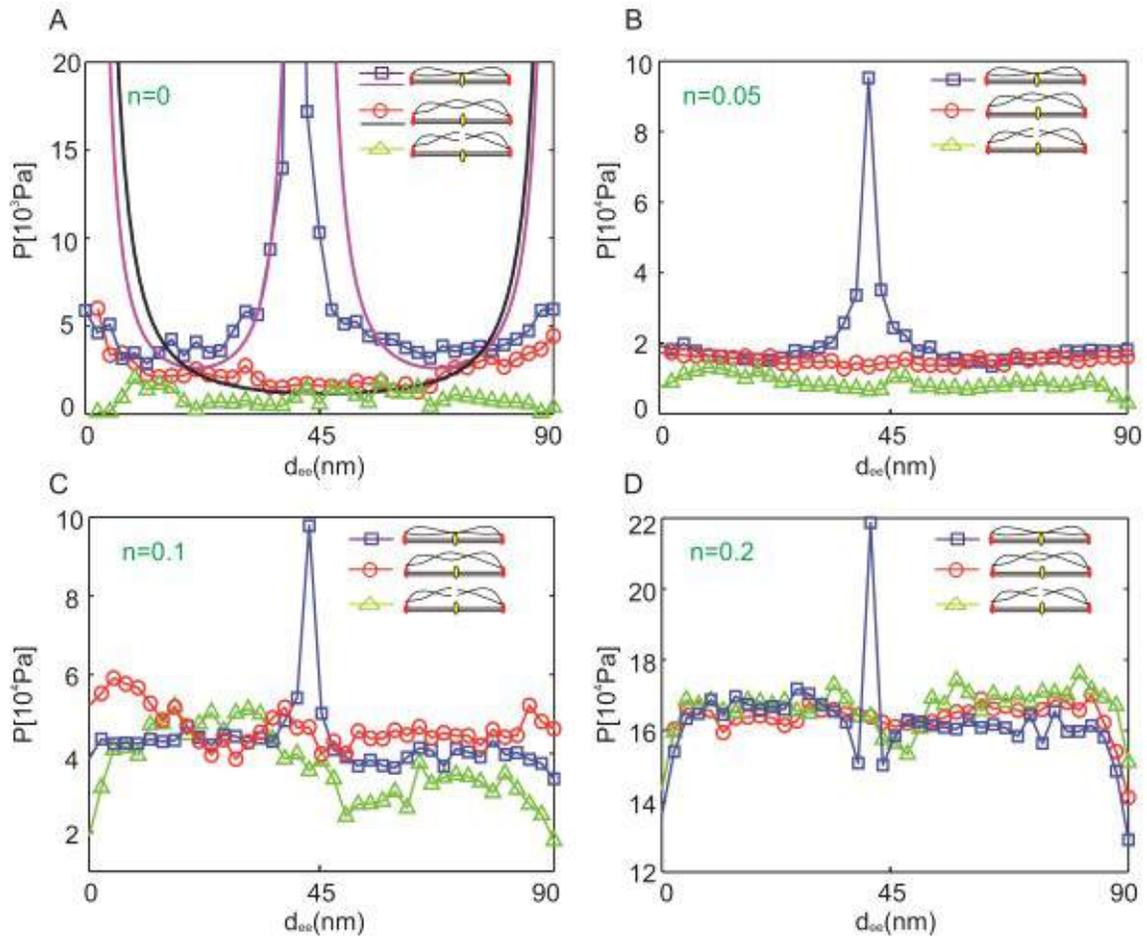

Figure 3.7 The pressure distribution exerted by the spectrin filaments on the lipid bilayer with attraction parameters (A) n = 0 (B) n = 0.05 (C) n= 0.1 (D) n = 0.2. $d_{ee}$ is the end to end distance of the spectrin filaments. The blue curve represents the pressure distribution measured from the membrane model. The purple curve represents the pressure distribution obtained from the analytical estimation for a normal membrane (118). The red curve represents the pressure distribution applied on the membrane with ankyrin protein defects. The black curve represents the pressure distribution obtained analytically for a membrane with ankyrin protein defects (118). The green curve represents the pressure distribution measured from the membrane with spectrin protein defects.

Next, we investigate the effect that the introduced attractive forces between the spectrin filament and the lipid bilayer have on the pressure applied to the lipid bilayer by the spectrin filaments. The parameter *n* in the Eq. (3.2) is selected to be 0.05, 0.1 and 0.2, as the spectrin-lipid interaction is relatively weak compared to the two primary connections between the cytoskeleton and lipid bilayer (158-165). We chose 0.2 as the maximum value of *n* because at this value the spectrin filaments are completely attached to the lipid bilayer. The pressure plots corresponding to *n* = 0.05 and 0.1 (blue curves in the Fig. 3.7B and C) show that the pressure profiles are similar to the profile obtained when *n* = 0, but the magnitudes of the pressure increase as *n* increases. The increased attractive forces reduce the average distances between the filaments and the lipid bilayer, resulting in more frequent interactions between the spectrin particles and the lipid bilayer, and subsequently in a larger pressure on the lipid bilayer. When *n* = 0.2, the attractive forces are large enough to cause attachment of the entire filament to the lipid bilayer, and thus generate a dramatic increase in the pressure (Fig. 3.7D). We note that the pressure is lower at the two ends of the filament because these points are connected to actin junctional complexes and not directly to the membrane. Previous experimental studies (115, 147-150, 173) and analytical modeling (116-118, 174-177) have shown that the pressure fence induced by the cytoskeleton hinders the lateral diffusion of mobile band-3 proteins and lipid molecules. Therefore, it is reasonable to predict that an increase in the attractive forces between the filament and lipid bilayer causes a reduction in the lateral diffusion of the band-3 proteins and lipid molecules.

Another question is how disruption of the connections between spectrin filaments and immobile band-3 proteins (vertical interactions in Fig. 3.1A) modifies the pressure field applied on the lipid bilayer by the spectrin network. This is relevant in HS where defects in proteins such as ankyrin and band-3 proteins, which play an important role in the coupling between the lipid bilayer and the cytoskeleton, result in a partial detachment between the network and the lipid bilayer. The red curve in the Fig. 3.7A shows that when only repulsive interaction between the lipids and the spectrin filaments is assumed ($n = 0$), the pressure is higher at the two ends of the filaments, compared to the middle portion of the filaments. This reduction is due to a decrease in the frequency of interactions between the middle sections of the filaments and the lipid bilayer. In particular, the pressure profile is consistent with the analytically estimated pressure distribution defined by Eq. (3.10) (see black curve in Fig. 3.7A) with the exception that the analytical pressure is much higher at the two ends of the filament. This is due to the assumption in the theoretical analysis that the two ends are directly anchored to the lipid bilayer while in our simulations they are connected to the junctional complexes. Therefore, we expect that a disruption of the connection between the filaments and the lipid bilayer at the band-3 anchoring points weakens the pressure fence. This causes an increase in the probability for band-3 particles to cross the boundaries of the filament-formed compartment, thus it enhances the diffusion of band-3 proteins in HS (112, 115, 118, 178). The introduction of small attractive forces between the lipid bilayer and the spectrin network ($n = 0.05$ and $n = 0.1$), when the connection between the filament and the lipid at the band-3 site is severed, results in a constant pressure along the filament. We note that the pressure is higher than in the case of only repulsive interactions between the lipid bilayer and the network. When the attractive force is large ($n = 0.2$), and the entire filament is in contact with the surface of the lipid bilayer, the pressure is similar with the pressure field developed in the normal membrane with the exception of the middle segment of the filament. In conclusion, the disruption of the anchoring of the spectrin filaments to the lipid bilayer results in a lower and nearly constant pressure along the filament. Since the spectrin filament-induced pressure acts as a pressure fence, hindering the lateral movement of mobile band-3 proteins and lipid molecules, it is predicted that the lower pressure facilitates the diffusion of lipids and band-3 proteins across the spectrin filaments. This results in a higher mobile band-3 diffusion coefficient compared to the band-3 diffusion coefficient in the normal RBC membrane.

Finally, we explore how the dissociation of a spectrin filament to two filaments affects the pressure applied to the membrane. Here, we aim to simulate defective spectrin dimer-dimer interactions that most commonly happen in HE (145). The pressure applied from both filaments on the membrane is shown as green curves in Fig. 3.7A-D for different attractive forces between the filaments and the membrane. We observe that when no attractive forces ($n = 0$) or small attractive forces ($n = 0.05$) are assumed, the pressure is lower than in the normal membrane or in a membrane with defects in the vertical interactions. It is reasonable to assume that a weaker pressure fence on the lipid bilayer causes an increase in the probability of band-3 particles to cross the spectrin filaments, justifying the observed higher diffusion coefficients of band-3 proteins in HE (112, 115, 118, 178). We also note that our simulations show that the pressure applied on the lipid bilayer in the case of dimer-dimer disruption, which corresponds to HE, is lower than the pressure measured in the case of disruption of vertical interactions, which corresponds to HS. Based on this result, we conjecture that our model predicts greater band-3 diffusion in HE than in HS, in agreement with experimental observations (112). Of course direct study of band-3 diffusion is required to validate these predictions. When the attractive forces are large ($n = 0.2$), the difference between the pressure measured in the normal membrane and in the membrane with protein defects becomes negligible with the exception of the area close to the band-3 binding point in the middle where the pressure is large in the normal membrane. This means that if the attractive force between the spectrin network and the lipid bilayer are large, the diffusion coefficients of band-3 should have been similar in normal RBCs and in RBCs from patients with HS or HE. However, since the band-3 diffusion coefficients measured in normal RBCs are smaller than the ones measured in RBCs in HS and HE (112), we conjecture that the attractive forces between the spectrin filaments and the lipid bilayer should be low and the parameter $n$ that defines the strength of the attractive forces cannot be larger than $n = 0.1$, which justifies our selection of $n$ in Section 2.

## 3.5. Summary

We introduce a particle-based model for the erythrocyte membrane that accounts for the most important structural components of the membrane, including the lipid bilayer, the spectrin network, and the proteins that play an important role in the anchoring of the spectrin cortex to the lipid bilayer, as well as the band-3 proteins. In particular, five types of CG particles are used to represent actin junctional complexes, spectrin, glycophorin, immobile band-3 protein, mobile band-3 protein and an aggregation of lipids. We first demonstrate that the model captures the fluidic behavior of the lipid bilayer and then that it reproduces the expected mechanical material properties of bending rigidity and shear modulus of the RBC membrane. The timescale of our simulations, which is found to be $t_s \sim 3\times10^{-6}s$, is inferred by comparing the viscosity of the membrane model to experimentally measured values. Then, the self-consistency of the model with respect to the timescale is tested by comparing the computed vibration frequency of the spectrin filaments and lipid membrane to analytically obtained values. We confirm that vibration frequency of the spectrin filaments and lipid membrane measured from the proposed membrane model, are also in agreement with experimental values. At last, we study the interactions between the cytoskeleton and the lipid bilayer, and measure the pressure applied on the membrane by the spectrin filaments. We also investigate how disruption of the connection between the spectrin network and the lipid bilayers, which simulates defects in HS, and rupture of the dimer-dimer association, which simulates defects in HE, affect the pressure exerted on the lipid bilayer. We show that overall the introduction of defects in the spectrin network or in the vertical connection between the lipid bilayer and the cytoskeleton results in lower pressure, which is consistent with prediction in (155). In addition, we find that the defects related to HE have a stronger effect than the defects related to HS. This result implies that diffusion of band-3 proteins in RBCs from patients with HS and HE is enhanced compared to the normal RBCs. Moreover, elliptocytes exhibit more prominent diffusion of band-3 proteins than spherocytes. Both conclusions are supported by experimental results. The level of attraction forces between the lipid bilayer and the membrane cytoskeleton is another important parameter that regulates the pressure applied by a spectrin filament to the lipid bilayer. We show that as the attractive force increases, it causes an overall increase in the pressure and it diminishes the differences in the pressure generated by membrane protein defects and, consequently, the differences in the diffusion of band-3 proteins in normal and defective erythrocytes. Since this finding is not supported by experimental results, we conjecture that the attractive force between the lipid bilayer and the spectrin filaments should be low, resulting in a membrane model where the filaments are not completely attached to the lipid bilayer. A detailed study of the band-3 diffusion in this model and direct comparison with experimental results is necessary in order to form a more accurate picture of how the model regulates diffusion. Because of the explicit representation of the lipid bilayer and the cytoskeleton, the proposed model can be potentially used in the investigation of a variety of membrane related problems in RBCs in addition to diffusion. For example, membrane loss through vesiculation and membrane fragility in spherocytosis and elliptocytosis, interaction between hemoglobin fibers and RBC membrane in sickle cell disease, and RBC adhesion are problems where the applications of the proposed membrane model could be beneficial.

# Chapter 4.

# Modeling of the Band-3 Proteins Diffusion in the Normal and Defective Red Blood Cell Membrane

**Abstract**

We employ a two-component red blood cell (RBC) membrane model to simulate the diffusion of band-3 proteins in the normal RBC and in the RBCs with protein defects. We introduce protein defects which reduce the connectivity between the lipid bilayer and the cytoskeleton or reduce the connectivity of the cytoskeleton and these defects are associated with the blood disorders of hereditary spherocytosis (HS) and elliptocytosis (HE), respectively. We first measure the band-3 diffusion coefficients in the normal RBC membrane and in the RBC membrane without cytoskeleton. Comparison of these two coefficients demonstrates that the cytoskeleton limits the band-3 lateral mobility. Second, we study band-3 diffusion in defective RBC membranes and quantify the relation between the band-3 diffusion coefficients and the percentage of protein defects in HS and HE RBCs. By comparing the diffusion coefficients measured in these two cases, we conclude that the band-3 mobility is primarily controlled by the cytoskeleton connectivity. Third, we study how the band-3 anomalous diffusion exponent depends on the percentage of protein defects in the membrane cytoskeleton. Our measurements show that the effect of the cytoskeleton connectivity on the anomalous diffusion exponent is small. Finally, we show that introduction of attraction between the lipid bilayer and the spectrin network can reduce band-3 diffusion. By comparing our measurements to experimental data, we predict that the attractive force between the spectrin filament and the lipid bilayer is at least 20 times smaller than the binding forces at the two membrane major binding sites at band-3 and glycophorin.

**4.1. Introduction**

Diffusion of membrane proteins has been studied for a long time because of its importance in various cellular processes (179-182). The early fluid mosaic model developed for the plasma membrane postulated that proteins were freely diffusing in the plane of the membrane and the proteins followed a normal diffusion pattern, meaning that the mean square displacement (MSD) of the diffusing protein is proportional to time, MSD ~ $t$ (183). However, as the experimental techniques advanced, the protein diffusion in the membrane was found to be more complicated and it was observed that proteins undergo anomalous diffusion (subdiffusion) in biological cells (184-190). The MSD of the diffusing protein is proportional to a fractional power of time, MSD ~ $t^\alpha$, where $\alpha < 1$ is called anomalous diffusion exponent. Anomalous diffusion of proteins in the cell membrane mainly originates from the interactions between the diffusing protein and the immobile and mobile particles in the membrane (191, 192). In addition, the diffusing proteins may stick to or be hindered by the cell cytoskeleton, causing the deviation from the normal diffusion (116, 175, 193-195). RBC membrane cytoskeleton consists of spectrin tetramers connected at actin junctional complexes, forming a two-dimensional (2D) six-fold triangular network tethered to a lipid bilayer, as shown in Fig. 4.1A. While the main function of the cytoskeleton is to maintain the mechanical properties and integrity of the RBC, experimental studies have demonstrated the cytoskeleton sterically hinders the lateral diffusive motion of the band-3 proteins (147, 149, 176, 178, 196). In the RBC membrane, approximately one third of the band-3 proteins bind to the cytoskeleton and they are considered to be immobile band-3. The rest of the band-3 proteins, which are not connected to the cytoskeleton, are mobile band-3 proteins. Although mobile band-3 proteins do not directly bind to the cytoskeleton, their lateral diffusive motion in the membrane is hindered by the presence of the cytoskeleton (115-118, 147-150, 174-177, 197). In the short time range (~10 ms), the motions of the mobile band-3 proteins are constrained within the compartments formed by the spectrin network underneath the lipid bilayer (178). In the long time range, the band-3 proteins occasionally hop from one compartment to a neighboring compartment due to the fluctuation of the spectrin network (117, 118, 177, 198) and the fluctuation of lipid bilayer (199). Therefore, two types of diffusive motions are defined, namely the microscopic diffusion, which describes the motion of band-3 protein within a compartment, and the macroscopic diffusion, which describes the global motion of band-3 proteins over the surface of the cell.

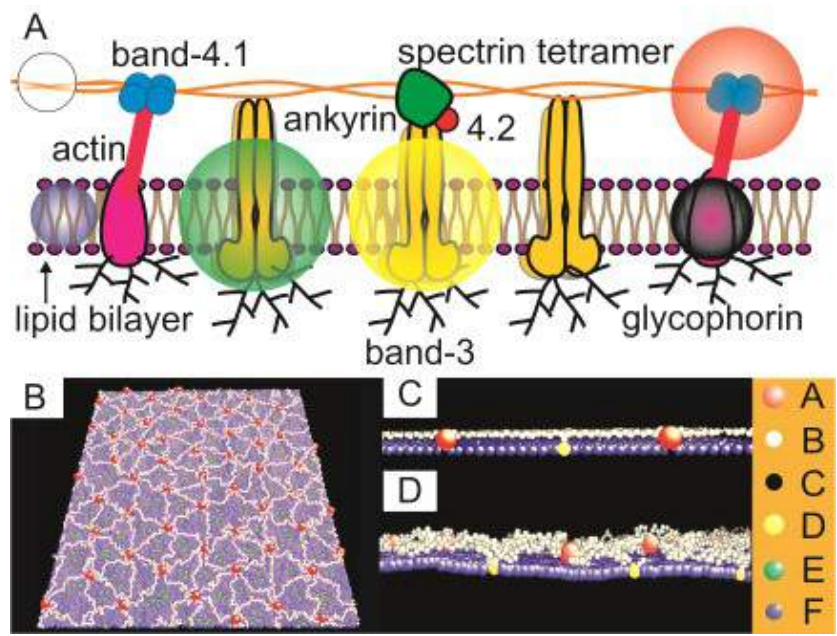

Figure 4.1. (A) Schematic of the human RBC membrane. Two-component human RBC membrane model (B) top view at equilibrium configuration, (C) side view at initial configuration (D) side view at equilibrium configuration. "A" type particles represent actin junctions. "B" type particles represent spectrin particles. "C" type particles represent glycophorin proteins. "D" type

particles represent band-3 complexes that are connected to the spectrin network (immobile band-3). "E" type particles represent band-3 complexes that are not connected to the network (mobile band-3). "F" type particles represent lipid particles.

Although experimental measurements clearly demonstrate that the band-3 proteins undergo hop diffusion in the RBC membrane, the exact mechanisms of how the band-3 proteins escape from one compartment to a neighboring one are still not well-understood because of the limited length and time scales that can be explored by experiments. Therefore, a number of analytical models were developed to investigate the mechanism of the hop diffusion (116, 118, 174, 175, 198-201). In general, three mechanisms have been proposed. The first mechanism considers that band-3 molecules escape from one compartment and move to the neighboring compartment due to the remodeling of the cytoskeleton (116, 174, 175, 198). According to the second and third mechanism, the thermal fluctuation of the spectrin filaments and the thermal fluctuation of the lipid bilayer respectively allow for the band-3 hop diffusion (118, 199, 201). In addition, the binding forces between the spectrin and specific type of lipid molecules, which were reported in a number of studies (158-163, 202-205), may enhance the confinement of the cytoskeleton on the lateral motions of the band-3 proteins. However, since the measurement of the forces between the spectrin filaments and lipid bilayer is not an easy task, studies on the effects of the binding force between the spectrin filaments and the lipid bilayer on the band-3 diffusion are still incomplete.

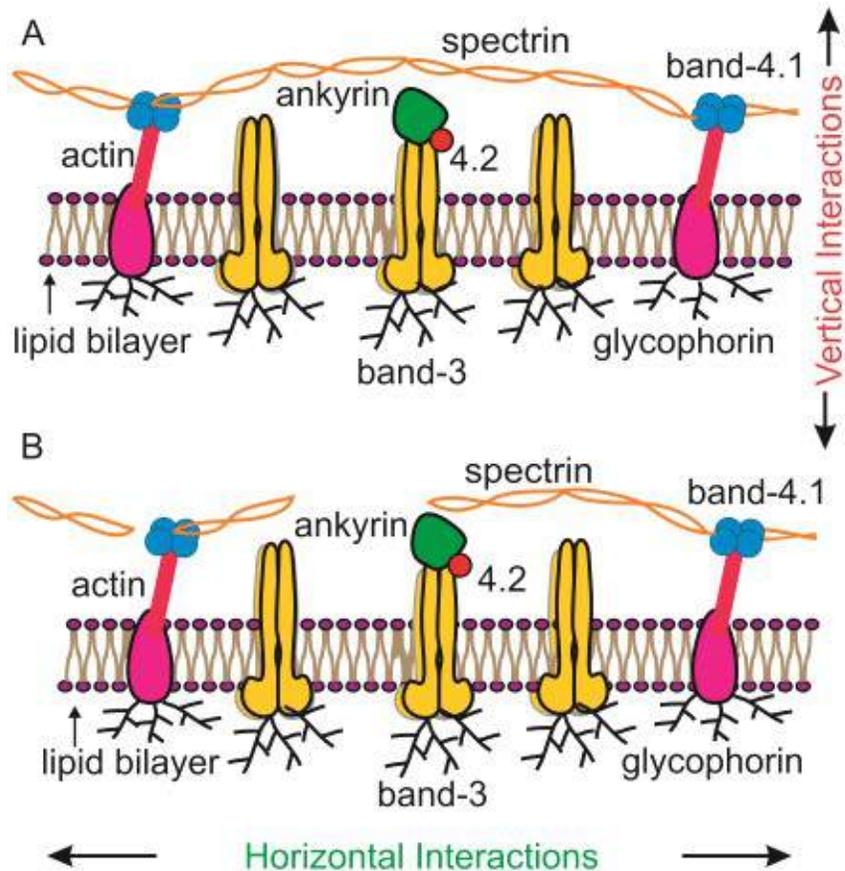

Figure 4.2. (A) Schematic of the HS RBC membrane. (B) Schematic of the HE RBC membrane.

Study of the diffusion of band-3 protein in the hemolytic blood disorders raises important questions. Band-3 diffusion measurements in the HS and HE RBCs showed that the mobile band-3 diffusion was accelerated and the number of mobile band-3 proteins was increased (112, 115, 197). More importantly, the abnormal band-3 diffusive motion may relate to the instability of RBC membrane and membrane vesiculation in the blood disorders (16, 26, 145, 206). In HS, there are protein defects in the vertical interactions (see Fig. 4.2A) between the cytoskeleton and the lipid bilayer (3, 16, 19, 21-24, 207), whereas in HE, protein defects occur in the horizontal interactions (see Fig. 4.2B) at the spectrin dimer-dimer connecting sites or at the actin-spectrin junctions of the cytoskeleton (3). Disruptions of either the vertical interaction or horizontal interaction change the lateral diffusion of the mobile band-3 proteins (112, 115, 116, 118, 197, 208, 209). While numerous experimental measurements and analytical modeling were conducted on the band-3 diffusion in the RBC membrane, only a limited number of papers quantitatively relate the band-3 diffusion coefficient to the percentage of the protein defects in the cells with blood disorders. In these experimental measurements, protein defects can only be identified by measuring the ratio between band-3 and spectrin content in HS and HE (115, 197). As for the analytical modeling, Saxton (116) studied the dependence of the band-3 diffusion coefficients on the cytoskeleton connectivity by performing Monte Carlo simulations. His simulation results showed that the band-3 diffusion coefficients fall significantly as the cytoskeleton connectivity was increased. However, the effects of the spectrin filaments and lipid bilayer fluctuations were not considered in these Monte Carlo simulations. Auth and Gov (118) proposed an analytical model, where the effect of the spectrin filaments is simulated by static pressure field. This model was capable of showing the band-3 diffusion in the normal RBC membrane and in the membrane with ankyrin protein defects, but protein defects in the cytoskeleton were not considered.

In this work (210), we simulate the band-3 hop diffusion in the RBC membrane due to the thermal fluctuation of the spectrin filaments and the thermal fluctuation of the lipid bilayer. Second, we quantify the relation between the percentage of membrane defects and band-3 diffusion coefficients in HS and HE. Third, we compute the band-3 anomalous diffusion exponents $\alpha$ in HE RBCs, by fitting the MSD measured from our simulation to the MSD ~ $t^\alpha$ relation for the appropriate time scale. The calculated exponents are compared with exponents obtained from the fixed obstacles model and picket fence model. Finally, we explore the effects of the attractive forces between the spectrin filaments and the lipid bilayer on the band-3 diffusion. In order to achieve our goals, we apply a recently developed two-component coarse-grained molecular dynamics (CGMD) model for the human RBC membrane which explicitly represents the cytoskeleton and the lipid bilayer by coarse-grained (CG) particles (211). Previously developed CGMD models explicitly simulate either only the RBC cytoskeleton with implicit representation of the lipid bilayer (80, 81, 109, 110, 142) or only the lipid bilayer (84, 86, 91, 92, 95, 108) with implicit spectrin network (111). Absence of either the explicit lipid bilayer or the cytoskeleton limits their application in studying the protein diffusion in the RBC membrane. Recently, a model consisting of two layers of 2D triangulated networks, where one layer represents the cytoskeleton while the other one represents the lipid bilayer, was introduced (154). Although this approach simulates two-component RBC membrane with high computational efficiency, it cannot be used to model the protein diffusion. The two-component RBC membrane model applied in this work explicitly comprises both the lipid bilayer and the cytoskeleton, and thus can be used to simulate interactions between the cytoskeleton and the lipid bilayer as well as interactions between the cytoskeleton and diffusing membrane proteins. In addition, the two-component model naturally simulates the band-3 hop diffusion due to the combined effects of the thermal fluctuation of the spectrin filaments and of the lipid bilayer. Furthermore, the explicit representation of the cytoskeleton allows us to easily incorporate a variety of proteins defects into this model and introduce the attractive force between the spectrin filaments and lipid bilayer.

## 4.2. Model and methods

In the applied RBC membrane model, three types of coarse-grained (CG) particles are introduced to represent lipid bilayer and two types of transmembrane proteins (see Fig. 4.1A-D), including the blue color particles signifying a cluster of lipid molecules, the black particles denoting glycophorin proteins, the yellow particles representing the band-3 proteins that are connected to the spectrin network through the ankyrin proteins (immobile band-3), and the green particles representing the band-3 proteins that are not attached to the spectrin network (mobile band-3). To simplify our model, the connections between spectrin and immobile band-3 are assumed to be fixed. The cytoskeleton consists of spectrin filaments connected at actin junctions forming a canonical hexagonal network. The actin junctions are represented by the red particles in Fig.4.1 A-D, and are connected to the lipid bilayer via glycophorin. The spectrin filament is represented by 39 spectrin particles (white particles) connected by unbreakable springs. The details of the potentials between the CG particles applied in the model, along with the mass and dimensions of the CG particles can be found in the authors' previous work (211). The interactions between the cytoskeleton and the lipid bilayer are represented by the repulsive part of the L-J potential

$$u_{LJ}(r_{ij}) = \begin{cases} 4\varepsilon\left[\left(\dfrac{\sigma}{r_{ij}}\right)^{12} - \left(\dfrac{\sigma}{r_{ij}}\right)^{6}\right] + \varepsilon & r_{ij} < R_{cut,LJ} \\ 0 & r_{ij} > R_{cut,LJ} \end{cases} \qquad (4.1)$$

where $\varepsilon$ is the energy unit and $\sigma$ is the length unit. $r_{ij}$ is the distance between the CG particles. Given the diameter of the lipid particles $r_{eq}^{l-l} = 2^{1/6}\sigma = 5$ nm, one finds $\sigma$ = 4.45 nm. The cutoff distance of the potential is chosen to be $R_{cut,LJ} = 2^{1/6} r_{eq}$, where $r_{eq}$ is the equilibrium distance between different types of CG particles. More information about the model can be found in authors' previous work (211). In our simulations, the dimension of the membrane is approximately ~ 0.8 × 0.8 μm². The system comprises $N$ = 36606 CG particles. By applying the Nose-Hoover thermostat, the temperature of the system is maintained at $k_BT/\varepsilon$ = 0.22. The model is implemented in the NAT ensemble (131). Because the model is solvent-free and the membrane is a two-dimensional structure, we controlled the projected area instead of the volume. The projected area is adjusted to result in zero tension at the equilibrium state. The time step is selected to be $0.01t_s$, where $t_s$ is the time scale. The time scale $t_s$ in our simulation will be determined in the following section by comparing the band-3 diffusion coefficients measured in our model with the former experimental results.

## 4.3. Results and discussion

### 4.3.1. Band-3 diffusion in RBC membrane with perfect cytoskeleton, in the normal RBC membrane and in the lipid membrane

In this section, we measure the diffusion coefficients of the mobile band-3 particles in a RBC membrane model with perfect cytoskeleton, in the normal RBC membrane and when only the lipid bilayer is considered. The diffusion coefficient of the mobile band-3 is calculated by measuring the slope of the MSD with respect to time. MSD is defined as $\frac{1}{N_{band-3}}\left\langle \sum_{j=1}^{N_{band-3}}\left[\mathbf{x}_j(t)-\mathbf{x}_j(0)\right]^2 \right\rangle$ (212), where $N_{band-3}$ is the number of the mobile band-3 particles. First, we simulated the diffusion of band-3 particles in the membrane with perfect cytoskeleton, meaning that the cytoskeleton is considered as a perfect 2D six-fold triangular structure with a fixed connectivity. Fig. 4.3A displays a representative trajectory of a mobile band-3 particle hopping to the neighboring compartment during $2\times10^6$ time steps. The trajectory of the band-3 particle clearly illustrates the steric hindrance effect of the cytoskeleton on the lateral motion of band-3 particles. Plot of the MSD of the mobile band-3 particles with respect to the time in Fig. 4.3C (black curve) shows that MSD rises rapidly for $t<t_{micro}=10000t_s$, where $t_{micro}$ is the timescale for the microscopic diffusion, indicating that band-3 particles diffuse within the compartments for $t < t_{micro}$. For $t > t_{micro}$, MSD increases linearly with time although the slope of MSD against time becomes much smaller than the slope in the microscopic diffusion. This means that the mobile band-3 particles explore areas larger than the triangular compartments for $t > t_{micro}$ and the increase of MSD results from the intercompartment hop diffusion (macroscopic diffusion). We extract the macroscopic diffusion coefficient of the mobile band-3 particles by fitting a straight line to the MSD for large times and then applying the definition of the diffusion coefficient $D = \lim_{t=t_{micro} \to \infty} (1/4t\, MSD)$ (212). The band-3 diffusion coefficients mentioned in the rest of the chapter all refer to the macroscopic diffusion coefficient. The diffusion coefficient for the band-3 particles in the RBC membrane with perfect network is found to be $D_{perfect} = 1.23 \times 10^{-4} \sigma^2/t_s$. It is noted that the value of the MSD at $t = t_{micro} = 10000 t_s$ in our simulation is about $55\sigma^2$ ($\Box 1130 nm^2$), which corresponds to a circular area with diameter of $8.5\sigma \sim 38$ nm. This estimation is in agreement with the numerical results obtained in (118) and 2 times smaller than the value observed in the single particle tracking experiment (178). The discrepancy between our result and the experimental value is probably caused by the application of perfect 2D hexagonal spectrin network in our model. For the normal RBC membrane, electron microscopy images revealed that although most of actin junctions were linked with other 6 actin junctions, a small amount of actins junctions were linked with 5 or 7 actin junctions (137, 138). In addition, Li et.al (142) demonstrated that the biconcave shape of RBCs can be obtained only if the spectrin network undergoes remodeling of the cytoskeleton to relax the in-plane elastic energy. The metabolic remodeling of the spectrin network is affected by the adenosine 5′-triphosphate (ATP) contents in the RBC(140, 151). The defects in the cytoskeleton of the normal

RBC membrane can effectively increase the area of the compartments. To obtain the connectivity of the cytoskeleton in our membrane model that corresponds to the normal RBC membrane, we match the numerical value of the shear modulus measured from our RBC membrane model with the experimental values of ~8.5μN/m obtained in (141, 213). Based on our previous work in (111, 211), the shear modulus of our membrane model falls within the range of the experimentally measured values when the cytoskeleton connectivity is approximately 90%. Therefore, we select the membrane with cytoskeleton connectivity of 90% to represent the normal RBC membrane. We reduce the cytoskeleton connectivity to 90% by randomly disconnecting the spectrin filaments in the middle and repeat the diffusion measurement performed for the perfect cytoskeleton. We find that the MSD (see Fig. 4.3C, blue curve) goes through a fast increase until $t = t_{micro} \approx 25000 t_s$, and then it increases slowly and linearly with respect to time. Although the general behavior of the MSD is similar to the behavior of the MSD measured in the perfect cytoskeleton case, the timescale for the microscopic diffusion and the corresponding area explored by the band-3 particles during microscopic diffusion are increased due to the reduced cytoskeleton connectivity. The corresponding diffusion coefficient of the mobile band-3 particles in the membrane with cytoskeleton connectivity of 90% is found to be $D_{normal} = 1.29 \times 10^{-4} \sigma^2 / t_s$. The experimentally measured band-3 diffusion coefficient in the RBC membrane is $D \approx 4 \times 10^{-11} \text{cm}^2/\text{s}$ (214). By comparing our numerical result to the experimental value, the timescale of our simulation is determined to be $t_s \approx 0.7$ μs. This estimation is comparable to the timescale calculated using the viscosity of the RBC membrane in the authors' previous work (211). Then, the timescale for the microscopic diffusion of the mobile band-3 particles in the normal RBC membrane is computed to be approximately $t_{micro} = 25000 t_s = 17.5$ ms, which is consistent with the experimentally measured timescale of ~ 10-20 ms for the microscopic diffusion in (178, 196). At last, we remove the cytoskeleton from the membrane model and simulate the band-3 diffusion in the lipid bilayer without the obstruction from the spectrin filaments. Fig. 4.3B shows that the trajectory of the band-3 particle covers, in the same time as before, larger distance than in the cases where the spectrin cytoskeleton is implemented. This result demonstrates that the cytoskeleton has a clear confining effect in the diffusive motion of the band-3 particles. The MSD of the band-3 particles, plotted in Fig. 4.3C, increases linearly with respect to time and the corresponding diffusion coefficient is measured to be $D_{no\_cyto} = 52 \times 10^{-4} \sigma^2 / t_s$, approximately 40 times larger than $D_{normal} = 1.29 \times 10^{-4} \sigma^2 / t_s$. This measurement is comparable with experimental results showing that band-3 diffuses about 50 times faster in mouse erythrocytes that lack major components of the cytoskeleton than in the healthy erythrocytes (147).

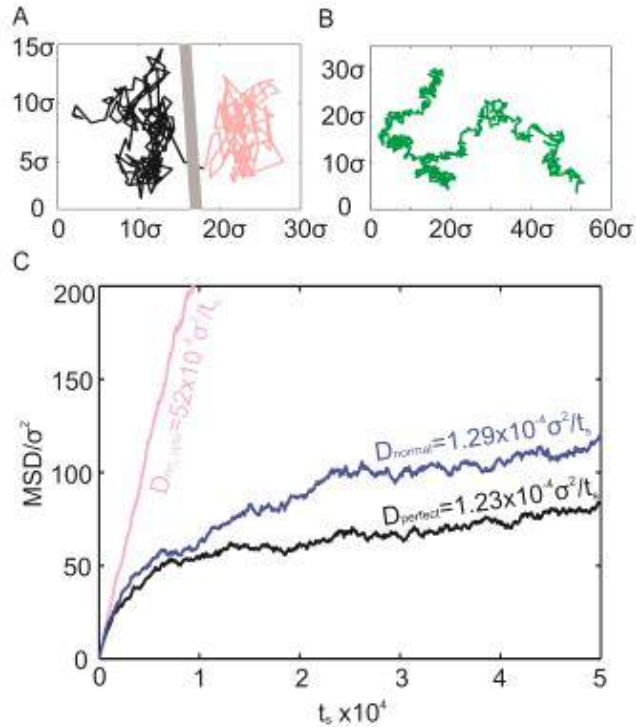

Figure 4.3. Trajectories of a mobile band-3 particle (A) hopping to a neighboring compartment in the normal RBC membrane and (B) diffusing in a lipid bilayer without cytoskeleton. The trajectories are extracted from simulation with duration of $2\times10^6$ time steps and the positions of mobile band-3 particles were recorded every 2000 time steps. (C) MSDs against time and corresponding diffusion coefficients of the mobile band-3 diffusing in the RBC membrane with perfect cytoskeleton, in the normal RBC membrane and in membrane without cytoskeleton.

**4.3.2. Band-3 diffusion in RBC membrane with protein defects in the vertical interactions**

In this section, we reduce the connectivity between the spectrin filaments and immobile band-3 particles in our model to simulate band-3 diffusion in the RBC membrane with defective proteins in the vertical interactions. This situation is encountered in the RBCs of patients suffering from HS. In this case, defects in proteins that play a direct or indirect role in the binding of the cytoskeleton to the lipid bilayer, such as band-3, 4.2, and ankyrin proteins, result in local detachments of the lipid bilayer (3, 145). Experimental measurements showed that the diffusion of the mobile band-3 was accelerated and mobile band-3 portion was increased in the HS cells (112, 115, 197). Auth and Gov (118) demonstrated, by using a pressure fence analytical model, that removal of all the connections between the spectrin filaments and immobile band-3 proteins leads to a 20-fold increase of the mobile band-3 diffusion coefficient. To simulate the diffusion of band-3 particles in the HS RBC membrane, we reduce the connectivity between the spectrin filaments and the immobile band-3 particles. We then compute the band-3 diffusion coefficient and validate it by comparing the numerical diffusion coefficients with experimentally measured values. In addition, we quantified the relation between the percentage of the protein defects and band-3 diffusion coefficients. To examine the effects of protein defects in the vertical connections only, the cytoskeleton connectivity is conserved at 100% for all the measurements in this section. In our simulations, the connectivities between the spectrin filament and band-3 particles are selected to be $C_{vertical}$ = 70%, 50% and 0%, respectively. The MSD of band-3 particles with respect to time are plotted in Fig. 4.4. When $C_{vertical}$ = 70%, the diffusion coefficient is found to be $D_{C_{vertical}=70\%} = 2.5 \times 10^{-3} \sigma^2/t_s$, approximately 2 times larger than $D_{perfect}$. The increase in the band-3 diffusion coefficient is attributed to the increased average distance between the spectrin filaments and the lipid bilayer as the connections between the filaments and band-3 particles are removed. Larger distance between the filaments and lipid bilayer causes an increase in the probability of band-3 particles passing to the neighboring compartments and thus results in a larger band-3 diffusion coefficient. When $C_{vertical}$ is further decreased to 0%, the band-3 diffusion coefficient increases to $D_{C_{vertical}=0\%} = 9.6 \times 10^{-3} \sigma^2/t_s$ approximately 8 times larger than $D_{perfect}$. By applying the timescale of our simulation $t_s \approx 0.7\ \mu s$, we find that the mobile band-3 particle diffusion coefficient in the membrane with protein defects in the vertical interactions is approximately $D \approx 0.8-3 \times 10^{-10}\ cm^2/s$. This result is consistent with the experimental values of the band-3 diffusion coefficients in HS RBCs (115, 197), suggesting that the band-3 diffusion coefficient increases, but not significantly, in HS cell (112).

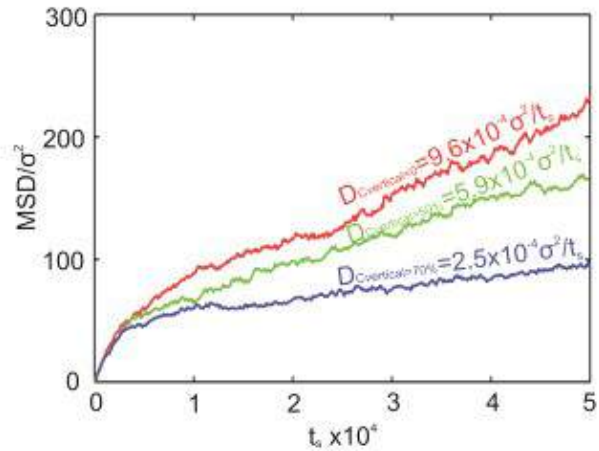

Figure 4.4. MSDs against time and corresponding diffusion coefficients of the mobile band-3 diffusing in the RBC membrane with different vertical connectivities.

### 4.3.3. Band-3 diffusion in RBC membrane with protein defects in the horizontal interactions

In this section, we disrupt the spectrin filaments in our membrane model to simulate the band-3 diffusion in the membrane with defective proteins in the horizontal interactions. This situation is encountered in the RBCs of patients suffering from HE where defects in α-spectrin, β-spectrin, and/or protein 4.1 cause dissociation of the spectrin tetramer into dimers or disconnection of the spectrin filaments from the actin junctions (3, 145). As a result, the diffusion coefficient of the mobile band-3 is largely increased in the HE RBCs (112, 115). Saxton performed Monte Carlo simulation based on the percolation analysis and showed that the mobile band-3 diffusion coefficients strongly depends on the cytoskeleton connectivity (116). To simulate the HE RBC membrane, we disconnect the spectrin chains in the middle. The cytoskeleton connectivities are selected to be $C_{horizontal}$ = 80%, 70%, 60%, 50% and 30%, respectively. Because of the ruptured spectrin filaments, band-3 particles can travel easily between the compartments with disrupted boundaries, resulting in an enhanced lateral mobility. The recorded band-3 trajectory when the cytoskeleton connectivity is $C_{horizontal}$ = 70% (see Fig. 4.5A) shows that band-3 particles trajectories cover larger areas (red and green curves) comparing to band-3 particles in the normal RBC membrane (blue curve). This means that the steric hindrance effects of the cytoskeleton are reduced as the cytoskeleton connectivity decreases. The MSDs and corresponding diffusion coefficients of band-3 particles measured with a variety of $C_{horizontal}$ (see Fig. 4.5B) show that as the $C_{horizontal}$ is decreased from 80% to 30% the band-3 diffusion coefficients increase significantly from $D_{C_{horizontal}=80\%} = 3.27 \times 10^{-4} \sigma^2/t_s$ to $D_{C_{horizontal}=30\%} = 50 \times 10^{-4} \sigma^2/t_s$, indicating that the cytoskeleton connectivity plays an important role in controlling band-3 diffusion. Particularly, when $C_{horizontal}$ = 30%, the band-3 diffusion coefficient increases to $D_{C_{horizontal}=30\%} = 50 \times 10^{-4} \sigma^2/t_s$, which is comparable to the band-3 diffusion coefficients measured from the membrane model without cytoskeleton $D_{no\_cyto} = 52 \times 10^{-4} \sigma^2/t_s$. This means that at low connectivity the cytoskeleton loses the function of hindering the band-3 lateral diffusion. Comparison between the band-3 diffusion coefficients measured in the HS and HE RBC membrane models shows that the cytoskeleton connectivity is the major determinant of the lateral diffusivities of band-3. This explains the experimental observation that band-3 diffusion coefficients measured in HE RBC are larger than those measured in HS RBCs (112).

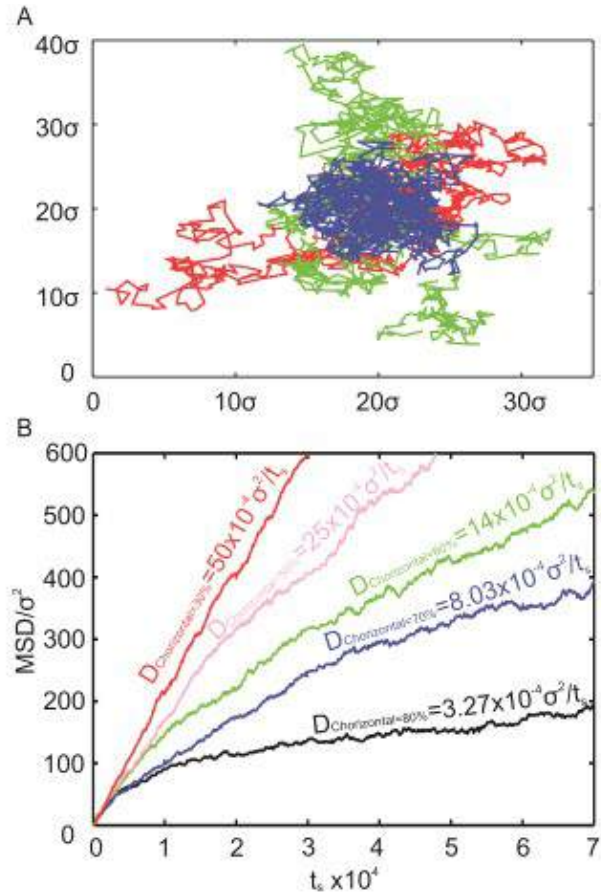

Figure 4.5. (A) Comparison between the trajectory of a band-3 particle diffusing within the compartment of an RBC membrane without defects (blue line) and the trajectories of two mobile

band-3 particle diffusing in RBC membrane with defects in the cytoskeleton (green and red lines). The trajectories are extracted from simulation with duration of $2\times10^6$ time steps and the positions of mobile band-3 particles were recorded every 2000 time steps. (B) MSDs against time and corresponding diffusion coefficients of the mobile band-3 diffusing in the RBC membrane with different cytoskeleton connectivities.

**4.3.4. Anomalous diffusion of the band-3 in the membrane with protein defects in the horizontal interactions**

The MSDs of the band-3 particles obtained in the previous section can be used to calculate the anomalous diffusion exponent in the case of the RBC membrane with protein defects in the horizontal interactions. Single particle tracking experiments have clearly demonstrated that mobile band-3 proteins undergo subdiffusion in the RBC membrane (112, 176, 178, 190), meaning that MSD of the band-3 follows the relation MSD ~ $t^\alpha$, where $\alpha < 1$ is the anomalous diffusion exponent. Several mechanisms have been hypothesized to explain protein subdiffusion in the cell membrane, including confinement by the membrane skeleton (116, 148, 150, 174-176, 215), obstruction by the lipid rafts or protein domains (192, 216-218), and binding of the diffusing proteins to the cytoskeleton (193, 219-221). Although the subdiffusion of the proteins in the cell membrane has been studied for a long time, quantitative relations between the degree of the obstruction and the anomalous diffusion exponents are difficult to be established due to limitations of the experimental techniques. Only few analytical models have been developed to study the dependence of the anomalous diffusion exponent on the membrane protein defects (192, 195). As seen in Fig. 4.5B, the MSDs of the mobile band-3 particles measured at high cytoskeleton connectivities do not increase proportionally with respect to time, illustrating a typical anomalous diffusion pattern. The anomalous diffusive behavior of the band-3 particles in our model results from the steric hindrance effect of the cytoskeleton, therefore, as the cytoskeleton connectivity decreases, the band-3 diffusion approaches to normal diffusion, as shown in Fig. 4.5B. To determine the quantitative relation between the band-3 anomalous diffusion exponents $\alpha$ and the cytoskeleton connectivity $C_{\text{horizontal}}$, we plot the MSDs of mobile band-3 particles against time with a variety of cytoskeleton connectivities in log-log coordinates. The log-log plots of the MSD against time exhibit a linear relationship with slight changes in slopes from 1 (normal diffusion) to 0.905 when the $C_{\text{horizontal}}$ is increased from 0 % to 70% (Fig. 4.6A). We observe that in the cases of $C_{\text{horizontal}}$ = 90% and 100%, the dependence of the $\log(MSD)$ on $\log(t)$ deviates from linearity at large time scales. This is because at large time scales the band-3 particles undergo macroscopic diffusion where the increase of the MSD is due to the band-3 hop diffusion. The small diffusivities in the macroscopic diffusion cause the deviation of the MSD curves from the MSD ~ $t^\alpha$ relation. These observations suggest that the anomalous diffusion relation MSD ~ $t^\alpha$ does not hold for the long time when describing the cytoskeleton-induced subdiffusion of band-3 proteins, which is in agreement with observations from previous analytical simulations (192, 195). At time scales where the anomalous diffusion relation holds, Fig. 4.6B shows that when the cytoskeleton connectivity ($C_{\text{horizontal}}$) increases from 0% to 100%, the corresponding anomalous diffusion exponents $\alpha$ varies from 1 to 0.86. Thus, anomalous diffusion exponents of band-3 particles induced by the cytoskeleton lie in the range of 1-0.86, which is larger than the exponents range of 1–0.94 induced by picket fence but smaller than the exponents range of 1–0.65 induced by lipid rafts or protein domains (195). The cytoskeleton-induced obstruction only has a small effect on the band-3 anomalous diffusion exponent because the anomalous diffusion relation MSD ~ $t^\alpha$ holds true only during short time scales (microscopic diffusion) and during the transition period from the microscopic diffusion to the macroscopic diffusion. The decrease in the cytoskeleton connectivity barely influences band-3 microscopic diffusion. In contrast, fixed obstacles and lipid rafts, which are not considered here, can largely alter band-3 microscopic diffusion motion in the microscopic diffusion time scale.

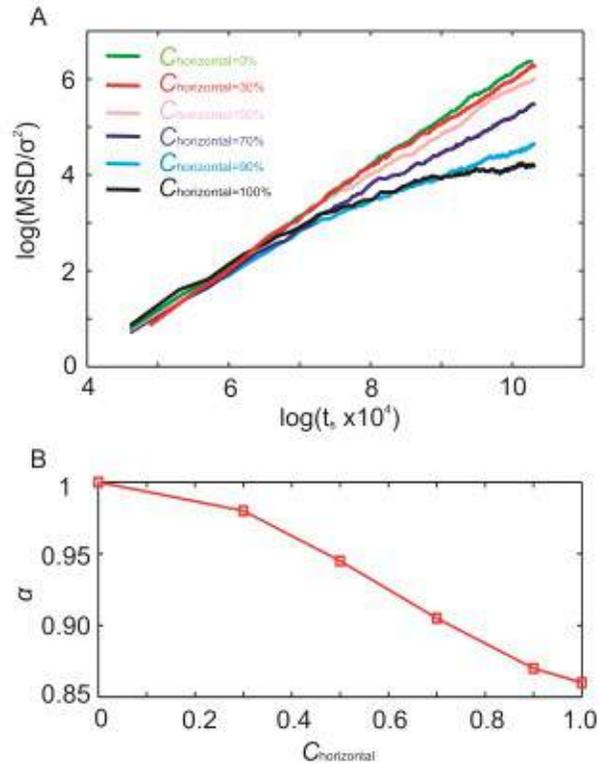

Figure 4.6. (A) log-log plot of the MSDs against time for the RBC membrane with cytoskeleton connectivities $C_{horizontal}$ = 0%, 30%, 50%, 70%, 90% and 100%. (B) The corresponding anomalous exponent $\alpha$ for the previously mentioned horizontal cytoskeleton connectivities.

**4.3.5. Effects of attractive forces between the spectrin filaments and lipid bilayer on the band-3 diffusion**

Here, we apply attractive force between the spectrin filaments and lipid bilayer to investigate its effect on the mobile band-3 diffusion in the normal RBC membrane and in the membrane with defective vertical and horizontal interactions. Previous studies of the lipid–spectrin interactions showed that the spectrin binds to the negatively charged lipid surfaces with association constants of 2-10×$10^6$ $M^{-1}$ (158-163, 202-205), whereas the association constants at the lipid bilayer-cytoskeleton major binding sites, such as spectrin-ankyrin, ankyrin-band-3, spectrin-protein 4.1-actin were 2×$10^7$ $M^{-1}$, 2×$10^8$ $M^{-1}$ and 2×$10^{12}$ $M^{-2}$, respectively (164, 165). Comparison between the association constants suggests that the spectrin-lipid affinity is not as strong as the primary binding sites where the spectrin filaments are anchored to the lipid bilayer. But the binding force between the spectrin and lipid molecules could reduce the average distance between the lipid bilayer and cytoskeleton and thus enhance the steric hindrance effect on the band-3 diffusion. Since the binding force between the spectrin filaments and lipid bilayer is difficult to measure, we apply the attractive part of the L-J potential to represent the effective binding forces between the spectrin particles and lipid particles,

$$u_{att}(r_{ij}) = \begin{cases} 4n\varepsilon\left[\left(\dfrac{\sigma}{r_{ij}}\right)^{12} - \left(\dfrac{\sigma}{r_{ij}}\right)^{6}\right] + n\varepsilon & r_{ij} > r_{eq}^{l-s} \\ 0 & r_{ij} < r_{eq}^{l-s} \end{cases} \qquad (4.2)$$

where $n$ is a parameter to tune the attractive energy between the spectrin filaments and lipid bilayer. $r_{eq}^{l-s} = 5$nm is the equilibrium distance between the spectrin particles and lipid particles. Our previous simulations (211) showed that the cytoskeleton was completely attached to the lipid bilayer when $n \geq 0.2$, which contradicts with experimental observations. Therefore, in the current simulations, the parameters *n* applied in the Eq. (4.2) are selected to be *n* = 0, 0.05, 0.1 and 0.15, respectively.

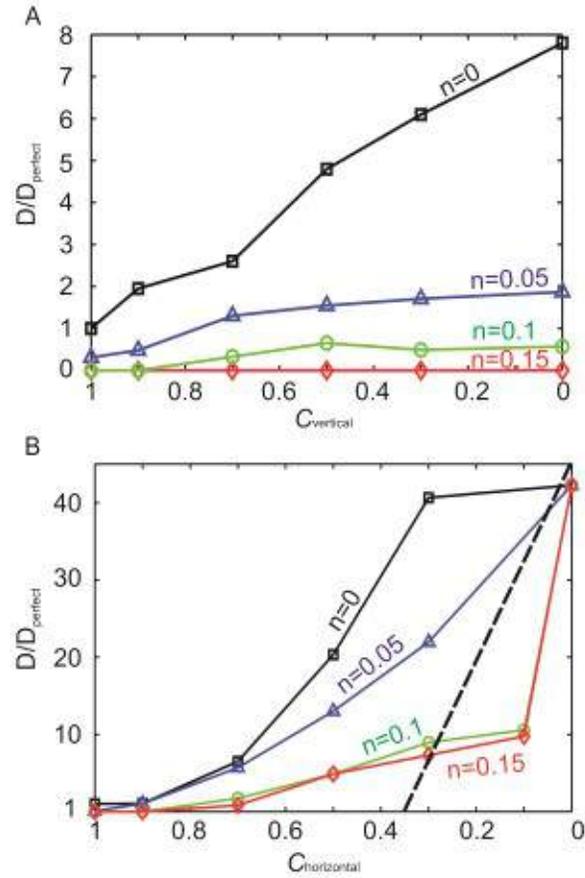

Figure 4.7. Ratios of the band-3 diffusion coefficients *D* measured in the RBC membrane with protein defects to $D_{perfect}$ at a variety of attractive forces between the spectrin filaments and the lipid bilayer (*n* = 0, 0.05, 0.1 and 0.15). We test membrane models with (A) vertical connectivities of $C_{vertical}$ = 100%, 90%, 70%, 50%, 30%, and 0%, and with (B) horizontal cytoskeleton connectivities of $C_{horizontal}$ = 100%, 90%, 70%, 50%, 30%, and 0%. The black dot line represents the results predicted by the percolation theory in (222).

First, we examine the effects of the attractive forces between the spectrin filaments and lipid bilayer on the band-3 diffusion in a RBC membrane with perfect cytoskeleton and in a membrane with protein defects in the vertical interactions. The connectivities between the spectrin filaments and immobile band-3 $C_{vertical}$ are selected to be 100%, 90%, 70%, 50%, 30% and 0%, respectively. The ratios of the measured band-3 diffusion coefficients $D$ to the band-3 diffusion coefficients in a RBC membrane with perfect cytoskeleton $D_{perfect}$ are plotted in the Fig. 4.7A. The black curve in the Fig. 4.7A shows that when no attractive force is applied ($n = 0$), the ratios of diffusion coefficients gradually rise to 8 with decreasing $C_{vertical}$, while it only grows to 2 with application of a small attractive force $n = 0.05$ (blue curve in Fig. 4.7A), meaning that the attractive force enhance the steric hindrance effect on the band-3 diffusion. This conclusion can be further proven if we apply a larger attractive force $n = 0.1$ (green curve in Fig. 4.7A). With this attractive force, no band-3 hop diffusion is observed when $C_{vertical} = 100\%$ and 90% because the distance between the filaments and the lipid bilayer is reduced such that the band-3 particles cannot pass the boundaries of the compartments. As $C_{vertical}$ is decreased to 70%, band-3 hopping diffusion is observed, but the band-3 macroscopic diffusion is still largely suppressed, resulting in smaller diffusion coefficients than $D_{perfect}$. Even as $C_{vertical}$ is reduced to 0%, the ratio of diffusion coefficients is only 0.6. When the attractive parameter $n$ is further increased to 0.15, the spectrin filaments are found to be fully attached to the lipid bilayer, preventing the band-3 particles from hopping to neighboring compartments. Thus, the band-3 diffusion coefficients become zero for all values of vertical connectivities (red curve in Fig. 4.7A). Based on the above discussion, we conclude that the attractive forces between the spectrin filaments and the lipid bilayer can reinforce the steric hindrance effects of the cytoskeleton on the band-3 motion and thus reduce the band-3 diffusion coefficients. In particular, small attractive forces diminish the effects of defective vertical interactions on increasing the band-3 diffusivity while large attractive forces almost totally negate the effects of the defective vertical interactions between the cytoskeleton and lipid bilayer on band-3 diffusivity.

Next, we study the effect of the attractive forces between the spectrin filaments and lipid bilayer on the band-3 diffusion in membrane with protein defects in the horizontal interactions. The connectivities of the cytoskeleton $C_{horizontal}$ are selected to be 100 %, 90%, 70%, 50%, 30% and 0% (lipid bilayer only), respectively. The corresponding ratio of the measured band-3 diffusion coefficients $D$ to $D_{perfect}$ are plotted in the Fig. 4.7B. When no attractive force is applied ($n = 0$), the band-3 diffusion coefficients are boost dramatically with decreasing $C_{horizontal}$ (black curve in the Fig. 4.7B), as $C_{horizontal}$ plays the major role on regulating the band-3 diffusion motion. When attractive forces are applied, lower band-3 diffusion coefficients are measured, as shown by the blue, red and green curves in the Fig 4.7B. These observations suggest that the attractive forces between the spectrin filaments and lipid bilayer can also diminish the enhancement of the band-3 diffusion due to the protein defects in the horizontal interactions. From the band-3 diffusion measurements in membrane with defective vertical interaction, we discover that when $n \geq 0.1$, the band-3 can barely pass the boundaries of the compartments, due to the small distances between the filaments and lipid bilayer, leading to small or zero band-3 diffusion coefficients (see Fig. 4.7A). However, Fig. 4.7B shows that even when $n$ is increased to 0.1 and 0.15, significant band-3 diffusion coefficients are measured at $C_{horizontal} \leq 50\%$. This difference is attributed to the fact that in the HE RBC membrane, the band-3 particles can travel to the neighboring compartments by passing the broken boundaries, even though the spectrin filaments are attached to the lipid bilayer. This also explains why the red and green curves measured at $n = 0.1$ and $0.15$ overlap in the Fig. 4.7B and demonstrates that the band-3 diffusion coefficients do not differ for attractive forces with $n \geq 0.1$.

At last, we compare the band-3 diffusion coefficients measured at large attractive forces ($n \geq 0.1$) to the band-3 diffusion coefficients computed by using Monte Carlo simulation where the authors studied the dependence of the protein diffusion coefficients on the cytoskeleton connectivity (116, 222). Both of these two measurements ignore the effect of the thermal fluctuation of the lipid bilayer and spectrin filaments on the band-3 diffusion. In these Monte Carlo simulation, diffusing particles are assumed to travel between sites connected with the unblocked bonds (corresponding to the broken spectrin filaments in our present simulations). The percolation analysis suggested that if the connections between the travelling sites are static, the diffusion coefficients dropped linearly down to zero as $C_{horizontal}$ was increased from 0 to 0.35 (see the dot lines in Fig. 4.7B), which is the percolation threshold (116). When $C_{horizontal} > 0.35$, the diffusion coefficients of the travelling particles become zero. The diffusion coefficients measured in our simulations at $n = 0.1$ and $0.15$ show that when $C_{horizontal} = 100\%$, $90\%$ and $70\%$, the band-3 diffusion coefficients are approximately zero, in agreement with percolation analysis. However, a notable band-3 diffusion coefficient of $6 \times 10^{-4} \sigma^2 / t_s$ is measured at $C_{horizontal} = 50\%$, contradicting with the percolation analysis. This discrepancy is believed to result from the fact that the band-3 particles in our simulation have not explored all the accessible areas in the membrane (Can you explain this). The time for the band-3 particles to fully explore the RBC membrane with $C_{horizontal} = 50\%$ cannot be accessed by our simulation. When $C_{horizontal} = 30\%$, which is below the percolation threshold, the diffusion coefficient measured from our simulation is again consistent with the percolation analysis. But when $C_{\text{horizontal}}$ is reduced down to 10%, the diffusion coefficients measured from our simulation are lower than the ones predicted by percolation analysis. This difference is probable caused by the fact that in the percolation analysis the diffusing particles can pass the broken boundaries of the compartments without interfering with the broken spectrin filaments whereas in our model, the broken spectrin filaments can still interact with the band-3 particles and hinder their lateral motions. Although the band-3 diffusion coefficients measured at large attractive forces ($n \geq 0.1$) are generally consistent with the percolation analysis, we note that our diffusion measurements only agree with the experimental measurements on the HS and HE cells (112, 115, 197) for the cases of $n = 0$ and 0.05 where the band-3 diffusion coefficients increase, but not significantly in the HS, and largely boosted in the HE (112). Therefore, it is reasonable to predict that the effective attractive force between the spectrin filaments and lipid bilayer is at least 20 times smaller than the binding forces at the two major binding sites between the cytoskeleton and the lipid bilayer. in agreement with the experimental measurements(204, 205).

## 4.4. Summary

We apply a two-component RBC membrane model to study band-3 diffusion in the normal RBC membrane and in the membranes with defective vertical and horizontal interactions. We measure the band-3 diffusion coefficients and quantify the relation between the band-3 diffusion coefficients and percentage of protein defects. Our measurements show that the band-3 diffusion coefficients are increased by 8 times when connectivity between the spectrin filaments and immobile band-3 particles $C_{\text{vertical}}$ is decreased from 100% to 0%, whereas the band-3 diffusion coefficients can be boosted by 20-40 times with significantly reduced cytoskeleton connectivity $C_{horizontal}$. These results are in agreement with the experimental measurements of the band-3 diffusion in the HS and HE RBCs (112, 115, 147, 197). By comparing the measured band-3 diffusions coefficients, we demonstrate that cytoskeleton connectivity is the major determinant of the lateral diffusivity of band-3. In addition, we quantify the relations between the anomalous diffusion exponent and the percentage of protein defects in the horizontal interactions. We find that the cytoskeleton has a small effect on the anomalous diffusion exponent, comparing to the obstructions due to the lipid rafts or protein domains. At last, we introduce attractive forces between the spectrin filaments and the lipid bilayer, and study the effects of the attractive forces on the band-3 diffusion. We find that application of the attractive forces slows down the band-3 diffusion in the normal and defective RBC membrane. Especially, large attractive forces can prevent the band-3 hop diffusion in the normal RBC membrane and in the membrane with protein defects in the vertical interactions. But, in the cases of the membrane with protein defects in the horizontal interactions, the band-3 diffusion coefficients become independent of the attractive forces when attractive forces are large. This is because the band-3 particles travel to different cytoskeleton compartments through the broken spectrin filaments. Moreover, we show that the band-3 diffusion coefficients measured at the large attractive forces generally agree with the percolation analysis in (116). By comparing the band-3 diffusion coefficients from our simulation with the experimental measured band-3 diffusion coefficients in HS and HE (112, 115, 147, 197), we estimated the scale of the effective attractive force between the spectrin filaments and lipid bilayer is at least 20 times smaller than the binding forces at the two major binding sites at the immobile band-3 proteins and the glycophorin C. The approach used here for the simulation of band-3 diffusion in the defective RBC membrane provides a basis for the future study of the mechanism of the membrane vesiculation in HS and HE.

# Chapter 5.

# Modeling of Vesiculation in Healthy and Defective Human Erythrocyte Membrane

**Abstract**

Red blood cells-derived microvesicles play important roles in regulating cell communications and modulating immune response and other patho-physiological processes. In the blood disorders, microvesiculation in the defective red blood cell (RBC) membrane becomes complicated due to the various protein defects. In addition, membrane domains which change the local curvature of the RBC membrane could also affect the microvesiculation. In this work, we simulate the vesiculation in the RBC membrane by applying a two-component coarse-grained molecular dynamics (CGMD) RBC membrane model. We study the effect of the spontaneous curvature of the membrane domain on the microvesiculation induced by the compression on the RBC membrane. Furthermore, we study the vesiculation in RBCs from patients suffering from the blood disorders of hereditary spherocytosis (HS) and elliptocytosis (HE). Our simulation results show that the large spontaneous curvature of a membrane domain can induce vesicles made of homogeneous composition and the diameters of the vesicles are less than 50 nm. On the other hand, the compression on the membrane can cause the formation of vesicles with heterogeneous composition and the sizes of the obtained vesicles are similar to the size of the cytoskeleton corral. When both effects are considered, the compression on the membrane tends to facilitate the formation of vesicles originated from the curved membrane domain. In the HS and HE, we find that the sizes of microvesicles become more diverse comparing to the sizes of the microvesicles released from the healthy RBC, due to the reduced confinement from the cytoskeleton on the lipid bilayer. When the vertical connectivity between the lipid bilayer and the cytoskeleton is elevated, multiple vesicles, with sizes similar to the cytoskeleton corral dimension, are shed from the compressed membrane, whereas membrane with low vertical connectivity tends to release larger vesicles under the same compression ratio as above. It is noted that vesicles released from the HE RBCs could contain cytoskeleton components due to the fragmentation of the cytoskeleton while vesicles released from the HS RBCs are depleted of cytoskeleton components.

## 5.1. Introduction

Human RBC circulates in human body for about half million times during its 120 days lifespan to deliver oxygen from the lung to all tissues and transport carbon dioxide from tissues back to the lung. During its circulation, RBC loses surface area and the lipid content by approximately 20%, mainly by shedding of hemoglobin-contained vesicles (223, 224). RBCs, which constitute 40% of the total blood volume, are one of the major vesicle-secreting cells in the circulating blood (225), particularly in the course of some pathological conditions such as HS and sickle cell disease (SCD)(226). Membrane released vesicles are commonly grouped into two categories: namely nanovesicles and microvesicles, distinguished by the size of the vesicle. Nanovesicles have a mean diameter of approximately 25 nm (15, 227, 228), whereas microvesicles have an approximate diameter in a range of 60-300 nm (229). The formation of nanovesicles is caused by the aggregation of specific types of the membrane proteins or lipids which have mutual affinity and capability of forming membrane domains (229). Membrane domains with significant spontaneous curvatures can bud off from the RBC membrane and form vesicles (229). The compositions of the nanovesicles mainly contain the lipids or the proteins that constitute the membrane domains. Due to the small sizes of the nanovesicles (~25 nm) comparing to the size of the corrals formed by the cytoskeleton (~90 nm), the formation of the nanovesicles is not affected by the integrity of cytoskeleton. On the other hand, the formation of microvesicles originates from the lipid bilayer area that uncouples from the cytoskeleton during the aging of the RBC (230). Previous studies have shown that the RBC cytoskeleton is under stretch when it is tethered to lipid bilayer and therefore exerts a compression force on the lipid bilayer (13, 120). This compression force is balanced by the bending force induced by the membrane curvature (13, 118). As the cell ages, the stiffness of the cytoskeleton is increased by 20% (231) and its density is raised by 30-40% (30). As a result, the cytoskeleton imposes larger compressive forces on the cell membrane. Thus, the curvature of the membrane has to increase to accommodate the enhanced compression forces, leading to membrane buckling and buds formation at the scale of the cytoskeleton corral-size. The buds could grow in size with further increased compression forces, and subsequently detach from the membrane in order to release the bending stress (118, 151). The compositions of the microvesicles are more heterogeneous than those of the nanovesicles. For example, the lipid composition of the microvesicles is similar to that of the parent RBCs, but the components of the membrane cytoskeleton, such as the spectrin and actin, are in general not detected (232-234). In addition, the microvesicles contain the major membrane integral proteins in approximately the same relative amounts as the parent cell membrane (235).

Vesiculation from the RBCs of the patients suffering from blood disorders, such as HS and HE, raises interesting questions as the erythrocyte-released vesicles were found to be increased in the blood content of the patients with blood disorders (3, 4, 16, 20, 26, 145, 236, 237). After losing vesicles, the RBCs transfer from discocytes to irregular-shaped RBC with the reduced cell deformability, resulting in early removal from circulation by the spleen (3, 25, 145, 207, 236, 238). In HS, the defects in the proteins that tether the cytoskeleton to the lipid bilayer, such as ankyrin, protein 4.2 and band-3 proteins, stiffen the cytoskeleton and produce an area in the RBC membrane which is not supported by the cytoskeleton, facilitating the vesiculation. A common feature of HS is loss of membrane surface area and resultant change in cell shape from discocytes to spherocytes. In HE, the protein defects, such as α-spectrin, β-spectrin and protein 4.1 defects, occur in membrane cytoskeleton and disrupt the cytoskeleton. Since the cytoskeleton is the main load carrier of the RBC (237, 239, 240), the HE RBCs become vulnerable in the high shear-stress blood flow. As a result, they shed vesicles and undergo progressive shape transformations from the biconcave to the elliptical shapes. Various protein defects in the defective RBC membrane not only impact the RBC morphology, but also determine the contents of released vesicles. For example, the microvesicles shed from the normal RBCs have similar compositions as the parent RBCs (241), while microvesicles released from spectrin/ankyrin-deficient RBCs are enriched in band-3 proteins (242). In addition, protein defects that cause the disruption of the cytoskeleton or weakened connections between the lipid bilayer and cytoskeleton could promote the protein diffusion in the membrane and potentially facilitate the formation of the membrane domains and nanovesiculation (112, 115, 118, 197).

Numerous experimental studies have been conducted to understand the mechanism of RBC membrane vesiculation and the change of the RBC morphology. For example, experimental measurements showed that increasing the intracellular calcium concentration could promote

RBC vesiculation and transition of the cell shape from discocyte to echinocyte (243, 244). Microvesicles were shed from the tip of echinocyte spicules (244). Calcium-induced vesicles are characterized by being devoid of phosphatidylcholine (PC), spectrin, actin, and glycophorin, but preserving the band-3 proteins. Depletion of adenosine triphosphate (ATP) in the RBC is found to induce similar morphological changes as boosting calcium content, but the protein compositions in the released vesicles are different (234, 244, 245). In addition, membrane vesiculation is found to serve as a way of removing the nonfunctional proteins and auto-antigens through releasing the damaged membrane patches (224). Recently, de Vooght et al (246) proposed that mature RBCs are probably the only cells that do not secrete exosomes and shed only microvesicles. Although the experimental measurements facilitate the studies of the membrane vesiculation, they are not sufficient to fully understand the mechanism of the vesiculation. Therefore, analytical studies were performed to explain the cell morphology changes and vesicle formation (11, 13, 15, 74, 129, 151, 247-249). Sheetz and Singer first raised a bilayer-couple hypothesis which assumed the sequential transition of stomatocyte–discocyte–echinocyte was induced by the area difference between the two leaflets of the plasma membrane alone (250). However, the area–difference–elasticity (ADE) model developed based on the bilayer-couple hypothesis showed that as the area difference increased, the model tended to form buds instead of the echinocytosis (251). This impasse was solved by introducing the elastic energy of the cytoskeleton into the ADE model (252-254). Since the cytoskeleton is largely stretched or sheared during the budding formation, the energy corresponds to the budding configuration is significantly increased, leaving the echinocyte to be the preferred shape. Lim et al (13) model the stomatocyte–discocyte–echinocyte sequence by varying only the area difference between the two leaflets in the ADE model with the effect of the cytoskeleton. The authors also noted that area difference had equivalent effects as the spontaneous curvature on changing the cell shape because the spontaneous curvature naturally originated from the area difference between the two leaflets. The spontaneous curvature of the RBC membrane could result from the molecule asymmetry (9), area mismatch between two leaflets of the lipid bilayer *(13, 249)*, or protein-lipid hydrophobic mismatch, or protein-assisted curvatures *(255-257)*. It can be enhanced by incorporating the effects of increased anionic amphipaths, high salt, high pH, ATP depletion and cholesterol enrichment. Although the ADE model with consideration of the cytoskeleton successfully describes the cell shape transition, other mechanisms for the regulating the shape of cells were proposed. Another popular mechanism states that the budding and vesiculation in the membrane is provoked by compression on the lipid bilayer induced by the stiffened cytoskeleton. The increased stiffness of the cytoskeleton may result from the depleted ATP as RBC ages (151, 229), or cause by the protein defects which weaken the vertical connections between the lipid bilayer and cytoskeleton, such as in the HS RBCs (3, 145, 236). Spangler et al. (258) introduced a lipid membrane sphere model by combining particle-based lipid bilayers to an elastic meshwork to study cytoskeleton-induced blebbing. Their simulation results clearly demonstrate that membrane blebbing results from a localized disruption or a uniform contraction of the cytoskeleton. The enhanced vesiculation observed in the HS RBC can be explained by the Spangler et al. (258)'s simulation and thus favor the mechanism that the budding and vesiculation in the membrane are induced by cytoskeleton-induced compression. Microvesiculation in the RBC membrane is more likely to result from the spontaneous curvature and cytoskeleton-induced compression simultaneously because the increased intracellular calcium or depleted ATP not only promote asymmetry of leaflets, but also activate sproteolytic enzymes like calpain which breaks down the tethering points between the membrane cytoskeleton and the lipid bilayer (236). Therefore, the exact underlying mechanisms of RBC vesiculation are not completely deciphered.

Although previously developed model succeeded in explaining the cell morphology change and membrane budding and vesiculation, there are still questions that remain unanswered. For example, the spontaneous curvatures and boundary line tension of the lipid rafts or protein-enriched membrane domain, which play important roles in the RBC signaling, trafficking and material transport (259-262), can results in a domain morphological transition from a flat to a dimpled state (263). The effects of the elastic properties on formation of the membrane domain as well as the interactions between the domains were studied in (263, 264). However, how does the spontaneous curvature of the membrane domain affect the microvesiculation is not clear. Also, the increase in raft lipids and sphingolipids could excites a lipid raft coalescence process by clustering the raft components, which potentially promotes the vesiculation from the membrane domain (265). In addition, mechanism of the vesiculation in RBCs in the pathological conditions such as HS and HE is still not well understood due to the complication induced by the different protein defects. In order to answer the above questions, we apply our previously developed two-component CGMD RBC membrane model (see Fig. 5.1*A-C*) to simulate the RBC membrane vesiculation in the normal RBC membrane and in the RBC membrane with protein defects. An important feature of the applied membrane model lies in introducing the rotational degrees of freedom into the interactions between the lipid CG particles, glycophorin and band-3 proteins, which allow us to introduce spontaneous curvatures into the lipid bilayer domains. In addition, the applied model explicitly represent the lipid bilayer, band-3 proteins and the cytoskeleton by CG particles which can be used to study the interactions between the lipid bilayer and cytoskeleton and the function of the cytoskeleton on maintaining the integrity of the RBC membrane. First, we simulation the vesiculation induced by the spontaneous curvature of the membrane domain and vesiculation induced by the compression on the lipid bilayer, respectively. Second, we simulate vesiculation in the RBC membrane due to the spontaneous curvature of the membrane domain and due to the compression on the lipid bilayer. At last, we study the vesiculation in the RBC membrane with protein defects, such as in HS and HE cells and investigate the effects of protein defects on the vesiculation(266).

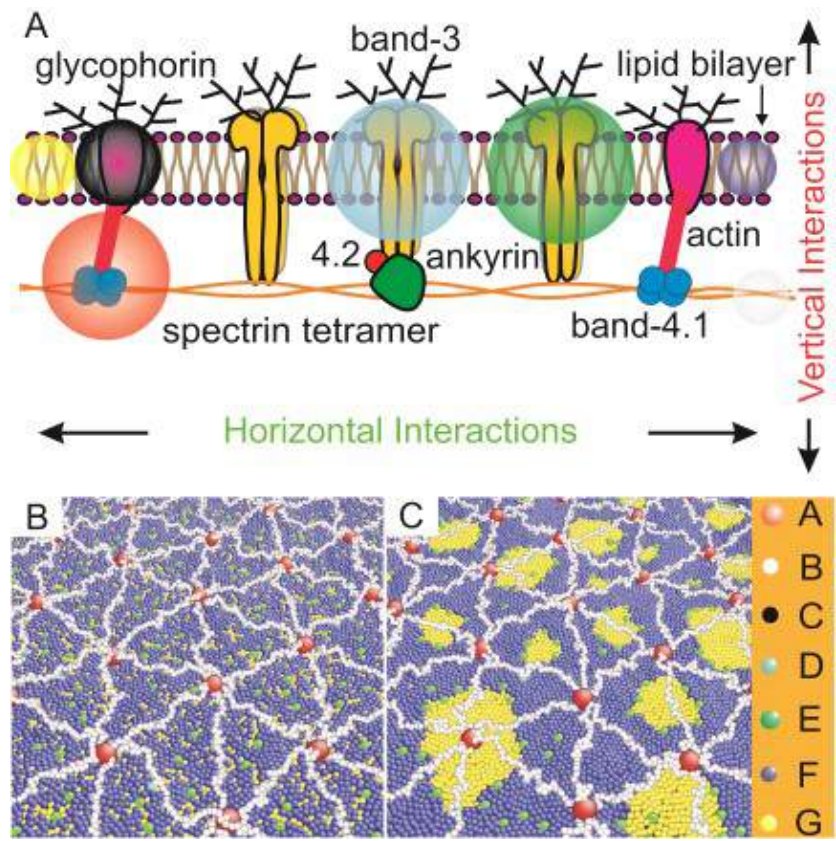

Figure 5.1. (*A*) Schematic of the human RBC membrane. The blue and yellow spheres represent two types of lipid particles corresponding to membrane areas with different spontaneous curvatures. The red sphere signifies an actin junctional complex. The white sphere represents a spectrin particle and the black sphere represents a glycophorin particle. The light blue circle corresponds to a band-3 complex that is connected to the spectrin network while green circle

represents a band-3 complex that does not connect to the lipid bilayer. (*B*) Top view of the membrane model with two types of lipid CG particles completely mixed. (*C*) Top view of the membrane model after the G type particles aggregate into membrane domains. "A" type particles represent actin junctional complexes, "B" type particles represent spectrin proteins, "C" type particles represent glycophorin proteins, "D" type particles represent a band-3 complex that are connected to the spectrin network ("immobile" band-3), "E" type particles represent band-3 complex that are not connected to the network ("mobile" band-3), "F" type particles represent majority of lipid particles and "G" type particles represent minority of the lipid particles.

## 5.2. Model and simulation method

The two-component CGMD human RBC membrane model applied in this work explicitly comprises both the lipid bilayer and the cytoskeleton. Three types of particles are introduced to represent lipid bilayer (see Fig. 5.1*A*). The blue and grey color particles signify a cluster of lipid molecules, which have a diameter of 5 nm, a similar value to the thickness of the lipid bilayer. The difference between the two types of lipid particles will be introduced later. The black particles denote glycophorin proteins with the same diameter of the lipid particles. The third type of particles denotes band-3 complexes. One third of band-3 complexes, represented by the light bule particles, are connected to the spectrin filaments. The rest of the band-3 complexes can freely diffuse in the lipid bilayer and they are represented by green particles. The cytoskeleton is comprised of the hexagonal spectrin filaments connected at actin junctions. The actin junctions are represented by the red particles in Fig.5.1*A* and it is connected to the lipid bilayer via glycophorin. Spectrin is a protein tetramer formed by head-to-head association of two identical heterodimers which consists of an α-chain with 22 triple-helical segments and a β-chain with 17 triple-helical segments. In the proposed model, therefore, the spectrin filament is represented by 39 spectrin particles (white particles) connected by unbreakable springs. The details of the potentials applied in the cytoskeleton of the membrane model and the geometry of the CG particles can be found in the authors' previous work (211). The lipid particles, the glycophorin particles and band-3 particles carry both translational and rotational degrees of freedom ($\mathbf{x}_i$, $\mathbf{n}_i$), where $\mathbf{x}_i$ and $\mathbf{n}_i$ are the position and the orientation (director vector) of particle $i$, respectively. The rotational degrees of freedom obey the normality condition $|\mathbf{n}_i| = 1$. Thus, each particle effectively carries 5 degrees of freedom. $\mathbf{x}_{ij} = \mathbf{x}_j - \mathbf{x}_i$ is defined as the distance vector between particles $i$ and $j$. Correspondingly, $r_{ij} \equiv |\mathbf{x}_{ij}|$ is the value of the distance, and $\hat{\mathbf{x}}_{ij} = \mathbf{x}_{ij}/r_{ij}$ is a unit vector. The lipid particles, the glycophorin particles and the band-3 particles interact with one another via a pair-wise additive potential

$$u_{mem}(\mathbf{n}_i, \mathbf{n}_j, \mathbf{x}_{ij}) = u_R(r_{ij}) + A(\alpha, a(\mathbf{n}_i, \mathbf{n}_j, \mathbf{x}_{ij})) u_A(r_{ij}), \quad (5.1)$$

$$\begin{cases} u_R(r_{ij}) = \varepsilon\left((R_{cut,mem} - r_{ij})/(R_{cut,mem} - r_{eq})\right)^8 & \text{for } r_{ij} < R_{cut,mem} \\ u_A(r_{ij}) = -2\varepsilon\left((R_{cut,mem} - r_{ij})/(R_{cut,mem} - r_{eq})\right)^4 & \text{for } r_{ij} < R_{cut,mem} \\ u_R(r_{ij}) = u_A(r_{ij}) = 0, & \text{for } r_{ij} \geq R_{cut,mem} \end{cases} \quad (5.2)$$

where $u_R(r_{ij})$ and $u_A(r_{ij})$ are the repulsive and attractive components of the pair potential, respectively. $\alpha$ is a tunable linear amplification factor and it is chosen to be 1.55. The function $A(\alpha, a(\mathbf{n}_i,\mathbf{n}_j,\mathbf{x}_{ij})) = 1+\alpha(a(\mathbf{n}_i,\mathbf{n}_j,\mathbf{x}_{ij}) -1)$ which tunes the energy well of the potential, regulate the fluid-like behavior of the membrane. The interaction between two CG particles depends not only on their distance but also on their relative orientation via the function $a(\mathbf{n}_i, \mathbf{n}_j, \mathbf{x}_{ij})$ which varies from -1 to +1 and adjusts the attractive part of the potential. We specify that $a=1$ corresponds to the case when $\mathbf{n}_i$ is parallel to $\mathbf{n}_j$ and both are normal to vector $\mathbf{x}_{ij}$ $((\mathbf{n}_i \uparrow\uparrow \mathbf{n}_j) \perp \hat{\mathbf{x}}_{ij})$, and the value $a=-1$ to the case when $\mathbf{n}_i$ is anti-parallel to $\mathbf{n}_j$ and both are perpendicular to vector $\mathbf{x}_{ij}$ $((\mathbf{n}_i \uparrow\downarrow \mathbf{n}_j) \perp \hat{\mathbf{x}}_{ij})$. The former instance is energetically favored due to the maximum attractive interaction between particles $i$ and $j$, while the latter is energetically disfavored due to the maximum repulsive interaction. One simple form of $a(\mathbf{n}_i, \mathbf{n}_j, \mathbf{x}_{ij})$ that captures these characteristics is

$$a(\mathbf{n}_i, \mathbf{n}_j, \hat{\mathbf{x}}_{ij}) = \mathbf{n}_i \cdot \mathbf{n}_j + (\beta - \mathbf{n}_i \cdot \hat{\mathbf{x}}_{ij})(\beta + \mathbf{n}_j \cdot \hat{\mathbf{x}}_{ij}). \quad (5.3)$$

where $\beta$ is the parameter that adjusts the curvature of the lipid bilayer, which allow us to introduce a variety of spontaneous curvatures into lipid bilayer. The effects of function $A(\alpha, a(\mathbf{n}_i, \mathbf{n}_j, \mathbf{x}_{ij}))$ and the details of the applied potentials between the lipid CG particles were discussed in the previous works (111, 211). The system consists of $N = 32796$ CG particles. The dimension of the membrane is approximately 0.8 μm × 0.8 μm. The temperature of the system is maintained at $k_BT/\varepsilon = 0.22$ with the Nose-Hoover thermostat. The model was implemented in the NAT ensemble (95, 131, 132). Since the model is solvent-free and the membrane is a two-dimensional structure, we controlled the projected area instead of the volume. The projected area was adjusted to result in zero tension for both the lipid bilayer and the cytoskeleton at the equilibrium state. The details about the potentials applied in the membrane model, geometries of the CG particles and the timescales of the model can be found in the authors' previous work (111, 211).

### 5.3. Results and discussion

5.3.1 Vesiculation in health RBC membrane

5.3.1.1 Vesiculation due to spontaneous curvature of membrane domain

First, we simulate the vesiculation due to spontaneous curvature of membrane domain. Membrane domains, such as lipid rafts, have draw significant attentions as increasing evidences have shown that they play an crucial role in regulating cellular processes including cell polarity, protein trafficking and signal transduction (263, 267-272). Membrane domains consist of specific type of lipids or proteins which are more ordered and tightly packed than the surrounding bilayer. For example, lipid rafts in plasma membranes are enriched in sterol- and sphingolipids with size ranging from 10 to 200 nm (259-261). In order to create phase separation and form membrane domains in the applied model, we randomly assign the lipid CG particles as type F and G. The number ratio between component G (blue) and F (gray) is 2:8 and it is conserved throughout simulations. Type G particles are considered to be the minority and clusters of G particles are treated as domains. The line tension between two types of lipid particles F and G is simulated by assuming that the association energies between the F-F and G-G particles are the same, but the association energy between the F-G particles is only 70% of the association energy between the F-F and G-G. We start the simulation with two types of lipid particles completely mixed, as shown in Fig. 5.1*B*. Driving by the different association energies, the G particles are aggregating gradually to small islands and the islands can merge to become the domains (see Fig. 5.1*C*). The domain size ranges from 20 to 100 nm, consistent with the range of 10 to 200 nm observed in (259-261). Then, we introduce a variety of spontaneous curvatures to the G particles to mimic different domain curvatures which caused by the mismatch between two leaflets **or** protein-assisted curvatures. In our model, the spontaneous curvature can be controlled by tuning the parameter $β$ in the potential introduced in Eq.(5.3). Fig. 5.2$A$ shows that when $β = 0.1$, the domains only bud off from the membrane and no vesicles are formed, meaning that the energy resulting from the line tension and spontaneous curvature is not large enough to drive the vesiculation. As the spontaneous curvature increases, the buds become more curved. When the $β$ is increased to 0.18, vesiculation is observed and the released vesicles are made of the G type particles only (see Fig. 5.2$B$). This means the spontaneous curvature of membrane domain has to be large enough to drive the vesiculation and the vesicles induced by the spontaneous curvature has the same compositions as the membrane domains. As the $β$ continues to increase, more vesicles are formed from the membrane (see Fig. 5.2$C$). Since the size of the vesicle depends on the spontaneous curvature of domain of the G particles, the maximum size of the vesicles induced by the spontaneous curvature is obtained at $β = 0.18$, and the diameter of the obtained vesicles is about 40 nm, which is larger than the size of the nanovesicles but smaller than the lower end of the 60-300 nm range for the size of the microvesicles (118). With the spontaneous curvature increasing, the sizes of the vesicles can drop down to 20 nm, similar to the size of the nanovesicles. Owing to the smaller vesicle size comparing to the typical size of the cytoskeleton corral (~90 nm), the cytoskeleton probably does not affect the vesiculation induced by the spontaneous curvature.

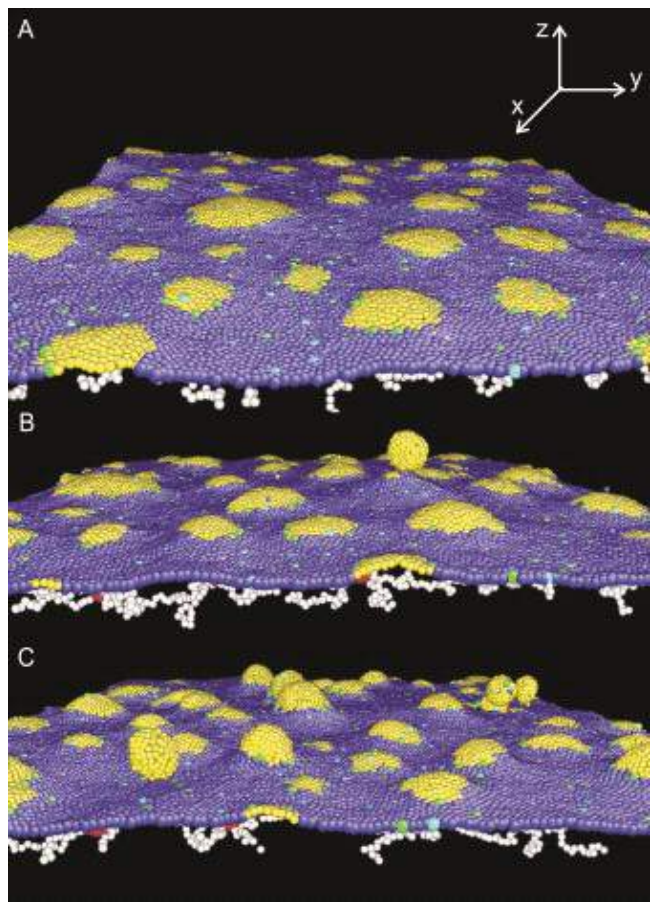

Figure 5.2. Vesiculation induced by the spontaneous curvature of the membrane domains. (*A*) When the spontaneous curvature is small ($\beta = 0.1$), the membrane domains only bud off from the membrane and no vesicle is formed. (*B*) When $\beta$ is increased to 0.18, vesiculation is observed and the released vesicles are made of the G type particles only. (*C*) When $\beta$ is increased to 0.24, more vesicles are formed from the membrane.

5.3.1.2 Vesiculation due to the compression on the lipid bilayer

In this section, we simulate vesiculation in the RBC membrane model driven by the compression on the lipid bilayer. When the cytoskeleton is tethered to the lipid bilayer, it is under tension because the cytoskeleton can shrink by 3-5 times after the lipid bilayer is removed (273). On the other hand, the cytoskeleton applies a compressive force on the lipid bilayer. As the RBC ages, the depleted ATP level in the cell leads to increased density and stiffness of the cytoskeleton, which impose a larger compressive force on the lipid bilayer and thus facilitate the vesicle formation (151). As a result, RBC constantly loses cell membrane through shedding vesicles during its circulation in the human body (151, 274). The size of RBC-released vesicles is similar to the size of the corral of the RBC cytoskeleton (275). In this section, we simulate the effects of the stiffened cytoskeleton by applying compression on the applied RBC membrane model. In the applied RBC membrane model, the soft cytoskeleton shrink after compression while the lipid bilayer is incompressible. Thus the applied compression on the membrane increases the area ratio between the lipid bilayer and cytoskeleton, similar to the method applied in a previous study (258). The compression ratios, which is defined as the percentage of the area reduced from initial projected area of the membrane, are selected to be $R_{compression}$ = 2%, 5%, 10%, 15% and 18%, respectively. In order to better understand the membrane vesiculation process, we explain our simulation results based on Helfrich's membrane model (9). It is known that the free energy function of non-sheared membrane is controlled by the membrane curvature induced by the thermal fluctuation, spontaneous curvature, membrane domain line tension and cytoskeleton stretching energy (ref.). The expression for the free energy is described by

$$F = \kappa_b/2 \int (C_1 + C_2 - C_0)^2 \, dA_l + \int \gamma \, dA_l + K_\alpha/2 \int \alpha^2 \, dA_c \qquad (5.4)$$

where $(C_1+C_2)/2$ is the mean curvature originated from the thermal fluctuation of the membrane, $C_0$ is the spontaneous curvature, $\gamma$ is the line tension at the boundary of the membrane domains, $K_\alpha$ is linear elastic moduli of the cytoskeleton and $\alpha$ is the cytoskeleton local area invariants (276). In addition, the lipid membrane and integral membrane proteins are constrained when they are moving towards the negative z-direction by applying a harmonic potential to mimic the effect of the cytoplasm inside the RBC. The potential is described as

$$V_{confine} = \frac{1}{2} K_{confine} z^2 \tag{5.5}$$

where $K_{confine}$ is the confinement coefficient and $z$ is the distance measured from the mid-plain of the supercell. $K_{confine}$ is chosen to be $\varepsilon/\sigma^2$.

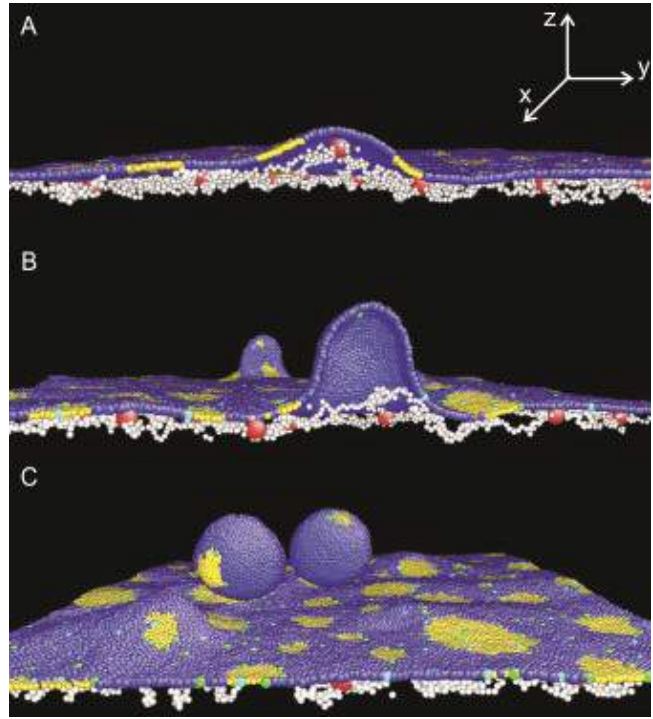

Figure 5.3. Vesiculation induced by compression on the lipid bilayer. (*A*) When the compression ratio $R_{compression}$ = 2%, only a small bud is created on the membrane and the cytoskeleton conforms to the bud. (*B*) When $R_{compression}$ is increased to 5%, the bud grows up and the cytoskeleton retracts from the budding area. (*C*) When $R_{compression}$ is increased to 15%, two vesicles are released from the membrane and they consist of both G and F lipid particles.

Fig. 5.3*A* shows that when $R_{\text{compression}}$ = 2%, only buds are created on the membrane and the cytoskeleton is observed to conforms to buds. Under this condition, the free energy of the membrane is controlled by the curvature energy of the lipid bilayer and the elastic energy of the cytoskeleton induced by the local stretch on the budding areas. As the compression ratio increases, the buds grow up and the cytoskeleton maintains conforming to the growing buds until the stretching force is strong enough to pull the cytoskeleton off from the budding areas. Then the budding area is unsupported from the cytoskeleton when $R_{\text{compression}}$ = 5% and 10%, as shown in the Fig. 5.3*B* and the cytoskeleton becomes stress-free. Therefore, the free energy of the membrane is only controlled by the bending energy of the lipid bilayer. Further increase in the compression ratio is balanced by the increased curvature in budding area. At $R_{\text{compression}}$ = 15%, the neck of the bud close itself and pinch the vesicle off from the membrane to release the bending energy, as shown in the Fig. 5.3*C*. It is noted that the released vesicles have a more heterogeneous composition (consists of both G and F lipid CG particles) compared to the vesicles that form due to domain spontaneous curvature. This observation indicates that the line tension induced by the different association energy between the type A and B lipid particles do not play an important role in the vesiculation caused by compression on the lipid bilayer. In addition, the sizes of the vesicles are ~90 nm, similar to the size of the cytoskeleton corral, larger than the vesicles induced by the membrane domain curvatures, which is consistent with the experimental observations (16, 206, 236). A larger compression ratio ($R_{\text{compression}}$ = 18%) in our simulation create multiple vesicles at the similar sizes, instead of a single larger size of vesicle. The larger-size microvesicles may result from the breakage between the band-3 and spectrin

filaments due to the depletion of ATP in the cell-aging process, which will be discussed in the section 3.2.2 below.

5.3.1.3 Vesiculation due to the combined effects

In the last two sections, we describe the membrane vesiculation induced by the spontaneous curvature of membrane domain and by the compression on the lipid bilayer, respectively. In the RBC vesiculation process, these two effects could co-exist. For example, the curved membrane domains or lipid rafts happen to locate in the membrane areas which are not supported by the cytoskeleton. Therefore, in this section, we simulate the membrane vesiculation under the combined effects of spontaneous curvature of membrane domain and compression. First, we simulate the conditions where the spontaneous curvature of the membrane domain is small so that the domains only form buds but not vesicles. The $R_{compression}$ are selected to 5% and 10%, respectively. We find that when the spontaneous curvature is very small ($\beta = 0.05$), the vesiculation processes at $R_{compression} = 5\%$ and 10%, only produce a big bud, as shown in Fig. 5.4*A*, are similar to the cases when membrane is under compression and no spontaneous curvature is introduced. When $R_{compression}$ is increased to 15%, vesicles consisting of two types of lipid particles are released. This means that small spontaneous curvatures do not affect the membrane vesiculation due to the compression. However, when the spontaneous curvature is increased to $\beta = 0.1$, vesicles made of only G lipid particles are observed at $R_{compression} = 5\%$ and 10% (see Fig. 5.4*B* and 5.4*C*). As shown in Fig.5. 2*A*, the spontaneous curvature of $\beta = 0.1$ alone cannot drive vesiculation. Therefore, the compression applied here promotes the vesiculation of the membrane domain. It is noted that the compression on the lipid bilayer prefer to stimulate the growth of the curved membrane domain, instead of promoting the buds from membrane patches made of two types of lipid particles, like the case of $\beta = 0.05$ under the same compression. When $R_{compression} = 5\%$ and $\beta = 0.1$, two vesicles with size of ~60 nm are observed (see Fig. 5.4B). When $R_{compression} = 10\%$, three vesicles are released from the membrane with sizes ranging from 40-60 nm, depending on the size of the originated membrane domains (see Fig. 5.4C). As $\beta$ increases, more vesicles made of only G lipid particles are released from the membrane at $R_{compression} = 5\%$ and 10%. For the cases of $\beta >= 0.18$, where the small-sized vesicles can be formed spontaneously without compression, our simulations show that when the membrane is compressed at $R_{compression} = 5\%$ and 10%, the self-driven small-sized vesicles are first released from the membrane and then larger size vesicles stimulated by the compression are observed.

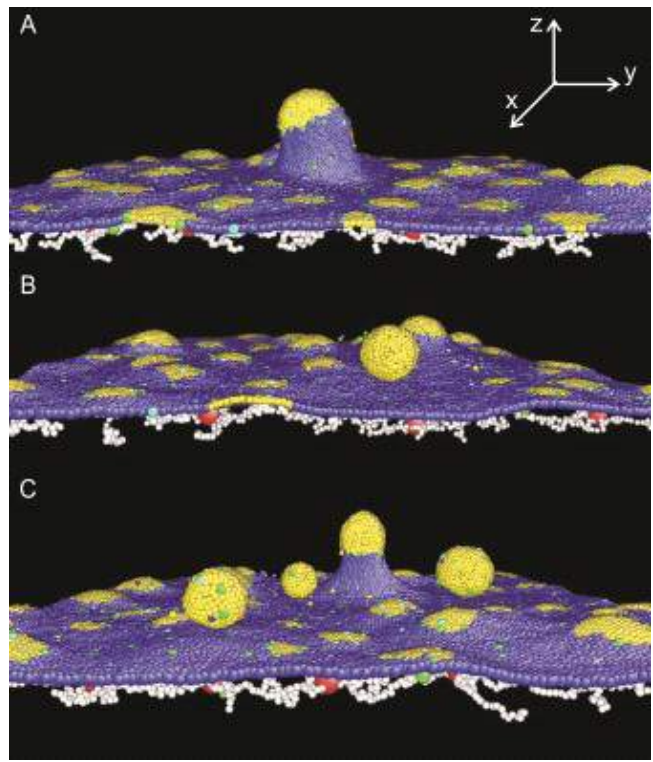

Figure 5.4. Vesiculation induced by the combined effects of spontaneous curvature and compression on the lipid bilayer. (A) When the spontaneous curvature is very small ($\beta = 0.05$), the vesiculation processes at compression ratio of $R_{compression} = 5\%$ and 10%, only produce a big

bud. (*B*) When $R_{compression}$ = 5% and $\beta$ = 0.1, one vesicle is released. The vesicle has a size of ~60 nm and consists of only G lipid particles. (*C*) When $R_{compression}$ = 10%, three vesicles are released from the membrane with sizes ranging from 40-60 nm.

5.3.2. Vesiculation in defective RBC membrane

In this section, we simulate the vesiculation in the RBC membrane with protein defects, such as in the HS and HE RBCs, and study the effects of the percentage of protein defects on the vesiculation. The defects in the cytoskeleton proteins or in the proteins that tether the cytoskeleton to the lipid bilayer can create an area in the membrane which is not supported by the cytoskeleton, and thus facilitate the RBC vesiculation. As a results, erythrocyte-released vesicles were found to be increased in the blood content of the patients with blood disorders (3, 4, 16, 20, 26, 145, 237, 277), leading to the altered cellular morphology, reduced cell deformability and early removal by the spleen (3, 25). Here, we introduce the protein defects into the applied membrane model and simulate the vesiculation process in the HS and HE RBC membrane, respectively.

5.3.2.1 Vesiculation due to the spontaneous curvature of the membrane domain in HS and HE cell membrane

First, we study the effects of the protein defects in the vertical interaction between the lipid bilayer and cytoskeleton, and the protein defects in the horizontal interaction within the cytoskeleton on the vesiculation originated from the membrane domain. We repeat the simulation conducted in the section 3.1.1, with the connectivity in the vertical interaction and horizontal interaction selected to be $C_{vertical}$ = 50% and $C_{horizontal}$ = 50%, respectively. We find similar results to the membrane vesiculation in the normal RBC membrane, meaning that the protein defects in the membrane do not affect the vesiculation caused by the spontaneous curvature. This probably results from the fact that the sizes of the vesicles induced by the spontaneous curvature of the membrane domain are smaller than the size of the corral in the cytoskeleton.

5.3.2.2 Vesiculation due to the compression in the HS cell membrane

Second, we simulate the vesiculation in the membrane with protein defects in the vertical interactions. It could be caused by the defective band-3, protein 4.2 and ankyrin proteins in the HS (3, 145, 236). In addition, the decreased $C_{vertical}$ also can be induced by the ATP depletion as RBC ages (151, 229). The effects of protein defects in the vertical interaction are represented by breaking the connections between band-3 and spectrin particles. The $R_{compression}$ are selected to be 5%, 10%, 15% and 18% of the initial projected area of the membrane. The band-3 and spectrin connectivity (vertical connectivity) $C_{vertical}$ between the lipid membrane and cytoskeleton is reduced from 100% to 0%. When the $R_{compression}$ is low ($R_{compression}$ = 5%), Fig. 5.5*A* shows that only one bud is created from the membrane and no vesicle is released from the membrane, similar to cases for the normal RBC membrane. When a higher $R_{compression}$ = 10% is applied, the unsupportive budding area becomes larger to accommodate the increased compression. Under this condition, vesiculation still does not occur for all $C_{vertical}$ and only two buds are formed. At large $R_{compression}$ (15% and 20%), we find vesiculation for all the $C_{vertical}$, as shown in the Fig 5.5*C*. However, vesiculation process is different for the low $C_{vertical}$ and high $C_{vertical}$. At high $C_{vertical}$, the vesiculation starts from the area within the corral of the cytoskeleton and thus creates vesicles with sizes similar to the corral (~90nm), while at the low $C_{vertical}$, buds can migrate and merge to larger buds and then form vesicles with sizes of ~400 nm, consistent with (229). This means that high vertical connectivity $C_{\text{vertical}}$ in the normal RBC membrane can slow down or prevent the migration of the buds, therefore favors the formation of vesicles with sizes similar to the size of corral. This is clearly demonstrated in the Fig. 5.5*C* which shows that at high $R_{\text{compression}} = 15\%$ and 18%, the membrane tends to release multiple corral-sized vesicles at high $C_{\text{vertical}}$, while when the $C_{\text{vertical}}$ is low, the sizes of the released become more diverse (from 90 nm to 400 nm) because there is less constrain from the cytoskeleton. These simulation results explain the various sizes of vesicles shed by RBC (16, 229, 236). In addition, our simulations illustrate that the cytoskeleton does not fragment during vesiculation, resulting in vesicles completely depleted of cytoskeleton components, as shown in the Fig. 5.5*B*. This result is consistent with the experimental observation on the development of the spherocytes from the RBCs with defective band 3 or protein 4.2, but normal amounts of spectrin in (278). The vesicles released from the HS RBC membrane all contain band-3 proteins (see Fig. 5.5*B*), which is consistent with the hypothesis raised by Eber and Lux (16). Since we do not assign band-3 particles structural functions on stabilizing the membrane in this model, the vesiculation in the HS RBC membrane here mainly result from membrane area which is not supportive by the cytoskeleton, instead of from band-3 deficient area.

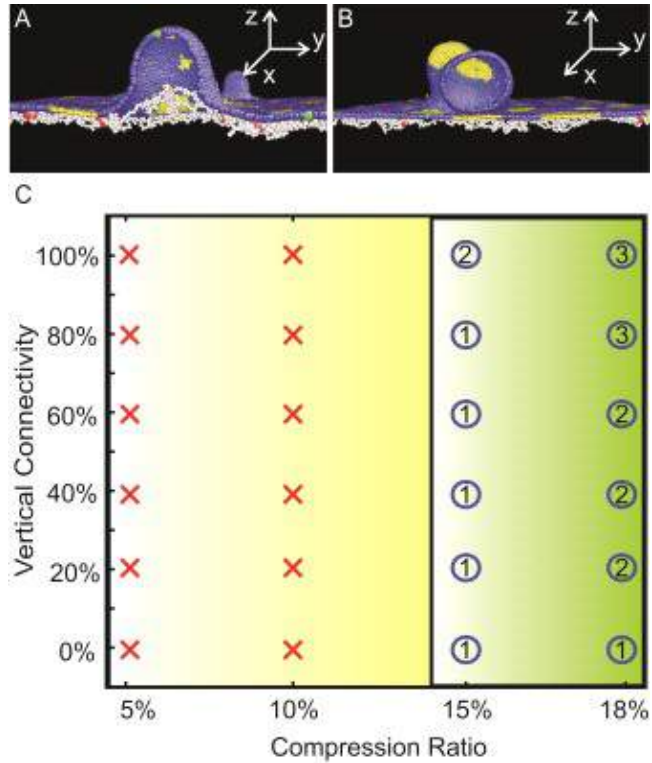

Figure 5.5. Vesiculation in the membrane with protein defects in the vertical interactions. (*A*) When the compression ratio $R_{compression}$ is low ($R_{compression}$ = 5%), only one bud is created from the membrane for all the selected $C_{vertical}$. (*B*) When $R_{compression}$ = 15% and vertical connectivity $C_{vertical}$ = 60%, one vesicle is created. (*C*) Summary of dependence of membrane vesiculation on vertical connectivity $C_{vertical}$ and compression ration $R_{compression}$. In the figure, × indicates no vesiculation occurs. ① indicates one vesicle is observed. ② indicates two vesicles are observed. ③ indicates three vesicles are observed.

5.3.2.3 Vesiculation due to the compression in the HE RBC membrane

At last, we simulate the vesiculation in the membrane with protein defects in the horizontal interactions, such as α-spectrin, β-spectrin and protein 4.1 defects in the HE cell membrane. The protein defects in the horizontal interaction are represented by dissociations of spectrin chains at the spectrin dimer-dimer interactions. The cytoskeleton connectivity (horizontal connectivity) $C_{horizontal}$ = 0%, 20%, 40%, 60%, 80% and 100% are tested, respectively. The compression ratios $R_{compression}$ are selected to be 5%, 10% and 15%, respectively. When the $R_{compression}$ is low ($R_{compression}$ = 5%), only one bud is created from the membrane and no vesicle is released from the membrane for all the selected $C_{horizontal}$, similar to cases for the normal RBC membrane and HS cell membrane. However, it is noted that when the $C_{horizontal}$ is low, such as the case of $C_{horizontal}$ = 20% shown in the Fig. 5.6A, the fragments of the cytoskeleton are observed to go with the bumped lipid membrane as the cytoskeleton is disrupted at this low $C_{horizontal}$. When a higher $R_{compression}$ = 10% is applied, the unsupportive budding area grows up to accommodate the increased compression. Vesiculation does not occur at $C_{horizontal} \geq 60\%$ and only two buds are formed. But as the $C_{horizontal}$ decreases, vesicles are observed, as shown in the Fig. 5.6B, as more membrane area becomes unsupported from the cytoskeleton. The vesiculation from the membrane with reduced horizontal connectivity $C_{horizontal}$ indicates that the cytoskeleton of the RBC has the functions of preventing the membrane from losing vesicles. Moreover, it is noted that at $R_{compression}$ = 10%, no vesiculation is observed from the HS cell membrane for all the selected $C_{vertical}$. The difference of the vesiculation in the HS and HE cells indicates that the cytoskeleton plays the major role in maintaining the stability and integrity of the RBCs. When $R_{compression}$ is increased to 15%, Fig. 5.6C shows that vesiculation occurs for all the selected $C_{horizontal}$. Similar to the results obtained from the HS cell membrane, two corral-sized vesicles are created at high $C_{horizontal}$ ($C_{horizontal} \geq 60\%$) while one larger vesicle is found at low $C_{horizontal}$ ($C_{horizontal} \leq 40\%$) due to the reduced confinement of the cytoskeleton on the lipid bilayer. It is interesting to see that the vesicles resulting from the HE RBC membrane may contain cytoskeleton fragments (see Fig. 5.6B), distinguished from the vesicles shed from the normal RBC and HS RBCs which are depleted from the cytoskeleton components.

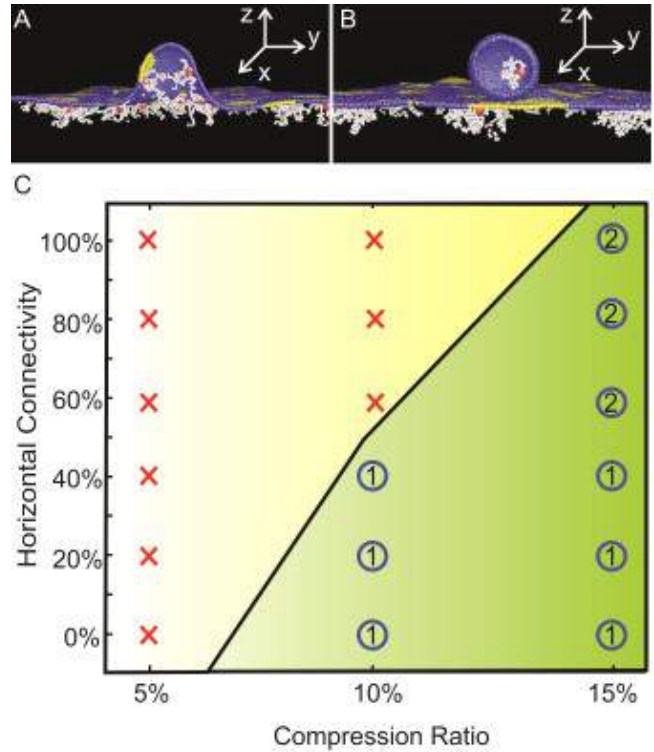

Figure 5.6. Vesiculation in the membrane with protein defects in the horizontal interactions. (*A*) When the compression ratio $R_{compression}$ = 5% and horizontal connectivity $C_{horizontal}$ = 20%, only one bud is formed from the membrane. (*B*) When $R_{compression}$ = 10% and $C_{horizontal}$ = 40%, one vesicle with cytoskeleton fragment inside is created (*C*) Summary of dependence of membrane vesiculation on horizontal connectivity $C_{vertical}$ and compression ratio $R_{compression}$. ① indicates one vesicle is observed. ② indicates two vesicles are observed. In the figure, × indicates no vesiculation occurs.

## 5.4 Summary

We applied a two-component CGMD RBC membrane model to simulate vesicle formation in the normal RBC membrane and in the HS and HE RCB membrane. We assumed vesiculation induced by differences in the spontaneous curvatures of membrane domains and vesiculation induced by compression on the lipid bilayer. Our results show that the difference in the spontaneous curvatures of the RBC membrane domains induce nanovesiculation or the formation of small-sized microvesicles that consist of same type of particles. The compression on the membrane caused the generation of cytoskeleton corral-sized vesicles with heterogeneous composition. When both effects were considered, the compression on the membrane tends to facilitate the formation of vesicles that were made of the same type of particles. Furthermore, we modeled the vesiculation in HS and HE RBC membrane by reducing the vertical and horizontal connectivities in our membrane model. We found that the sizes of the vesicles released from the HS and HE cell membrane are more diverse than the sizes of the vesicles released from the normal RBC membrane. This is due to the higher constrains induced by the cytoskeleton on the lipid bilayer in the normal RBC membrane compared to the defective RBC membranes. When the vertical or horizontal connectivity is high, multiple vesicles with sizes similar to the cytoskeleton corral size are created from the membrane under large compression ratios. In contrast, membrane with low vertical or horizontal connectivity tends to produce larger vesicles. Moreover, we found that under the same compression ratio, membrane with protein defects in the horizontal interactions are more likely to lose vesicles than in membranes with defects in the vertical interactions. This means that the cytoskeleton plays a major role in maintaining the stability and integrity of the RBCs. In addition, vesicles released from HE RBC membrane may contain fragments of cytoskeleton while the vesicles shed from HS RBC membrane are devoid of cytoskeleton components as the cytoskeleton is still intact.

# Chapter 6.

# Modeling Sickle Hemoglobin Fibers as Four Chain of Coarse-Grained Particles

**Abstract**

The intracellular polymerization of deoxy-sickle cell hemoglobin (HbS) has been identified as the main cause of sickle cell disease. Therefore, the material properties and biomechanical behavior of polymerized HbS fibers is a topic of intense research interest. A solvent-free coarse-grain molecular dynamics (CGMD) model is developed to represent a single hemoglobin fiber as four tightly bonded chains. A finitely extensible nonlinear elastic (FENE) potential, a bending potential, a torsional potential, a truncated Lennard-Jones potential and a Lennard-Jones potential are implemented along with the Langevin thermostat to simulate the behavior of a polymerized HbS fiber in the cytoplasm. The parameters of the potentials are identified via comparison of the simulation results to the experimentally measured values of bending and torsional rigidity of single HbS fibers. After it is shown that the proposed model is able to very efficiently simulate the mechanical behavior of single HbS fibers, it is employed in the study of the interaction between HbS fibers. It is illustrated that frustrated fibers and fibers under compression require a much larger interaction force to zipper than free fibers resulting to partial unzippering of these fibers. Continuous polymerization of the unzipped fibers via heterogeneous nucleation and additional unzippering under compression can explain the formation of HbS fiber networks and consequently the wide variety of shapes of deoxygenated sickle cells.

## 6.1. Introduction

A red blood cell (RBC) contains 25-30% hemoglobin whose main function is to carry oxygen from the lungs to tissues. Normal RBCs contain hemoglobin A (HbA) that has 2 subunits denoted α and 2 denoted β. However, RBCs of people suffering from sickle cell disease contain HbS in which a charged surface group glu at β6 is replaced by a hydrophobic group val (27). This replacement promotes polymerization of deoxygenated hemoglobin at high enough concentrations resulting to abnormal sickle-shaped RBCs (see Fig. 6.1A) which are less compliant and more adherent than normal RBCs. Because of increased stiffness and cell adherence to the endothelium, the circulation of sickle cells through the body's narrow blood vessels, such as arterioles, venules, and capillaries, is often obstructed resulting in infarctions and organ damage (23, 50, 279). In addition, overt stroke caused by occlusion of large cerebral arteries is one of the main complications of sickle-cell disease (62, 280-282). Since polymerized HbS fibers are stiff, they create protrusions and cause vesiculation of the RBC membrane (35). The increased stiffness of HbS fibers is considered to be the main reason for the wide variety of shapes that deoxygenated RBCs from patients with SCD acquire (27, 35, 283). Since HbS fibers play a very important role in causing sickle cell disease, the material properties and thermal behavior of HbS during the polymerization process have been widely studied for an extended period of time.

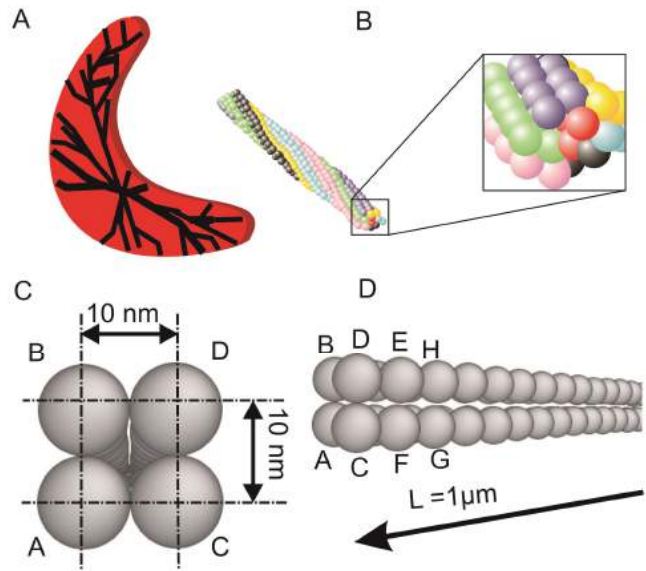

Figure 6.1. (A) A schematic of HbS fibers in a sickle cell (B) A single HbS fiber comprises 7 pairs of double strands of hemoglobin tetramers. The pairs are shown as same-color agents. It is drawn after (27). (C) Top view of the HbS fiber model. (D) Side view of the HbS fiber model.

Electron microscopy has revealed that a single HbS fiber consists of 7 double strands in a hexagonally shaped cross-section twisted about a common axis in a rope-like fashion retaining their atomic contacts (42, 43, 284-286). Single HbS fibers have an average radius of approximately 11 nm and a mean helical path of approximately 270 nm, defined as the length over which the fiber twists through 180° (39, 284, 285, 287). Fig. 6.1B, drawn after Fig. 1 in (27), illustrates the structure of a single HbS fiber. In vivo, HbS fibers are formed in an unusual fashion by two types of nucleation processes. Homogeneous nucleation entails the nucleation of a new fiber from an aggregate of monomers. In heterogeneous nucleation, the surface of a formed polymer assists the nucleation of a new polymer (46, 288). Because the homogeneous nucleation is very slow, there are only a few homogeneous nuclei in the deoxygenated state leading to a very low number of polymer domains via the dominant process of heterogeneous nucleation, which generates aligned polymers. In fact, most red blood cells gel in a single polymer domain (49, 289). Another important observation is that the red blood cell membrane considerably enhances the nucleation by a factor of 6 (50), and that HbS is associated with the red blood cell membrane and specifically with the cytoplasmic tail of the band-3 protein.

The properties of the HbS fiber have been extensively studied experimentally in the past two decades. Bending rigidities and persistence lengths of a group of HbS fibers with different thickness were measured by differential interference contrast (DIC) microscopy in (290). All the tested fibers were approximately 10 µm long while the variation in the thickness of HbS fibers was due to the bundling of single fibers. The persistence length of a single HbS fiber was found to be $l_p = 0.24$ mm and the associated bending rigidity was measured $\kappa = 8.3 \times 10^{-25}$ N m². In another approach, the bending and torsional rigidities of approximately 1 µm long single HbS fibers were obtained by cryo-electron microscopy (291, 292). The bending of each fiber was quantified by measuring the normal deviation of its middle point from the straight line joining its ends. In these experiments, frozen hydrated HbS fibers were used. The measurements showed that the bending rigidity of a single HbS fiber is approximately $\kappa = 5.2 \times 10^{-25}$ N m², which is consistent with the value measured by (290), while the torsional rigidity is approximately $6 \times 10^{-27}$ J m (292). While in homogeneous materials the bending and torsional rigidities are on the same order of magnitude, in single HbS fibers the bending rigidity is two orders larger than the torsional rigidity meaning that the material is anisotropic and the axial shear response is much softer than the extension (292). It is likely that the difference between the various types of contacts of the double strands is the source of the anisotropy. HbS fibers are among the stiffest filaments in a mammalian cell. For comparison, the persistence length of microtubules is approximately 5 mm, of intermediate filaments it is in the range of 200 nm to 1.3 µm, of F-actin it is approximately 10 µm, and of genomic DNA it is $53 \pm 2$ nm (126). Amyloid fibrils, which may be involved in chronic neurodegenerative diseases such as Alzheimer's disease, have persistence lengths that vary approximately from 5 µm to 0.3 mm, depending on their structure (293).

The interactions between sickle hemoglobin fibers were studied in (294). The authors found that HbS gels and fibers were fragile and easily broken by mechanical perturbation. They also observed different types of fiber cross-links in gel and the process of fibers zippering. In (44), the focus of the studies was on the interactions between two HbS fibers. They estimated that the attractive binding energy between two zippered HbS fibers is $u = 7.2\ k_B T\ \mathrm{µm}^{-1}$ with a corresponding characteristic length $l_c = k_B T/u \cong 140$ nm, which is significantly smaller than the persistence length of a single fiber. The fact that $l_c \ll l_p$ means that single HbS fibers are very unlikely to spontaneously separate because of thermal vibrations. However, this is true for non-stressed quiescent fibers. It will be shown that HbS fibers under bending can separate without additional external force other than the ones that cause bending.

So far, most of the studies on the HbS fibers were based on experimental observations. There is only limited molecular dynamics literature on HbS, which investigated the contact points of the double strands for the formation of the HbS fiber and in general the crystal structure of HbS (295, 296). Here, a coarse-grain (CG) model of a single HbS fiber is introduced. The parameterization scheme employed fits the CG potential to the experimentally measured material properties of single fibers, namely the bending and torsional rigidities (290, 292). It ignores the detailed structure of the fibers and the dynamics of the polymerization in order to reach a very large time scale, which is necessary in the study of the mechanical behavior of the fibers. A similar approach has been successfully used in the representation of the spectrin cytoskeleton of erythrocyte membrane (109), and in the study of cross-linked actin-like networks (297, 298). Essentially, the proposed model follows a top-down approach and the parameters are validated on the basis of experimental measurements (299). The approach is much simpler than multiscale coarse-graining strategies where molecular dynamics simulations considering atomistic details are performed and the CG force field is parametrized to match the atomistic forces and to predict the thermodynamic behavior (299-303) and references therein. Finally, the interactions between two HbS fibers and the zippering process are investigated.

## 6.2. Model and method

The single hemoglobin fiber studied here is modeled as four tightly bonded chains, each of which is composed of 100 soft particles. The cross-section of the HbS fiber model is shown in Fig. 6.1C. The distance between the centers of neighboring particles is 10 nm. The total length of the simulated HbS fiber is 1 μm and the radius of the fiber is approximately 10 nm (see Fig. 6.1D), which is consistent with the geometry of single hemoglobin fibers described in experimental studies (292). The displacements of all particles are governed by the Langevin equation (304)

$$m_i \frac{d^2 r_i}{dt^2} = \boldsymbol{F}_i - f \frac{dr_i}{dt} + \boldsymbol{F}_i^B, \qquad (6.1)$$

where $m_i$ is the mass of the ith particle, $f$ is the friction coefficient and it is identified to be 20 $m_i/t_s$, where $t_s$ is the time unit. $r_i$ is the position vector of the ith particle, and $t$ is time. $F_i = -\partial U/\partial r_i$ is the deterministic force produced by the implemented total potential $U$, $F_i^B$ is Gaussian white noise that obeys the fluctuation-dissipation theorem (98, 305)

$$\langle F_i^B \rangle = 0, \tag{6.2}$$

$$\langle F_i^B F_j^B \rangle = \frac{2k_B T f \delta_{ij}}{\Delta t}, \tag{6.3}$$

where $k_B$ is the Boltzmann's constant, $T$ is the absolute environmental temperature, $\delta_{ij}$ is the Kronecker delta, and $\Delta t$ is the time-step. The friction coefficient $f$ is selected to maintain the environmental temperature at 300 K. The Langevin thermostat is similar to Berendsen's thermostat but it introduces a damping effect that slows down the dynamics of the system (299, 306).

In the simulations, the energy unit is $\varepsilon/k_B T = 1$, and the length unit is $\sigma = 2^{-1/6} r_0$, where $r_0 = 10$ nm is the diameter of the particles. Since each hemoglobin tetramer has a molecular weight of approximately 68,000 Da, we assign to each coarse-grained particle a mass of about $m_i \cong 3.95 \times 10^{-2}$ kg. The time scale $t_s = (m_i \sigma^2/\varepsilon)^{1/2} \cong 2.7 \times 10^{-9}$ s of the simulation is set by the motion due to the deterministic force (130). The time step for the numerical solution of the Langevin equation is chosen to be $\Delta t = 0.001\ t_s$. However, the time scale $t_s$ in CGMD does not correspond to a real time since the particles do not represent real atoms. Only via comparison with a physical process, the correspondence between the simulation time and the "real" time can be established.

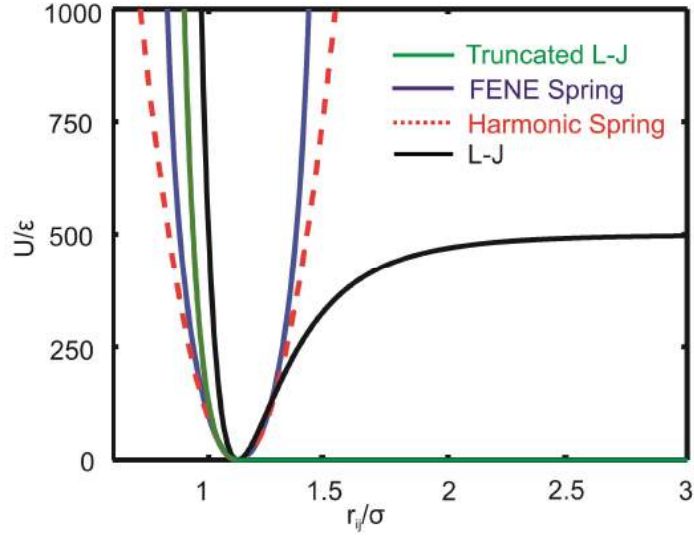

Figure 6.2. The finitely extendable nonlinear elastic (FENE) potential, the corresponding harmonic potential, the Lennard-Jones potential and the truncated Lennard-Jones potential employed in the HbS fiber model are shown.

A finitely extendable nonlinear elastic (FENE) potential (see Fig. 6.2) is introduced between adjacent particles that belong to the same chain (e.g., D and E in Fig. 6.1D) and between particles that belong to different chains that are initially in the same cross-sectional plane (A-B, A-C, A-D, B-C, B-D and C-D in Fig. 6.1C). The FENE potential is described by the equation

$$U_s = -\frac{1}{2} K_s \Delta r_{max} \ln\left[1 - \left(\frac{r_{ij} - r_0}{\Delta r_{max}}\right)^2\right], \tag{6.4a}$$

where $K_s$ is a parameter related to the constant $K_{sp} = K_s/\Delta r_{max}$ of the harmonic potential that has the same stiffness with the above potential at equilibrium, $r_{ij}$ is the distance between particles $i$ and $j$, and $r_0$ is the equilibrium distance between the particles. The FENE potential is harmonic at its minimum but the bonds cannot be stretched more than $\Delta r_{max}$. The force is given by the expression

$$F = -K_s \frac{(r_{ij}-r_0)/\Delta r_{max}}{1-[(r_{ij}-r_0)/\Delta r_{max}]^2} . \qquad (6.4b)$$

In these simulations, the maximum extension allowed between two particles is set to be $\Delta r_{max} = 0.3\ r_0$. The main role of the FENE potential is to bond the four chains, maintain the geometry of the model, and ensure that interacting fibers cannot move through one another. Since the proposed model ignores the detailed structure of the HbS fibers, only one spring constant is considered. The value of $K_s$ is determined based on the following argument: the Young's modulus of an elastic beam is related to its bending rigidity via the expression $\kappa = EI$, where $E$ is the Young's modulus of the beam and $I$ is the moment of inertia about the axis along the beam. However, in particle models, the spring potential does not generate the expected bending rigidity because of the free rotation of the neighboring points. Nevertheless, the generated overall Young's modulus has to be consistent with the bending rigidity. In the appendix A, we show that the value of $K_s$ that generates the Young's modulus $E$ which is consistent with the bending rigidity of the HbS fiber is approximately $K_s = K_{sp}\Delta r_{max} = \pi\ ER\Delta r_{max}/4 = 3600\ \varepsilon/r_0$.

A truncated Lennard-Jones (L-J) potential (see Fig. 6.2) is applied between two particles in the same fiber, but in different cross-sectional planes (A-E, A-F, B-E and B-F in Fig. 6.1D), to provide repulsive force when the two particles are in close range and thus prevent particle overlap. The expression of the truncated L-J potential is given by

$$U_{LJ}(r_{ij}) = \begin{cases} 4\varepsilon' \left[ \left(\frac{\sigma}{r_{ij}}\right)^{12} - \left(\frac{\sigma}{r_{ij}}\right)^{6} \right] + \varepsilon' & r_{ij} < r_{cut}^{tr} = 2^{1/6}\sigma \\ 0 & r_{ij} \geq r_{cut}^{tr} \end{cases} \quad (6.5)$$

where $r_{ij}$ is the distance between particles $i$ and $j$. The cut-off distance $r_{cut}^{tr}$ for truncated L-J potential is chosen to be $r_0$. In order to limit the number of the independent model parameters, we determined $\varepsilon' = K_s \sigma / (36 \cdot 2^{2/3})$ by requiring that the truncated L-J potential has the same curvature as the FENE potential at $r_{cut}^{tr}$.

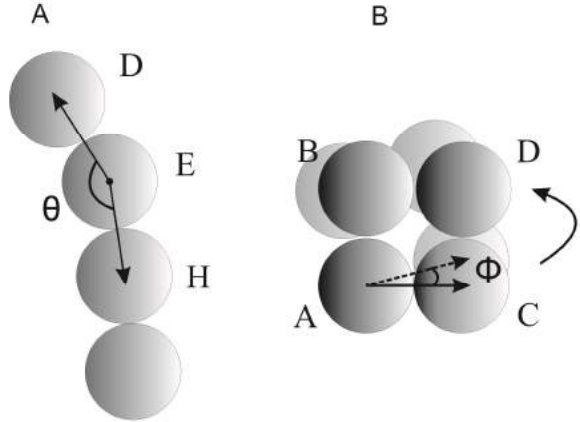

Figure 6.3. (A) Bending and (B) twist of the HbS fiber model.

A bending FENE potential $U_b$ is employed to describe the bending rigidity of the HbS fiber

$$U_b = -\frac{1}{2} K_b \Delta\theta_{max} \ln\left[1 - \left(\frac{\theta - \theta_0}{\Delta\theta_{max}}\right)^2\right], \quad (6.6)$$

where $K_b$ is the parameter that directly regulates the bending stiffness of the hemoglobin fiber, and $\theta$ is the angle formed by three consecutive particles in the same chain, as illustrated in Fig. 6.3A. $\theta_0$ is the equilibrium angle and it is chosen to be $\pi$, meaning that the three consecutive particles are initially located in-line. $\Delta\theta_{max}$ is the maximum allowed bending angle between two particles and is set to be $0.3\,\theta_0$. $K_b$ is selected via a trial and error process in order to produce a bending rigidity identical to the experimental value and it is determined to be $K_b = 5 \times 10^3 \varepsilon$.

The torsional rigidity of the proposed model is introduced via a torsional FENE potential between particles belonging to consecutive cross-sectional planes defined by four particles. The potential is expressed as

$$U_t = -\frac{1}{2} K_t \Delta\Phi_{max} \ln\left[1 - \left(\frac{\Phi}{\Delta\Phi_{max}}\right)^2\right], \tag{6.7}$$

where $K_t$ is the parameter that regulates the torsional stiffness of the HbS fiber, and $\Phi$ is the angle between two directional vectors defined in two consecutive cross-sectional planes, as it is illustrated in Fig. 6.3B. $\Delta\Phi_{max}$ is the maximum allowed twisting angle between two consecutive cross-sectional planes and is set to be $\cong 0.3\,\pi$. $K_t$ is tuned to match the model against experimentally obtained values of the torsional rigidity of single hemoglobin fibers (44) and it is chosen to be $K_t = 500\varepsilon$.

The L-J potential, as plotted in Fig. 6.2, is employed between particles that belong to different fibers to produce attractive interfiber forces when two fibers are close. The expression of L-J potential is given by

$$U_{LJ}(r_{ij}) = \begin{cases} k * 4\varepsilon \left[ \left(\dfrac{\sigma}{r_{ij}}\right)^{12} - \left(\dfrac{\sigma}{r_{ij}}\right)^{6} \right] + k\varepsilon & r_{ij} < r_{cut} \\ 0 & r_{ij} \geq r_{cut} \end{cases}, \quad (6.8)$$

where $r_{cut} = 2.5\sigma$ and k is a parameter used to adjust the interfiber forces between two fibers. The value of k that is used for the plot in Fig. 6. 2 is 500.

As it was mentioned earlier, the spring constant $K_{sp}$ is related to the Young's modulus of the fiber through the expression $K_{sp} = \pi E R/4$. By approximating the fiber as a circular cylindrical rod of radius R and cross-sectional moment of inertia $I = \pi R^4/4$, the relationship between the bending rigidity and the spring constant can be established as $K_{sp} = \kappa/R^3$. Then, a relevant time-scale for the vibration of the HbS fibers can be estimated as $t_c = 2\pi \sqrt{m/K_{sp}} = 2\pi\sqrt{mR^3/\kappa} \cong 176$ ps, which is approximately 15 times smaller than $t_s$ and this justified why the time step $\Delta t$ is 10 times smaller than the value usually employed in MD simulations.

Because the Langevin equation acts as a thermostat, the noise and friction are designed to maintain a given temperature. Deviations from that temperature are corrected with a decay time of $t_d = m_i/f$. By choosing $f = 20\ m_i/t_s$, the decay time becomes $t_d = t_s/20 = 5\ \Delta t$. This means that the system corrects the temperature variations in about 5 time steps. Another typical relevant time for the fluctuations of the HbS fibers is the duration of such fluctuations. This time can be estimated as $t_f = f/K_{sp}$ (307). By using that $f = 3\pi\eta_{\ 0}$, where $\eta = 8.9 \times 10^{-4}$ Pa s approximates the dynamic viscosity of the solvent, and the relation between the single spring constant and the overall bending rigidity of the fiber, the relaxation time can be estimated as $t_f = 3\pi\eta r_0^4/\kappa \cong 162$ ps. The fact that $t_c$ and $t_f$ have a ratio almost one means that the motion of the fibers is overdamped.

For the integration of Eq. (6.1), a modified version of the leapfrog algorithm is used:

$$\boldsymbol{v}_i(t + \Delta t/2) = \boldsymbol{v}_i(t) + \Delta t/2 \, \boldsymbol{a}_i(t), \qquad \text{6.9(a)}$$

$$\boldsymbol{r}_i(t + \Delta t) = \boldsymbol{r}_i(t) + \Delta t \, \boldsymbol{v}_i(t + \Delta t/2), \qquad \text{6.9(b)}$$

$$\tilde{\boldsymbol{v}}_i(t + \Delta t) = \boldsymbol{v}_i(t + \Delta t/2) + \Delta t/2 \, \boldsymbol{a}_i(t), \qquad \text{6.9(c)}$$

$$\boldsymbol{a}_i(t + \Delta t) = \boldsymbol{a}_i\big(\boldsymbol{r}_i(t + \Delta t), \tilde{\boldsymbol{v}}_i(t + \Delta t)\big), \qquad \text{6.9(d)}$$

$$\boldsymbol{v}_i(t + \Delta t) = \boldsymbol{v}_i(t + \Delta t/2) + \Delta t/2 \, \boldsymbol{a}_i(t + \Delta t). \qquad \text{6.9(e)}$$

Because the total force applied on a particle depends on the velocity of the particle, a prediction is made for the new velocity, which is denoted as $\tilde{\boldsymbol{v}}$, and it is corrected in the last step. A similar approach was employed for the modification of the velocity-Verlet algorithm used in (308).

### 6. 3. Results and discussion

6. 3.1 Measurements of material properties of HbS fiber

We show that the proposed model is able to simulate a single HbS fiber with the appropriate mechanical properties. In particular, by tuning the parameters $K_b$, and $K_t$, the model reproduces the experimental values for the bending rigidity $\kappa$, and the torsional rigidity $c$, of a HbS fiber (292). During the simulations, one end of the HbS fiber is fixed and it is used as the reference point for the central axis of the fiber.

6. 3.1.1 Bending rigidity

Thermally driven fluctuations of semi-flexible fibers can be used to obtain the bending moduli of the fibers (126). One method is to measure the mean-squared amplitude of each dynamical mode of vibrations and by using the principle of equipartition of energy to extract the bending rigidity for each mode independently (309). However, because HbS fibers are stiff with a large persistence length of approximately 120 μm and the simulated fibers have a total length of 1μm, they bend a little resulting to a large uncertainty when the motion is decomposed into Fourier modes. Here, a method introduced by (290) is followed. The bending rigidity of the HbS fiber is calculated from the equation

$$\kappa = k_B T l_p = \frac{k_B T L^3}{48 \langle (u_x(L/2))^2 \rangle}, \tag{6.10}$$

where $l_p$ is the persistence length of the HbS fiber, $L$ is the length of the fiber, and $u_x(L/2)$ is the spatial displacement of the projected fiber midpoint from the central axis. The central axis, from which the displacement of the fiber's middle point is measured, is defined as a line that is perpendicular to the cross-section of the fiber at the fixed point and it passes through the center of the cross-sectional area (dashed line in Fig. 6.4). Once the bending rigidity is obtained, the persistence length can be easily calculated by the expression

$$l_p = \kappa/k_B T .  \tag{6.11}$$

The measured bending rigidity of HbS fiber with respect to time is plotted in Fig. 6.5. The Initial fiber configuration at 0°K was a straight line. The free end of the HbS fiber model first began to fluctuate and then the kinetic energy was transferred in the fiber from the free end towards the fixed end. The system reached equilibrium at approximately $t \cong 10^4 t_s$ (see Fig. 6.5). The measured bending rigidity is close to the experimentally measured value $\kappa = 5.2 \times 10^{-25} \mathrm{Nm}^2$ and it corresponds to a persistence length of approximately $l_p \cong 121$ µm (292). Because of the large persistence length, the numerical results show that a single HbS fiber subjected to thermal forces behaves similarly to a stiff rod as it fluctuates about its central axis (see Fig. 6.4).

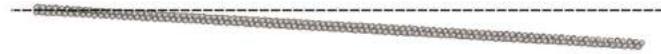

Figure 6.4. Characteristic configuration of the HbS fiber model.

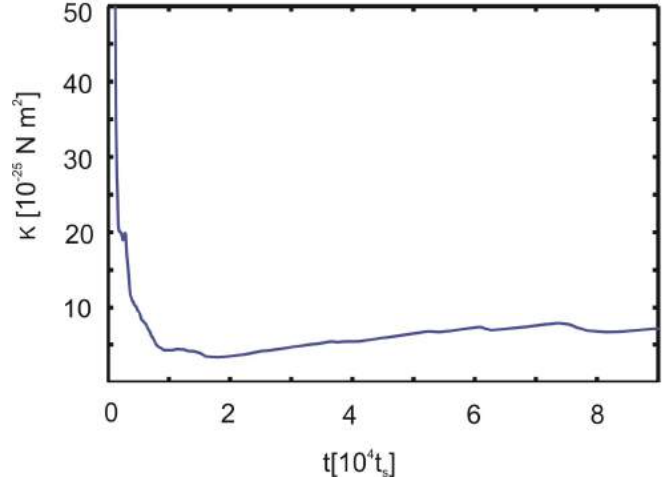

Figure 6.5. Variation with time of the bending rigidity of the HbS fiber model.

The probability density of the deviations at the middle point of the HbS fiber can be approximated by a normalized Gaussian distribution (solid curve in Fig. 6.6) given by

$$P(\delta u) = \frac{1}{\sqrt{2\pi \langle \delta u^2 \rangle}} e^{\left(-\frac{\delta u^2}{2 \langle \delta u^2 \rangle}\right)}, \qquad (6.12)$$

where the $\delta u$ is the HbS fiber midpoint displacement and $\langle \delta u^2 \rangle^{1/2} \cong 1.4\,\sigma$. The result is in agreement with the theoretical prediction that the probability density of the deviations of a semi-flexible rod follow the Gaussian distribution in the case of thermal fluctuations (310).

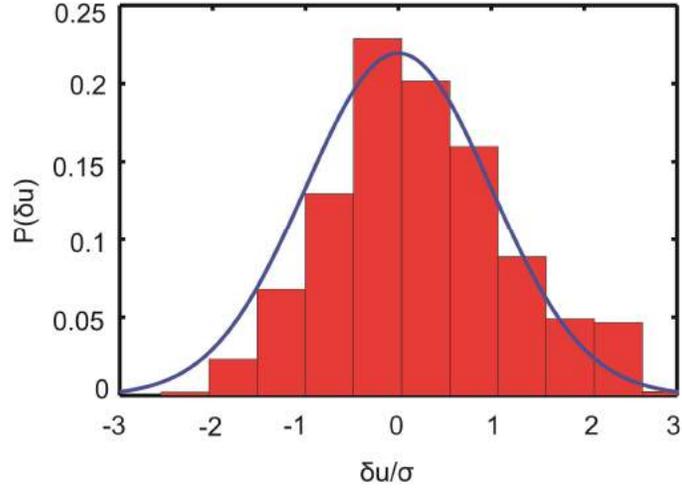

Figure 6.6. The distribution of HbS fiber midpoint displacements and the associated normalized Gaussian probability distribution.

6. 3.1.2 Torsional rigidity

As for the torsional rigidity, it has been shown (290) that

$$\langle \Delta\lambda^2 \rangle = \frac{k_B T \lambda^3}{c\pi^2} \left[1 - \frac{L}{\lambda}(1 - e^{-\frac{\lambda}{L}})\right], \tag{6.13}$$

where $\Delta\lambda = \lambda \Delta\theta/\pi$, and $\lambda = 135$ nm is half of the average pitch length for the HbS fiber, $\Delta\theta$ is the twisted angle in half pitch length, and $L$ is the length of the HbS fiber. By substituting $\Delta\lambda$ into Eq. (6.13), we obtain $c$ as

$$c = \frac{k_B T \lambda}{\langle \Delta\theta^2 \rangle} \left[1 - \frac{L}{\lambda}(1 - e^{-\frac{\lambda}{L}})\right]. \tag{6.14}$$

Fig. 6.7 shows that the selected value $K_t = 500\ \varepsilon$ results in a torsional rigidity of the hemoglobin fiber model $c \cong 6.5 \times 10^{-27}$ J m, which is very close to the experimentally measured value of $c = 6 \times 10^{-27}$ J m (292).

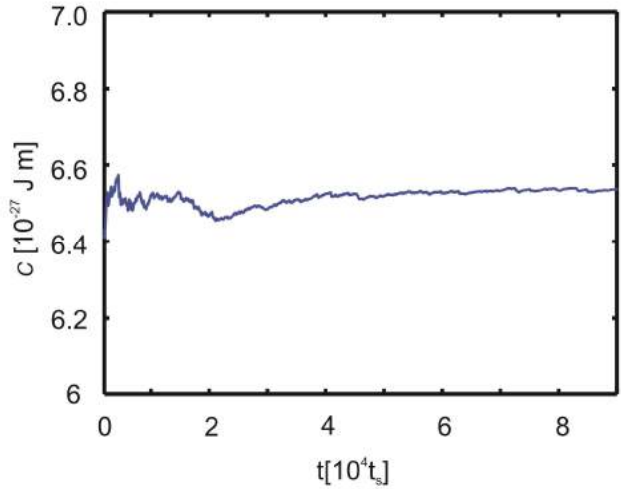

Figure 6.7. Variation with time of the torsional rigidity of the HbS fiber model.

6. 3.1.3 Effect of the bending and torsional potentials on the HbS fiber model

Finally, we ensure that the contribution to bending and torsional rigidity results mainly from the special potentials $U_b$ and $U_t$ and not from the FENE spring potential between the particles. When only the spring and the truncated L-J potentials are implemented, the shape of the HbS fiber is sinusoidal and strongly twisted (see Fig. 6.8). Also, it is found that the amplitude of the HbS fiber fluctuations is small and that the fluctuations are limited only about the central axis (marked with dashed line in the Fig. 6.8) of the fiber and no bulk motion is observed.

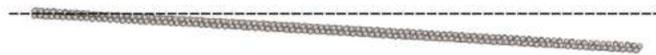

Figure 6.8. Example of the behavior of the HbS fiber model without applying the bending and the torsional potentials.

According to (290), Eq. (6.10) is derived for semi-flexible fibers which are relatively stiff. Since the fiber model without the special bending and torsional potentials is highly flexible, the Eq. (6.10) cannot be employed in the calculation of the bending rigidity of this fiber. Instead, we use the equation

$$\langle r_{ee}^2 \rangle = 2l_p L - 2l_p^2 [1 - exp\left(-\frac{L}{l_p}\right)], \tag{6.15}$$

to obtain the persistence length $l_p$ and then to calculate the bending rigidity from the expression (11) (126). $\boldsymbol{r}_{ee}$ is the end-to-end displacement vector, and $L$ is the length of the HbS fiber. Based on the Eqs. (6.14), and (6.15), the values of bending rigidity and torsional rigidity of HbS fiber model without the bending $U_b$ and torsional $U_t$ potentials are shown in Fig. 6.9A and Fig. 6.9B respectively.

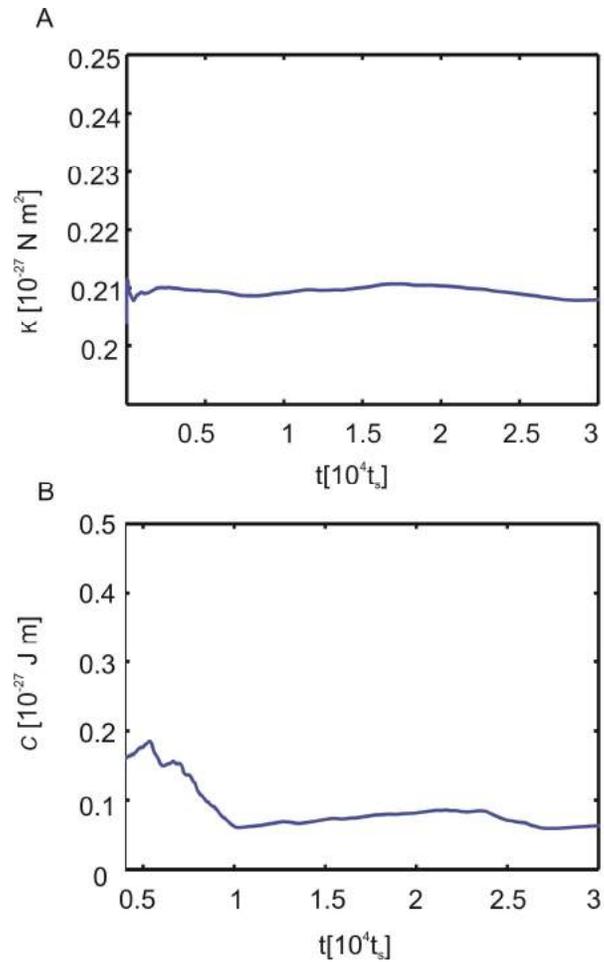

Figure 6.9. Variation with time of (A) the bending rigidity and (B) the torsional rigidity of the HbS fiber model without applying the bending and the torsional potentials.

The measured bending rigidity is approximately $0.21 \times 10^{-27}$ N m$^2$, which is about 1/2000 of the actual value of the HbS fiber ($\kappa = 5.2 \times 10^{-25}$ N m$^2$ (292)). The torsional rigidity is approximately $6 \times 10^{-29}$ J m, which is 1/100 of the experimental value ($c = 6 \times 10^{-27}$ J m (292)). The results above indicate that the spring potential has little effect on the model's bending and torsion rigidity. It is noted that the proposed HbS fiber model is strongly anisotropic since the experimentally measured torsional rigidity is significantly less than the bending rigidity, while for isotropic materials these properties are on the same order of magnitude.

6.3.2 Modeling the zippering of two HbS fibers

Individual HbS fibers interact with each other to form various X-shaped junctions, Y-shaped branches, and side-to-side coalescence ("zippering") cross-links (294, 311). We apply the developed model to study the formation of bundles by zippering fibers. Two cases of HbS fibers zippering are simulated. In the first case, two fibers are initially parallel to each other and then come into contact as shown in Fig. 6.10. In the second case, two fibers in a Y-shaped cross-link zippered from their contacting tips (see Fig. 6.13). The interfiber force is represented by a L-J potential applied between particles belonging to different fibers.

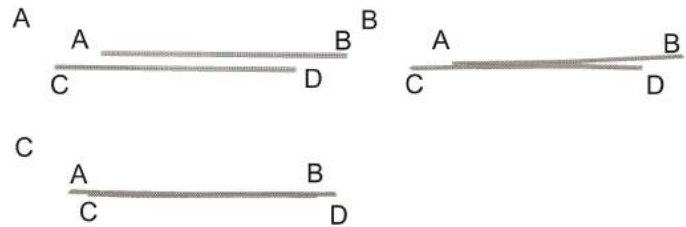

Figure 6.10. (A) Initial states of two parallel HbS fibers before zippering. (B) Intermediate state of two initially parallel HbS fibers zippering. (C) Equilibrium configuration of two initially parallel HbS fibers. The value of the L-J parameter k is 200.

The minimum value of the parameter k, which adjusts the interaction between the fibers, is determined by assuming that the interfiber forces have the same origin as the forces between the particles comprised by the fiber. In this case, the FENE potential between the particles that form the fiber and the L-J potential must have the same curvature at the equilibrium, meaning that $k\varepsilon = K_s\sigma/(36 \cdot 2^{2/3})$ and since $K_s = 3600 \; \varepsilon/r_0$ we obtain that $k \cong 60$. According to (44), there are two main contributions to the attraction energy between zippering fibers. One is due to Van der Waals forces and the second is due to depletion forces. The depletion energy was computed to be approximately two times the Van der Waals energy. Based on the previous estimations, we explored the interaction between two fibers for the L-J parameter k varying from a very small value of k = 5 to a very large value of k = 500.

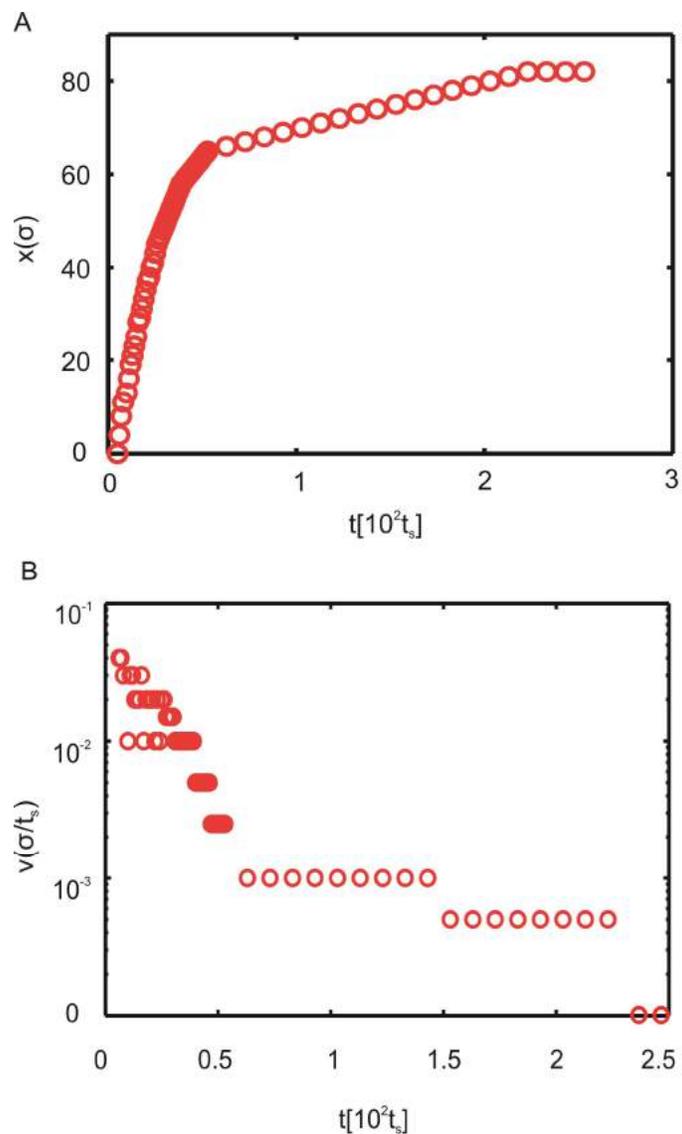

Figure 6.11. (A) Position $x$ and (B) speed of the fiber junction $v$ during the zippering process of two parallel HbS fibers.

In the first case, two fibers AB and CD are initially placed in parallel arrangement as seen in Fig. 6.10A with an interaction parameter $k = 200$, which ensures zippering. At the beginning of the simulation, they fluctuate under brownian forces. Fig. 6.10B shows that the tip A approaches and subsequently touches the body of the fiber CD. Then, the two fibers begin to progressively merge from the contact point to form a thicker fiber in Fig. 6.10C. The non-symmetrical behavior of the two ends A and D is due to random fluctuations. One of the two ends, in this case point A, approaches the neighboring fiber closer than $r_{cut}$ resulting to a broken symmetry. As for the position and speed of the fiber junction during the zippering process, Fig. 6.11A and B show that the speed of the zippering is relatively fast at the beginning of the zippering process and at a critical distance gradually decreases to zero when the two fibers reach the equilibrium state.

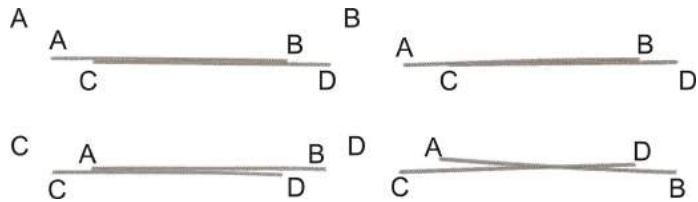

Figure 6.12. Equilibrium states of two parallel HbS fibers zippered under different interfiber forces. The L-J parameter k was selected to be (A) k = 100, (B) k = 50, (C) k = 10, (D) k = 5.

The behavior of the fiber pair for different values of parameter $k$ is explored next. Knowing that in the case of $k = 200$ the fibers zipper tightly, we decrease the value of k to $k = 100$. Again the two fibers finally move like a single fiber (see Fig. 6.12A). When $k = 50$, the fibers still "zipper" together but the attraction is not large enough to prevent relative rotation (see Fig. 6.12B). For k = 10 (see Fig. 6.12C), the two fibers still stay together but relative rotation and instantaneous dissociation are observed between the two fibers which no longer behave as a single fiber. When k is finally reduced to k = 5 (see Fig. 6.12D), no stable zippering is achieved and two fibers separate easily under thermal fluctuations. The results show that because zippered fibers are stable for k ≥ 100, the stability of a polymerized bundle of fibers is due to both Van der Waals and depletion forces.

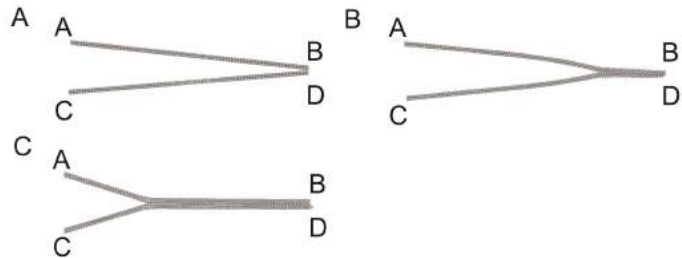

Figure 6.13. (A) Initial state of two HbS fibers before zippering. (B) Intermediate state of the two HbS fibers during the "Y" shape zippering. (C) Equilibrium configuration of the two HbS fibers

Next, we study how the strength of interactions drives zippering of HbS fibers in frustrated structures formed by three fibers. In the first simple case shown in Fig. 6.13, two fibers AB and CD are arranged to form a "V" shape junction. Points A and C represent the attachments of the two fibers along a third fiber and they are thus fixed, while their two ends B and D are free to move. The value of the parameter k, in the expression (8) of the L-J potential between different fibers is 500. The initial temperature increases and it reaches the final value of $T = 1$ at the $2 \times 10^4$ timestep. The attractive force between the two fibers is turned on after the $4 \times 10^4$ timestep to allow the system to reach the equilibrium. Then, the two fibers start zippering from points B and D. At the initial steps, the junction point quickly moves toward points A and C. However, as the zippering progresses (see Fig. 6.13B), the speed of the zippering reduces significantly and finally the system reaches the equilibrium state, as illustrated in Fig. 6.13C. After zippering, the two fibers move together as a single thicker fiber. The interfiber attractive forces are balanced by the bending energy stored in the two fibers (44, 294). The position $x$ of the junction point during zippering is shown in Fig. 6.14A as a function of time and the variation of the tip-velocity $v$ with time is shown in Fig. 6.14B. Initially, the two end points snap together and the tip advances very fast but gradually the velocity decreases and zippering ceases at approximately 55 $\sigma$. Next, the effect of the interfiber forces during the fibers zippering process is examined. The parameter k of the L-J potential is changed to 100, 200, 300 and 400 and the final configurations of the zippered fibers are shown in Figs. 6.15A, B, C and D respectively. Fig. 6.15E clearly demonstrates that when fibers are constrained in their relative motion then the interaction energy has to increase substantially to cause partial zippering.

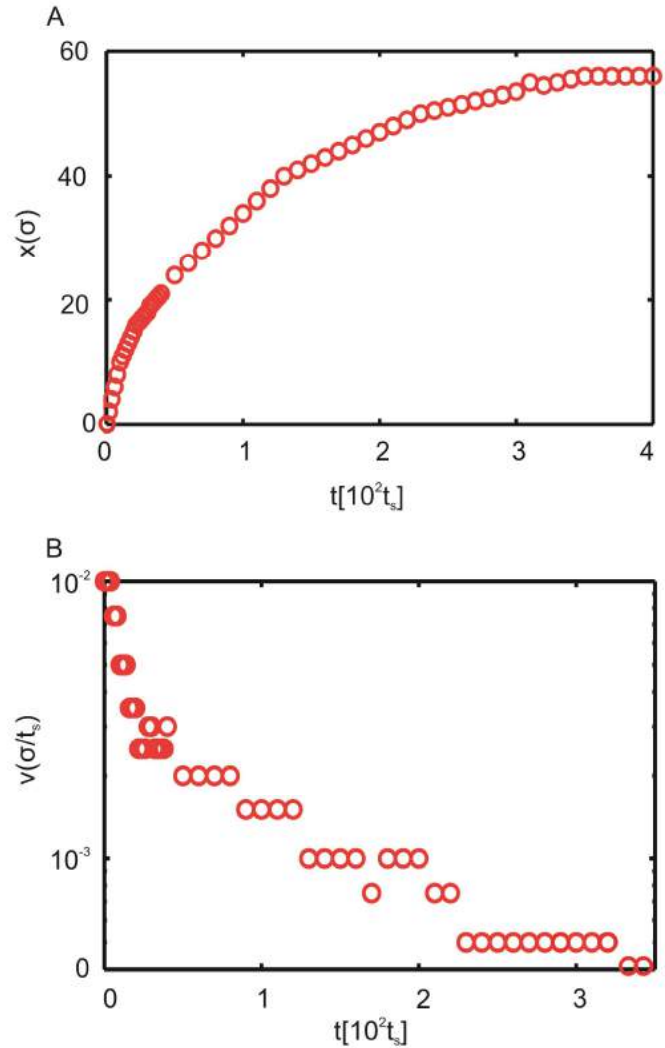

Figure 6.14. (A) Position $x$ and (B) speed of the fiber junction $v$ during the "Y" shape zippering process.

In a variation of the above numerical experiment, the two fibers are cross-linked to a third fiber while the attractive interfiber forces between the two inclined fibers and between an inclined fiber and the third fiber are governed by the same L-J potential (see Fig. 6.16). The values of the parameter k of the L-J potential are 200, 300, 400 and 500 and the final configurations of the zippered fibers are shown in Figs. 6.16A, B, C and D respectively. The behavior is similar as in the previous case where point A and C were completely immobilized. It is noted that zippering does not induce relative sliding of the attaching points between EF and AB, CD fibers. By comparing Figs. 6.15E and 16E, it is obvious that a higher attractive force is required to generate zippering in the three-fiber configuration. The reason is that the constrained points are not at the ends of the fibers resulting to a shorter effective fiber length. The cases shown in Figs. 6.15 and 6.16 demonstrate that frustration in HbS fibers can cause partial zippering and consequently the polymerization of HbS fibers can deviate from rod-like shapes and branch out to more irregular configurations resulting to the variety of shapes of deoxygenated sickle cells.

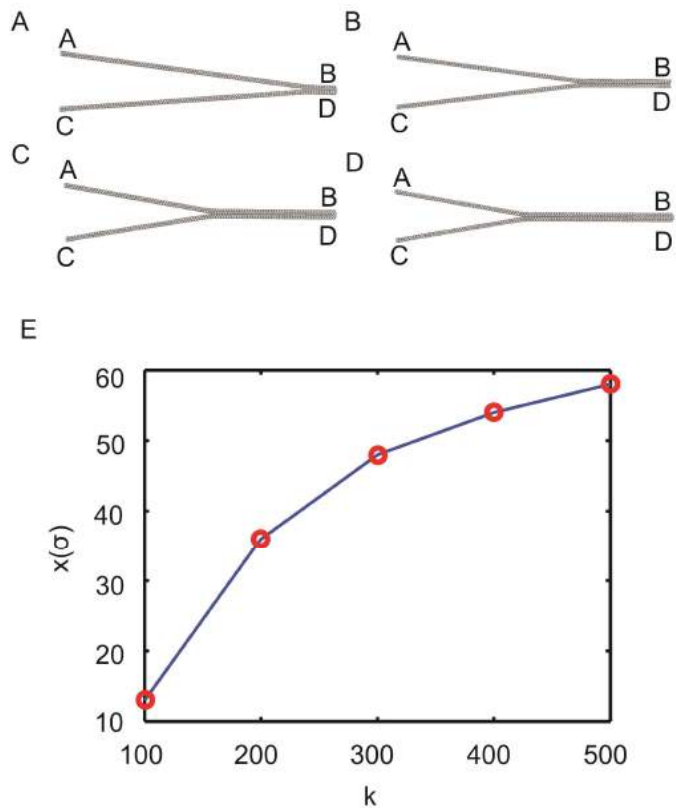

Figure 6.15. Equilibrium states of two HbS fibers zippered as "Y" shapes under different interfiber forces. The L-J parameter k is selected to be (A) k = 100, (B) k = 200, (C) k = 300, (D)

k = 400. (E) The final position of the fiber junction *x* measured from point D for different interfiber forces.

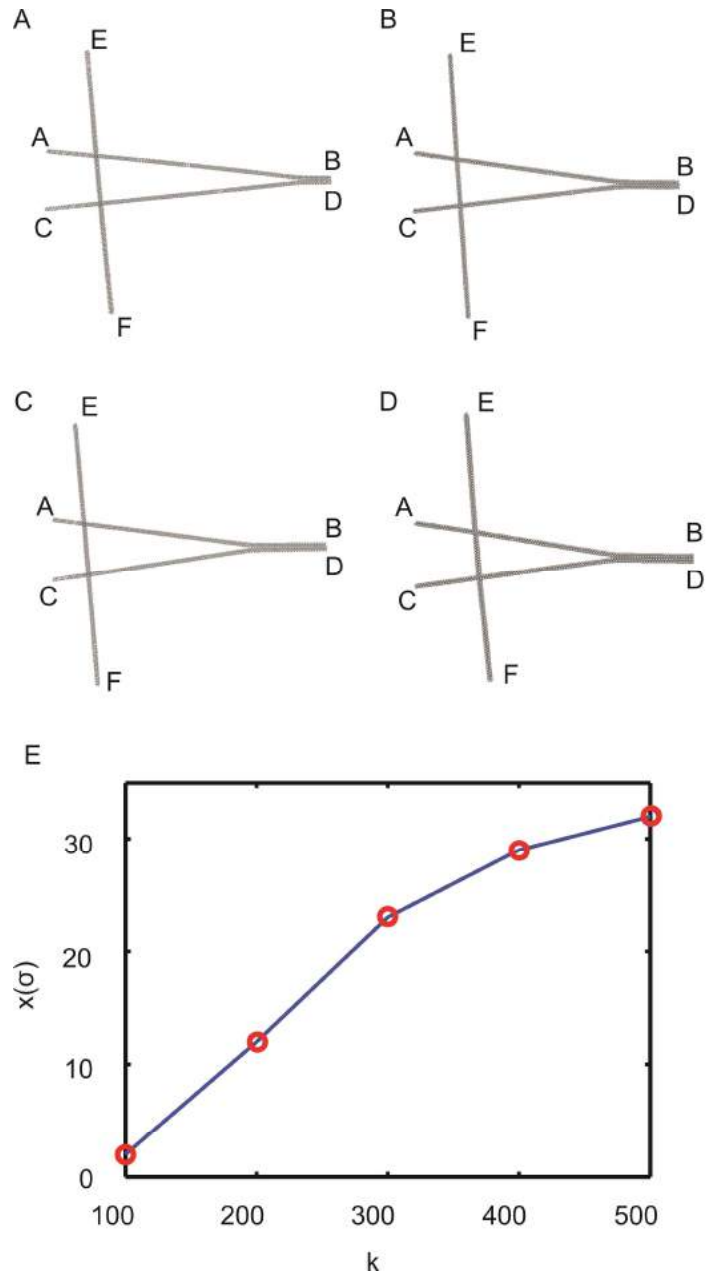

Figure 6.16. Equilibrium states of two HbS fibers zippered in a "Y" shape as a result of a third fiber crossing the other two fibers near to their ends for different interfiber forces. The L-J parameter k is selected to be (A) k = 200, (B) k = 300, (C) k = 400, (D) k = 500. (E) The final position of the fiber junction *x* measured from point D for different interfiber forces.

### 6.3.3 Bending of two zippered fibers

In this part, we show that bending of zippered HbS fibers can cause partial separation. This is another mechanism that could generate irregular shapes of deoxygenated sickle cells via polymerization of the partially separated fibers. The two zippered fibers are initially maintained at equilibrium state and then the two ends of one fiber (C and D) are compressed until their initial distance is reduced by 10% (see Fig. 6.17). Point B of the fiber AB is forced to move together with point D while point A is free. At the beginning of the compression, the two fibers are bent together because the bending force is balanced by the attractive force. When the maximum attractive force, which is regulated by the parameter k, becomes equal to the bending force required for fiber AB, the free end A separates from the fiber CD, and the two fibers begin to unzipper. Figs. 6.17A, B, C and D correspond to k equals 400, 300, 200 and 100 respectively. Fig.6.17E illustrates the dependence of the degree of separation of the two fibers on the magnitude of the attractive force between them. In another approach shown in Fig. 6.18, the parameter k is kept constant (k = 200) while the end points of the fiber CD are gradually compressed causing buckling. As the compression increases, the separation between the two fibers increases non-linearly showing that the expected interaction energy cannot restrain bent fibers from separation. At 20% decrease of the initial distance between the end points C and D, the two fibers are almost completely separated. As a result, polymerization of initially free fibers can cause bending and consequently partial separation and generation of irregular fiber networks and sickle cell shapes.

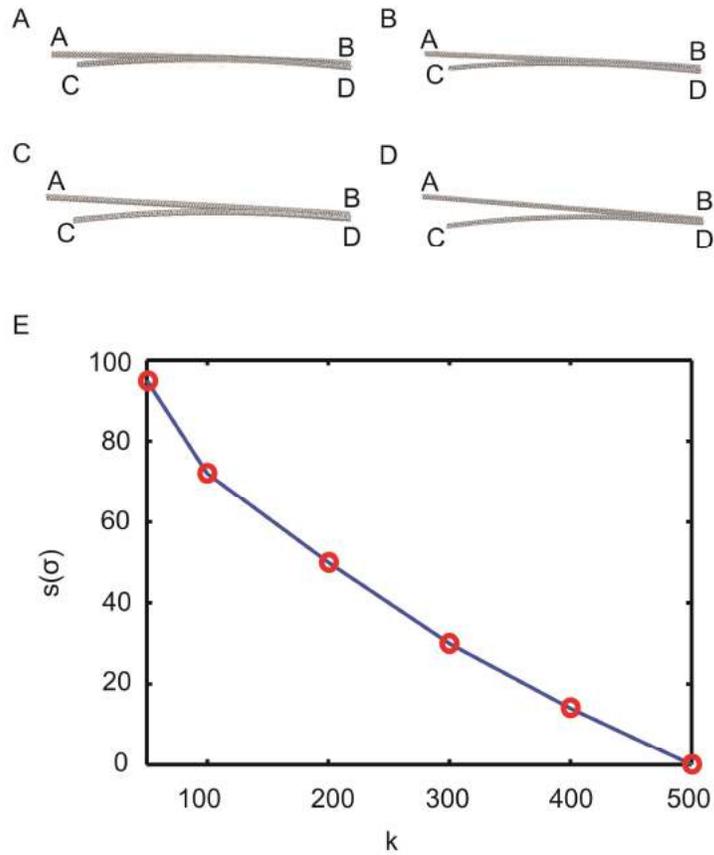

Figure 6.17. Equilibrium states of two zippered HbS fibers under compression for the same final deformation ratio e = 0.2 and at different interfiber forces. The L-J parameter k is selected to be (A) k = 400, (B) k = 300, (C) k = 200, (D) k = 100. (E) Separations of the two fibers measured from point C, at different interfiber forces.

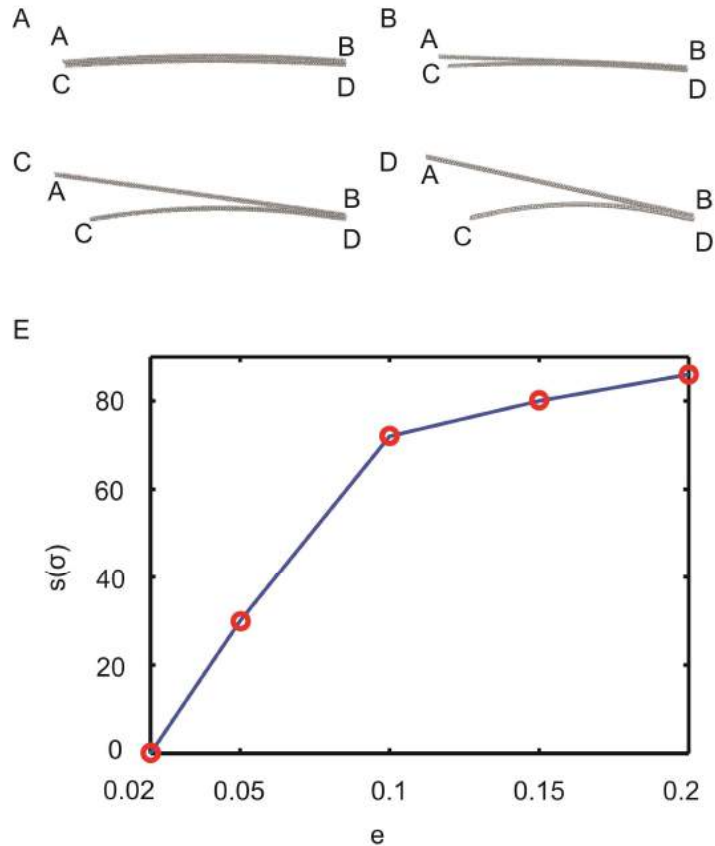

Figure 6.18. Equilibrium states of two zippered HbS fibers under compression at different final deformation ratios and for the same interfiber force (k = 200). The deformation ratio e is selected to be (A) e = 0.02, (B) e = 0.05, (C) e = 0.15, (D) e = 0.2. (E) Separations of the two fibers measured from point C at different deformation ratios.

## 6. 4. Summary

We model single HbS fibers using a quadruple chain of particles and the hexagonally shaped cross-section of the HbS fiber is coarse grained into four particles. The motions of all the particles are governed by the Langevin equation. In order to simulate the thermal behaviors of HbS fibers, five different interaction potentials are applied between the particles, namely a FENE potential, a truncated L-J potential, a bending potential, a torsional potential and a L-J potential.

By employing these five potentials, the proposed model is able to derive the experimentally measured bending rigidity, and the torsional rigidity of a single HbS fiber. Then, the model is used in the study of the zippering process of initially parallel free fibers and of frustrated fibers. Finally, the behavior of zippered fibers under compression is explored. The results show that while low interaction energy between free fibers is enough to cause zippering, frustrated fibers or fibers under compression require much higher interaction energy to remain zippered. This means that fiber frustration and compression can result to partial unzippering of bundles of polymerized HbS fibers. Continuous polymerization of the unzippered fibers via heterogeneous nucleation and additional unzippering under compression can explain the formation of HbS fiber networks and consequently the wide variety of shapes of deoxygenated sickle cells.

**Appendix 6.A**

The single hemoglobin fiber is modeled as four tightly bonded chains, each of which is composed of 100 soft particles (Fig. 6.1D). Adjacent particles that belong to the same chain (e.g., D and E in Fig. 6.1D) and particles that belong to different chains that are initially in the same cross-sectional plane (A-B, A-C, A-D, B-C, B-D and C-D in Fig. 6.1C) interact via the FENE potential

$$U_s = -\frac{1}{2} K_s \Delta r_{max} \ln\left[1 - \left(\frac{r_{ij}-r_0}{\Delta r_{max}}\right)^2\right], \tag{6.A.1}$$

which generates a force

$$F = -K_s \frac{(r_{ij}-r_0)/\Delta r_{max}}{1-[(r_{ij}-r_0)/\Delta r_{max}]^2}. \tag{6.A.2}$$

In the simulation, the deformations of the HbS fibers are small. In this case the FENE potential can be approximated by the expression

$$U_s = \frac{1}{2} K_s / \Delta r_{max} \left(r_{ij} - r_0\right)^2, \qquad (6.A.3)$$

Which corresponds to a harmonic potential with a spring constant of

$$K_{sp} = K_s / \Delta r_{max}. \qquad (6.A.4)$$

For small deformations, it can be assumed that the relation between the applied stress $\tau$ and the applied strain $\varepsilon$ is linear

$$\tau = E\varepsilon, \qquad (6.A.5)$$

where E is the Young's modulus. The Eq. (6.A.5) can be rewritten as

$$F/A = E\Delta L/L, \qquad (6.A.6)$$

where $F$ is the applied force along the fiber, $A = \pi r_0^2$ is the cross-sectional area, $\Delta L$ is the change of the spring's length, and $L = N r_0$ is the total length of the fiber, where $N$ is the number of particles along one of the four chains. The Young's modulus is expressed as

$$E = FL/A\Delta L . \tag{6.A.7}$$

Approximating the FENE potential with the harmonic potential for small deformations, the force can be calculated by

$$F = K_{sp}^t * \Delta L , \tag{6.A.8}$$

where $K_{sp}^t$ is the total spring constant of the whole fiber. Since the fiber model comprises 4 groups of springs in parallel and each group has $N$ springs in series. The total spring constant of the whole fiber is approximately

$$K_s^t = 4K_{sp}/N . \tag{6.A.9}$$

Substituting the Eqs. (6.A.9) and (6.A.8) into the Eq. (6.A.7), we obtain

$$K_{sp} = E\pi r_0/4. \tag{6.A.10}$$

If the fiber is approximated by a cylindrical rod, then the bending rigidity is related to the Young's modulus by

$$E = \kappa/I = 4\kappa/\pi r_0^4 , \tag{6.A.11}$$

where $I = \pi r_0^4/4$ is the cross-sectional moment of inertia, and $\kappa$ is the bending rigidity. Substituting Eq. (6.A.11) into Eq. (6.A.10), we obtain the relation between the spring constant and $\kappa$

$$K_{sp} = \kappa/r_0^3 . \tag{6.A.12}$$

The expected value of the bending rigidity is approximately $\kappa = 5 \times 10^{-25} \text{Nm}^2$, which corresponds to

$$\kappa \cong 12000 \, \varepsilon r_0, \tag{6.A.13}$$

where $\varepsilon = k_b T$ is the energy unit. Therefore the spring constant is

$$K_{sp} = 12000 \, \varepsilon/r_0^2. \tag{6.A.14}$$

By substituting the Eq. (6.A.14) into the Eq. (6.A.4) and by using that $\Delta r_{max} = 0.3 r_0$, we obtain that

$$K_s = K_{sp} \Delta r_{max} = 3600 \, \varepsilon/r_0 . \tag{6.A.15}$$

# Chapter 7.

# Modeling Sickle Hemoglobin Fibers as One Chain of Coarse-Grained Particles

**ABSTRACT**

Sickle cell disease (SCD) is caused by a single point mutation in the beta-chain hemoglobin gene, resulting in the presence of abnormal hemoglobin S (HbS) in the patients' red blood cells (RBCs). In the deoxygenated state, the defective hemoglobin tetramers polymerize forming stiff fibers which distort the cell and contribute to changes in its biomechanical properties. Because the HbS fibers are essential in the formation of the sickle RBC, their material properties draw significant research interests. Here, a solvent-free coarse-grain molecular dynamics (CGMD) model is introduced to simulate single HbS fibers as a chain of particles. First, we show that the proposed model is able to efficiently simulate the mechanical behavior of single HbS fibers. Then, the zippering process between two HbS fibers is studied and the effect of depletion forces is investigated. Simulation results illustrate that depletion forces play a role comparable to direct fiber-fiber interaction via Van der Waals forces. This proposed model can greatly facilitate studies on HbS polymerization, fiber bundle and gel formation as well as interaction between HbS fiber bundles and the RBC membrane.

## 7.1. Introduction

The erythrocytes of the SCD patients contain defective HbS, in which a charged surface group glu at β6 is replaced by a hydrophobic group val, inducing polymerization of deoxygenated HbS at high enough concentrations (27, 283). The stiff polymerized HbS fibers (see Fig. 7.1A) can create protrusions and cause vesiculation of the RBC membrane (35). Electron microscopy and X-ray crystallography studies revealed that a single HbS fiber consists of 14 filaments arranged in 7 double helical strands twisted about a common axis in a rope-like fashion (see Fig. 7.1B) (312). The average radius of a single HbS fiber is approximately 11 nm. The mean helical path along the fiber, defined as the length over which the fiber twists through 180º, is about 270 nm. Measurements of the bending rigidities and persistence lengths of HbS fibers were conducted via the differential interference contrast (DIC) microscopy (290) and cryo-electron microscopy (291, 292), which showed that the bending rigidity of a single HbS fiber was approximately $5.2 \times 10^{-25}$ N m$^2$ and the torsional rigidity was approximately $6 \times 10^{-27}$ J m (292).

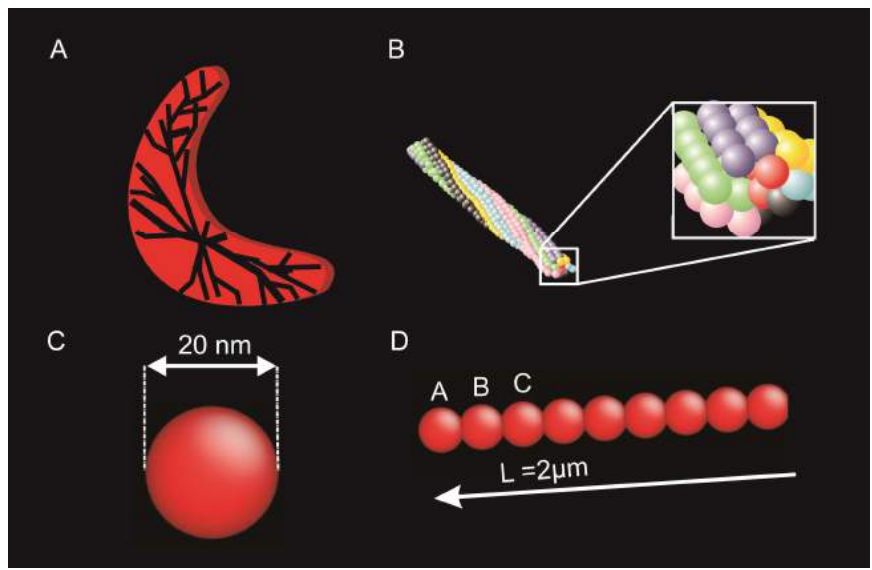

Figure 7.1. (A) Schematic of HbS fibers in a sickle cell. (B) A single HbS fiber comprises 7 pairs of double strands of hemoglobin tetramers. The pairs are shown as same-color agents. It is drawn after (27). (C) Each bead represents the entire cross-section of a single HbS fiber. (D) Side view of the HbS fiber model.

The formation of X-shaped and Y-shaped cross-links, along with zippering of HbS fibers affects the rheology and stiffness of deoxygenated sickle RBCs and consequently the pathophysiology of SCD (313). Various analytical schemes based on Van der Waals and depletion forces have been proposed to estimate the lateral attraction between hemoglobin fibers and the results show that the depletion forces have an equivalent contribution with the Van der Waals forces. However, the experimentally obtained binding energy is lower than the analytically computed value (44, 314). This discrepancy was attributed to several possible reasons such as existence of fiber tensile stress, presence of non-steric repulsive forces between the fibers, aggregation of HbS monomers which reduces the solution osmotic pressure, steric frustration due to the helical nature of the fibers, and non-zero surface separation because of thermal fluctuations (44).

Research on the mechanical behavior of HbS fibers so far is mainly based on analytical methods and experimental observations. Molecular dynamics simulations have only been employed in investigations related to the contact points of the double strands in HbS fiber formation and the crystal structure of HbS (295, 296). Here, a CGMD model of single HbS fibers is introduced. The parameters used in the CGMD potentials are validated by the experimentally measured bending and torsional rigidities of HbS fibers (290, 292). The proposed model follows a top-down approach. It is much simpler than multiscale coarse-graining strategies which usually consider atomistic details while the CG force field is parameterized to match the atomistic forces and predict the thermodynamic behavior of the system (299-303). Similar approaches have been successfully used in the representation of the spectrin cytoskeleton of erythrocyte membrane (109) and in the study of cross-linked actin-like networks (297, 298). A four chain model for the hemoglobin filaments has been proposed earlier by the authors (315). While the overall strategy is similar with the model introduced here, there are significant differences that considerably improve the present model. The cross-section of the HbS fiber in the present model is represented by one CG particle instead of four particles. Also, the introduction of torsional rigidity is more straightforward while in the previous model the torsional rigidity originated from the interactions between the four chains. The present model is applied to investigate the interaction between two HbS fibers and the fiber zippering process. The effects of the Van der Waals force and the depletion force on the HbS fiber zippering are explored.

## 7.2. Model and method

A 2 μm long HbS fiber with diameter of 20 nm is modeled as one chain of 100 soft particles (see Fig. 7.1C and Fig. 7.1D). The diameter of each particle $d_0$ is 20 nm. The translational motion of the particles is governed by the Langevin equation

$$m_i \frac{d^2 r_i}{dt^2} = F_i - f \frac{dr_i}{dt} + F_i^B, \qquad (7.1)$$

where $m_i$ represents the mass of the ith particle, $f$ is the friction coefficient and it is identified to be 100 $m_i/t_s$, where $t_s$ is the time scale. $r_i$ is the position vector of the ith particle, and $t$ is time (304). $F_i = -\partial U/\partial r_i$ is the deterministic force produced by the implemented total potential $U$, which is introduced below. $F_i^B$ is related to the environmental Gaussian white noise and it obeys the fluctuation-dissipation theorem

$$\langle F_i^B \rangle = 0, \tag{7.2}$$

$$\langle F_i^B F_j^B \rangle = \frac{2k_B Tf \delta_{ij}}{\Delta t}, \tag{7.3}$$

where $k_B$ is the Boltzmann's constant, $T = 300$ K is the absolute environmental temperature, $\delta_{ij}$ is the Kronecker delta, and $\Delta t$ is the time-step (98, 305). The energy unit is $k_B T$. The time step for the numerical solution of the Langevin equation is chosen to be $\Delta t = 0.001 t_s$. Deviations from that temperature are corrected with a decay time $t_d = m_i/f$. By choosing $f = 100$ $m_i/t_s$, the decay time becomes $t_d = t_s/100 = 10\Delta t$, meaning that the system corrects the temperature variations in about 10 time steps.

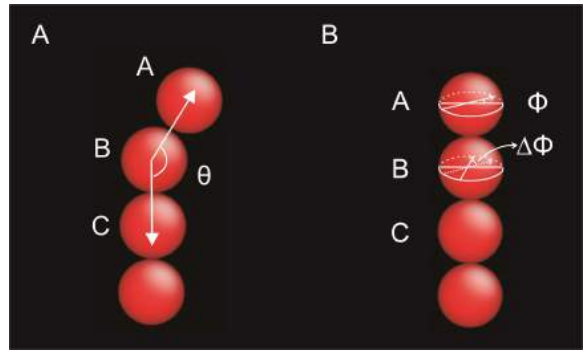

Figure 7.2 (A) Bending angle $\theta$ and (B) twist angle $\Delta\Phi$ implemented in the HbS fiber model.

The rotational motion of the particles is simplified. Each particle only rotates in the plane perpendicular to the vector defined by the centers of this particle and the next one towards the end of the fiber (see Fig. 7.2A). The rotational angle $\Phi$ (see Fig. 7.2B) is controlled by the Langevin equation for the rotational Brownian motion

$$\frac{dM_i(t)}{dt} = T_i - \zeta\omega_i(t) + \lambda_i(t), \tag{7.4}$$

where $M_i = I\omega_i$ is the angular momentum of the ith particles (316), $I$ is the moment of inertia and $\omega = d\Phi(t)/dt$ is the angular velocity. $T_i$ is the torque generated by the torsional potential, which is introduced below. $\zeta\omega_i(t)$ and $\lambda(t)$ are the frictional and white noise torque, respectively. $\zeta$ is the friction coefficient and it is chosen based on the assumption that the Langevin equation for the translational and rotational motion should have the same decay time, $t_d = m_i/f = I_i/\zeta$. Then, the friction coefficient is computed to be $\zeta = 10m_i\sigma^2/t_s$, where the length unit is $\sigma = 2^{-1/6}d_0$. The white noise torque follows the properties below,

$$\langle \lambda_i(t) \rangle = 0 \tag{7.5}$$

$$\langle \lambda_i(t)\lambda_j(t) \rangle = \frac{2k_BT\zeta\delta_{ij}}{\Delta t}. \tag{7.6}$$

A FENE potential (see Fig. 7.3) is introduced between adjacent particles that belong to the same chain (e.g., A and B in Fig. 7.1D). The FENE potential is expressed as

$$U_s = -\frac{1}{2}K_s\Delta d_{max}\ln\left[1-\left(\frac{d_{ij}-d_0}{\Delta d_{max}}\right)^2\right], \tag{7.7}$$

where, $K_s$, is related to the stiffness $K_{sp} = K_s/\Delta d_{max}$ of the harmonic potential which has the same stiffness with the FENE potential at equilibrium. The value of $K_s$ that generates the Young's modulus of the HbS fiber is approximately $K_s = K_{sp}\Delta d_{max} = 28800k_BT/d_0$. $d_{ij}$ and $d_0$ are the distance and the equilibrium distance between particles $i$ and $j$, respectively. In the simulations, the maximum extension allowed between two particles is set to be $\Delta d_{max} = 0.3d_0$.

$$U_b = -\frac{1}{2}K_b\Delta\theta_{max}\ln\left[1-\left(\frac{\theta-\theta_0}{\Delta\theta_{max}}\right)^2\right], \tag{7.8}$$

where $K_b$ is the parameter that directly regulates the bending stiffness of the HbS fiber, and $\theta$ is the angle formed by three consecutive particles in the same chain, as illustrated in Fig. 7.2A. $\theta_0$

is the equilibrium angle and it is chosen to be $\pi$, meaning that the three consecutive particles are initially located in-line. $\Delta\theta_{max}$ is the maximum allowed bending angle between two particles and is set to $0.3\theta_0$. $K_b$ is selected via a trial and error process in order to produce a bending rigidity identical to the experimental value and it is determined to be $K_b = 8\times10^3\ k_BT$.

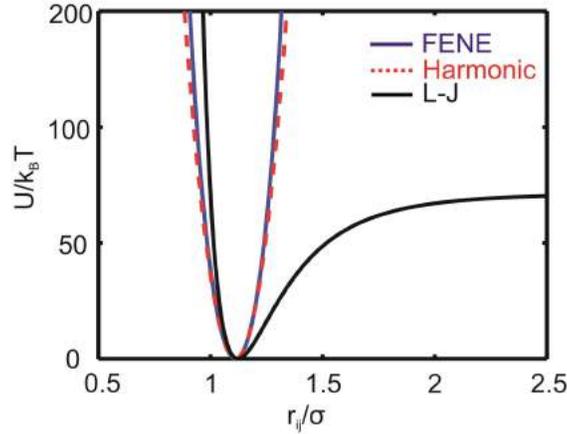

Figure 7.3  The finitely extendable nonlinear elastic (FENE) potential applied between consecutive fiber particles, the corresponding harmonic potential, and the Lennard-Jones potential applied between particles belonging to different HbS fibers. The additional bending and torsional FENE potentials employed in the simulations have similar behavior with the FENE potential shown here.

The torsional rigidity of the proposed model is introduced via a torsional FENE potential between neighboring particles in the chain. The potential is expressed as

$$U_t = -\frac{1}{2}K_t \Delta\Phi_{max} \ln\left[1-\left(\frac{\Delta\Phi}{\Delta\Phi_{max}}\right)^2\right], \qquad (7.9)$$

where $K_t$ adjusts the torsional stiffness of the HbS fiber, and $\Delta\Phi$ is the angle between two directional vectors defined in two consecutive particles, as illustrated in Fig. 7.2B. $\Delta\Phi_{max}$ is the maximum allowed twist angle between two consecutive particles and is set to be 0.3. $K_t$ is tuned to match the model against experimentally obtained values of the torsional rigidity of single HbS fibers (44) and it is identified to be $K_t = 8.1 \times 10^3 \, k_B T$.

The interactions between neighboring HbS fibers are represented by a L-J potential between particles from different fibers (see Fig. 7.3). The expression of L-J potential is given by

$$U_{LJ}(r_{ij}) = \begin{cases} 4k\varepsilon \left[ \left(\frac{\sigma}{r_{ij}}\right)^{12} - \left(\frac{\sigma}{r_{ij}}\right)^{6} \right] + k\varepsilon & r_{ij} < r_{cut} \\ 0 & r_{ij} > r_{cut} \end{cases}, \quad (7.10)$$

where $r_{ij}$ is the distance between particles $i$ and $j$. The parameter $k$ adjusts the force between two HbS fibers. The value $k = 70$ is used for the plot in Fig. 7.3. The distance between two particles, beyond which the attractive force is neglected, is $r_{cut} = 2.5\sigma$.

A coupling torque $T_c$ is introduced between two particles from different fibers to represent the resistance to the rotational motion when the two fibers are zippered. The torque is assumed to be

$$T_c = -T_0 \left(e^\alpha - 1\right), \quad (7.11)$$

where $T_0$ is chosen to be $10 \, k_B T$, which is on the same scale of the torque generated by the FENE torsional potential. $\alpha$ is the relative rotational angle between two neighboring bonded particles, measured after the instance of the two particles zippering.

The molecular weight of each hemoglobin tetramer is approximately 64.5 KDa. Thus, each coarse-grained particle has a mass of $m_i \approx 6 \times 10^{-21}$ kg representing a cluster of 56 hemoglobin tetramers. In the Langevin equation, the friction coefficient is chosen to be $f = 100 \, m_i/t_s$ while $f$ also can be expressed as a function of the particle diameter $d_0$ and the medium viscosity $\eta$ ($f = 3\pi\eta d_0$). The viscosity of Hb solution with concentration of c = 24.4g/dl used by (44) is found to be $\eta \approx 3\times10^{-3}$ kg/m s (317). By solving the equation 100 $m_i/t_s = 3\pi\eta d_0$, the timescale of the simulation is found to be $t_s \approx 1\times10^{-9}$ s. The spring constant for the harmonic potential $K_{sp}$ can be expressed in terms of the bending rigidity $\kappa$, as $K_{sp} = 16\kappa/d_0^3$. A relevant time-scale for the vibration of the HbS fibers can be estimated as $t_c = 2\pi\sqrt{m/K_{sp}} \cong 556.4\text{ps}$. The time-scale for the fluctuation of the rotational motion of the HbS fibers can be similarly obtained by the expression $t_r = 2\pi\sqrt{I/K_t \cdot \Delta\Phi_{max}} \cong 560.6\text{ps}$, which is close to $t_c$.

The numerical method employed for the integration of Eq. (7.1) is a modified version of the leapfrog algorithm and it is given by

$$v_i(t+\Delta t/2) = v_i(t) + \Delta t/2 a_i(t) \qquad 7.12(a)$$

$$r_i(t+\Delta t) = r_i(t) + \Delta t v_i(t+\Delta t/2) \qquad 7.12(b)$$

$$\tilde{v}_i(t+\Delta t) = v_i(t+\Delta t/2) + \Delta t/2 a_i(t) \qquad 7.12(c)$$

$$a_i(t+\Delta t) = a_i(r_i(t+\Delta t), \tilde{v}_i(t+\Delta t)) \qquad 7.12(d)$$

$$v_i(t+\Delta t) = v_i(t+\Delta t/2) + \Delta t/2 a_i(t+\Delta t). \qquad 7.12(e)$$

The total force applied on a particle depends on the velocity of the particle. Thus, a prediction is made for the new velocity, which is denoted as $\tilde{v}$, and it is corrected in the last step (106, 315). The same strategy was followed for the integration of the rotational Langevin equation.

### 7.3. Results and discussion

7.3.1 Measurements of material properties of HbS fiber

*7.3.1.1 Bending rigidity*

The bending modulus of semi-flexible fibers can be derived from the thermally driven fluctuations (126). One method is to measure the mean-squared amplitude of each dynamical mode of vibrations and by using the principle of equipartition of energy to extract the bending rigidity for each mode independently (318). However, since HbS fibers are stiff with a large persistence length of approximately 120 -240 µm while the simulated fibers have a total length of 2 µm, only small bending is observed in the simulations. This leads to a large uncertainty when the motion is decomposed into Fourier modes. In order to measure the bending rigidity of HbS fibers under thermal fluctuation, a method that relates the fiber bending rigidity with the normal deviation of the fiber end point through the expression

$$\kappa = \frac{k_B T L^3}{3\langle \delta u(L)^2 \rangle},  \quad (7.13)$$

is followed (44, 314). $L$ is the total length of the HbS fiber and $\delta u(L) = u(L) - \langle u(L) \rangle$ is the normal deviation of the fiber end measured from its average position during the simulation to eliminate the rigid body motion effects and to only account for thermal fluctuations. Once the bending rigidity is obtained, the persistence length can be easily calculated by the expression

$$l_p = \frac{\kappa}{k_B T}. \quad (7.14)$$

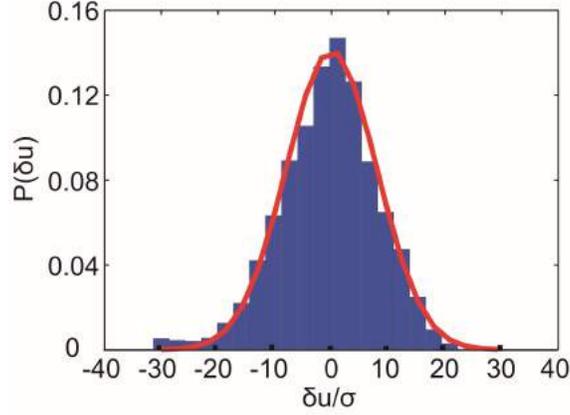

Figure. 7.4. The distribution of HbS fiber end displacements and the associated normalized Gaussian probability distribution.

The probability density of the deviations from the equilibrium at the end of the HbS fiber can be approximated by a normalized Gaussian distribution (solid curve in Fig. 7.4) given by

$$P(\delta u) = \frac{1}{\sqrt{2\pi \langle \delta u^2 \rangle}} \exp\left(-\frac{\delta u^2}{2\langle \delta u^2 \rangle}\right), \qquad (7.15)$$

where $\delta u$ is the HbS fiber end displacement, and $\langle \delta u^2 \rangle^{1/2} \cong 8\sigma$. The result is consistent with the theoretical prediction that the probability density of the deviations of a semi-flexible rod follows the Gaussian distribution in the case of thermal fluctuations (310) and it indicates that the HbS fiber has reached thermal equilibrium. The value of the bending rigidity calculated via the equation (16) is $\kappa = 5.3 \times 10^{-25}$ N m$^2$, which is close to the experimentally measured value of $\kappa = 5.2 \times 10^{-25}$ N m$^2$ (292). The obtained bending rigidity corresponds to a persistence length of approximately $l_p \cong 121$ μm. The numerical results show that, because of the large persistence length, a single HbS fiber subjected to thermal forces behaves similarly to a stiff rod as it fluctuates about its central axis (see Fig. 7.5). It is noted that when a longer fiber of 5 μm length is tested, the computed bending rigidity has the same value with the bending rigidity of the 2 μm long fiber, meaning that the result is size independent at least up to 5 μm.

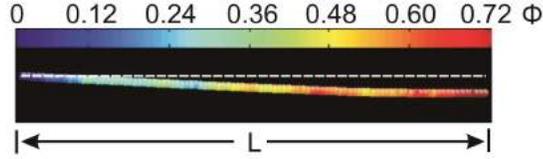

Figure 7.5. Characteristic configuration of the HbS fiber model. The colors of the particles represent the value of the rotation angle $\Phi$, the blue color corresponds to $\Phi = 0$ and the red color corresponds to $\Phi = 0.72$. The color bar is shown at the top of the figure.

*7.3.1.2 Torsional rigidity*

The torsional rigidity $c$ of the HbS fiber can be obtained via the expression

$$\langle \Delta\lambda^2 \rangle = \frac{k_B T \lambda^3}{c\pi^2}\left[1 - \frac{L}{\lambda}\left(1 - e^{-\lambda/L}\right)\right], \quad (7.16)$$

where $\Delta\lambda = \lambda\Psi/\pi$, $\lambda = 135$ nm is half of the average pitch length for the HbS fiber, $\Psi$ is the accumulated relative rotation between the first fixed particle and the $7^{th}$ particle located at approximately half of the pitch length, and $L$ is the length of the HbS fiber (290). By substituting $\Delta\lambda$ into equation (16), the torsional rigidity $c$ is found to be

$$c = \frac{k_B T \lambda}{\langle \Psi^2 \rangle}\left[1 - \frac{L}{\lambda}\left(1 - e^{-\lambda/L}\right)\right]. \quad (7.17)$$

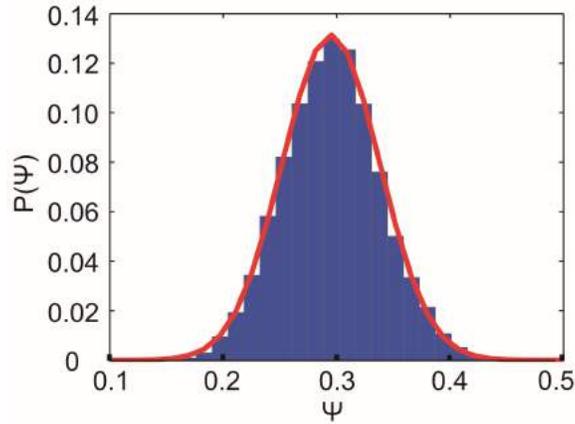

Figure. 7.6. The distribution of the accumulation of the twist angle at half of the pitch length (7$^{th}$ chain particle) $\Psi$ and the associated normalized Gaussian probability distribution

Fig. 7.6 shows that the twist angle through half of the pitch length $\Psi$ also follows the Gaussian distribution with a mean twist angle of 0.293, resulting in a torsional rigidity of the HbS fiber model $c \cong 6.5 \times 10^{-27}$ J m, which is very close to the experimentally measured value of $c \cong 6.0 \times 10^{-27}$ J m (292). As in the case of bending rigidity, it is found that a longer fiber of 5 μm exhibits the same torsional rigidity with the shorter one when the same parameters are used. It is also worth mentioning that the proposed HbS fiber model is strongly anisotropic since the torsional rigidity is significantly lower than the bending rigidity.

7.3.2 Modeling the zippering of two HbS fibers

The developed CGMD model is employed to study the effect of the Van der Waals and the depletion forces on the zippering of two fibers in a Y-shaped cross-link (see Fig. 7.7). The Van der Waals interaction is represented by a L-J potential applied between particles belonging to different fibers. The effect of the depletion force is measured by the contribution of HbS monomers on the fiber zippered length. The concentration of the hemoglobin solution is chosen to be $c = 24.4$ g/dl, which is the value used in the previous experimental work (44). The HbS monomer particles are assumed to have the same size ($d_0 = 20$ nm) and same mass ($6 \times 10^{-21}$ kg) of the fiber particles. Therefore, the concentration of the solution $c = 24.4$ g/dl corresponds to HbS monomers density of $\rho = 0.23$ particles/$\sigma^3$ in the simulation space (a box with the size of $140\sigma \times 50\sigma \times 30\sigma$). The height of the simulation space is chosen to be $30\sigma$ (about 540 nm), which is more than two times thicker than the fiber interaction space observed in the experiments (44, 313). In this section, we applied the Berendsen thermostat to control the temperature of the system. In addition, the effect of monomer aggregation on the fiber zippering is studied by reducing the density of the free monomer particles to $\rho = 0.115$ particles/$\sigma^3$.

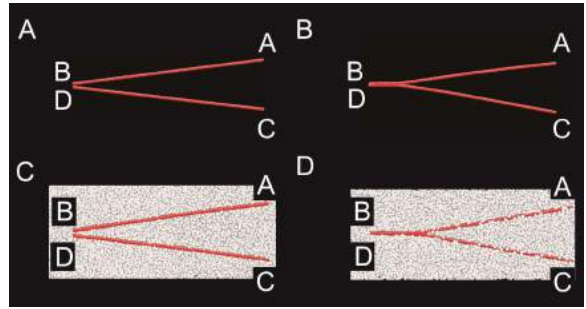

Figure 7.7. (A) Initial 3D configuration of two HbS fibers (red particles) arranged in "Y" shape. (B) Equilibrium 3D configuration of the two HbS fibers after zippering. The value of the L-J parameter is $k = 30$. (C) Initial 3D configuration of two HbS fibers arranged in "Y" shape surrounded by HbS monomer particles (white particles). (D) Equilibrium 3D configuration of the two HbS fibers surrounded by HbS monomer particles after zippering. The value of the L-J parameter is $k = 30$ and the density of the HbS monomers is $\rho = 0.23$ particles/$\sigma^3$. It is observed that the zippered length of two fibers is increased by $7\sigma$ compared to the (B) due to the presence of the HbS monomer particles.

The case of two fibers arranged initially in a "V" shape junction without any HbS monomers is first studied (see Fig. 7.7A). Points A and C represent the points where the measured fibers are bonded stably to a third fiber and they are thus fixed. The other two ends B and D are free to move. The HbS fibers interact through physical and chemical forces including direct electrostatic effects, Van der Waals forces and the entropic depletion forces between the free monomers and the fibers. In addition, there are small hydrophobic forces applied on patches on the surfaces of fiber aggregates (44). The attractive energy between two fibers is represented by a L-J potential (Eq. (7.10)) and it is adjusted by varying the value of the parameter $k$ in the expression (7.10). After zippering (see Fig. 7.7B), the two fibers move together as a Y-shaped branch. In the equilibrium state, the interfiber attractive forces are balanced by the bending energy stored in the two fibers (44, 313). The zippered lengths for various values of $k$ are illustrated by the black line in Fig. 7.8. It is observed that for low $k$ values, the zippering is not significant. As $k$ increases, the slope of the zippered length with respect to energy decreases since the effect of the constraints on the fiber ends A and C becomes more pronounced. Secondly, the entropic effect of the free HbS monomers on the zippering of the two HbS fibers is investigated (Figs. 7C and 7D). It is assumed that there are only repulsive forces between HbS monomers and between HbS monomers and fiber particles. The repulsive forces result from the repulsive part of the L-J potential (Eq. (7.10)) and it creates an excluded volume for HbS monomers and fibers. The effect of the depletion force on the zippered lengths of two fibers for two different densities of the free HbS monomers is shown in Fig. 7.8. The depletion force has a dramatic effect when the attractive force is low. For example, when $\rho = 0.23 \text{particles}/\sigma^3$ (red line), contributions of the depletion force to the fiber zippering is more than the Van der Waals force when $k <= 15$. As the attractive energy increases, the effect of the depletion force attenuates.

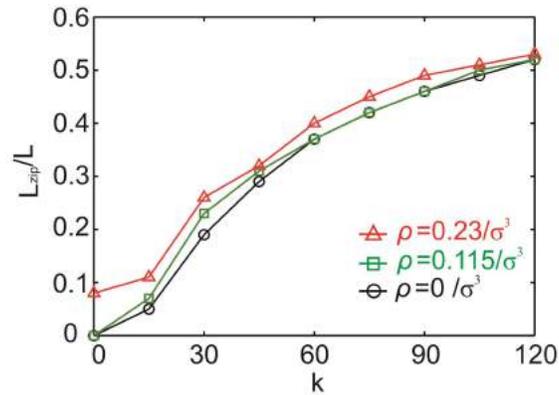

Figure 7.8. Zippered length ratios of two fibers ($L_{zip}/L$) for different attractive forces between the HbS fibers. The solid lines describe the ratios of the zippered length to the fiber length for different values of the parameter $k$. The solid lines in different colors represent different densities of HbS monomers.

A main question raised from experimental studies is how much the aggregation of the HbS monomers reduces the depletion force (44, 314). To investigate it, we assume that 50% of the monomers aggregate resulting to a reduced HbS monomer density $\rho = 0.115 particles/\sigma^3$. The corresponding zippering lengths of two fibers for various values of $k$ are illustrated by the green lines in Fig. 7.8. For $k < 30$, the reduction in the monomer density results in a decrease in the additional zippering caused by the depletion effect. As the $k$ increases, however, the drop of the monomer density has smaller effect on the reduction of the fiber zippering because the attractive forces between the two fibers become dominant. This result indicates that the effect of monomer aggregation on the zippered length is larger at lower values of the interaction force between the fibers.

In summary, a CGMD model with high computational efficiency is proposed to simulate single HbS fibers as one chain of particles. The model is able to derive the experimentally measured bending and torsional rigidities of single HbS fibers. In addition, it is employed to investigate the effect of the Van der Waals force and the depletion force on the zippering of two HbS fibers to improve our understanding on the HbS fiber aggregation in SCD. The present model has the potential to significantly facilitate the studies of HbS polymerization, fiber aggregation and gel formation. The present model could be combined with a hybrid method of dynamic coarse-graining, to simulate HbS polymerization. Since the HbS polymerization process is governed by monomer addition as opposed to oligomer addition (319), the polymerization has to be studied at the atomistic level. Therefore, the tip of a HbS fiber, where polymerization occurs, could be simulated at the atomistic level while the already polymerized HbS fiber could be simulated by the proposed coarse-grained model to reduce the overall computational cost. A similar technique has been introduced in phospholipid bilayer models (320, 321).

# Chapter 8.

# Epilogue

We have developed a two-component CGMD model for RBC membranes. In this model, the lipid bilayer and the cytoskeleton are considered as an ensemble of discrete particles that interact through a direction-dependent pair potential. Three types of coarse-grained particles are introduced, representing a cluster of lipid molecules, actin junctions and band-3 complexes. The large grain size extends the accessible length and time scales of the simulations to ~ μm and ~ ms. By tailoring only a few parameters of the inter-grain interaction potential, the model can be stabilized into a fluid phase manifested by the free diffusion of the particles and the reproduction of the essential thermodynamic properties of the RBC membrane. An important feature of the proposed model is the combination of the spectrin network with the lipid bilayer, which represents a more compete representation of the RBC membrane compared to one-component CGMD model and to most of the continuum models. This model allows us to study the behavior of the membrane under shearing. The behavior of the model under shearing at different strain rates illustrates that at low strain rates up to $0.001 d/t_s$, the developed shear stress is mainly due to the spectrin network and it shows the characteristic non-linear behavior of entropic networks, while the viscosity of the fluid-like lipid bilayer contributes to the resulted shear stress at higher strain rates. Decrease of the spectrin network connectivity results to significant decrease of the shear modulus of the membrane, which demonstrates that the cytoskeleton carries most of the load applied on the cell while the lipid bilayer only functions as 2D fluid. The values of the shear moduli measured from the membrane with reduced spectrin network connectivity are in a good agreement with previous experimental, theoretical, and numerical results.

We introduce a particle-based model for the erythrocyte membrane that accounts for the most important structural components of the membrane, including the lipid bilayer, the spectrin network, and the proteins that play an important role in the anchoring of the spectrin cortex to the lipid bilayer, as well as the band-3 proteins. In particular, five types of CG particles are used to represent actin junctional complexes, spectrin, glycophorin, immobile band-3 protein, mobile band-3 protein and an aggregation of lipids. We first demonstrate that the model captures the fluidic behavior of the lipid bilayer and then that it reproduces the expected mechanical material properties of bending rigidity and shear modulus of the RBC membrane. The timescale of our simulations, which is found to be $t_s \sim 3\times10^{-6}s$, is inferred by comparing the viscosity of the membrane model to experimentally measured values. Then, the self-consistency of the model with respect to the timescale is tested by comparing the computed vibration frequency of the spectrin filaments and lipid membrane to analytically obtained values. We confirm that vibration frequency of the spectrin filaments and lipid membrane measured from the proposed membrane model, are also in agreement with experimental values. At last, we study the interactions between the cytoskeleton and the lipid bilayer, and measure the pressure applied on the membrane by the spectrin filaments. We also investigate how disruption of the connection between the spectrin network and the lipid bilayers, which simulates defects in HS, and rupture of the dimer-dimer association, which simulates defects in HE, affect the pressure exerted on the lipid bilayer. We show that overall the introduction of defects in the spectrin network or in the vertical connection between the lipid bilayer and the cytoskeleton results in lower pressure, which is consistent with prediction in (155). In addition, we find that the defects related to HE have a stronger effect than the defects related to HS. This result implies that diffusion of band-3 proteins in RBCs from patients with HS and HE is enhanced compared to the normal RBCs.

Moreover, elliptocytes exhibit more prominent diffusion of band-3 proteins than spherocytes. Both conclusions are supported by experimental results. The level of attraction forces between the lipid bilayer and the membrane cytoskeleton is another important parameter that regulates the pressure applied by a spectrin filament to the lipid bilayer. We show that as the attractive force increases, it causes an overall increase in the pressure and it diminishes the differences in the pressure generated by membrane protein defects and, consequently, the differences in the diffusion of band-3 proteins in normal and defective erythrocytes. Since this finding is not supported by experimental results, we conjecture that the attractive force between the lipid bilayer and the spectrin filaments should be low, resulting in a membrane model where the filaments are not completely attached to the lipid bilayer. A detailed study of the band-3 diffusion in this model and direct comparison with experimental results is necessary in order to form a more accurate picture of how the model regulates diffusion. Because of the explicit representation of the lipid bilayer and the cytoskeleton, the proposed model can be potentially used in the investigation of a variety of membrane related problems in RBCs in addition to diffusion. For example, membrane loss through vesiculation and membrane fragility in spherocytosis and elliptocytosis, interaction between hemoglobin fibers and RBC membrane in sickle cell disease, and RBC adhesion are problems where the applications of the proposed membrane model could be beneficial.

We apply a two-component RBC membrane model to study band-3 diffusion in the normal RBC membrane and in the membranes with defective vertical and horizontal interactions. We measure the band-3 diffusion coefficients and quantify the relation between the band-3 diffusion coefficients and percentage of protein defects. Our measurements show that the band-3 diffusion coefficients are increased by 8 times when connectivity between the spectrin filaments and immobile band-3 particles $C_{vertical}$ is decreased from 100% to 0%, whereas the band-3 diffusion coefficients can be boosted by 20-40 times with significantly reduced cytoskeleton connectivity $C_{horizontal}$. These results are in agreement with the experimental measurements of the band-3 diffusion in the HS and HE RBCs (112, 115, 147, 197). By comparing the measured band-3 diffusions coefficients, we demonstrate that cytoskeleton connectivity is the major determinant of the lateral diffusivity of band-3. In addition, we quantify the relations between the anomalous diffusion exponent and the percentage of protein defects in the horizontal interactions. We find that the cytoskeleton has a small effect on the anomalous diffusion exponent, comparing to the obstructions due to the lipid rafts or protein domains. At last, we introduce attractive forces between the spectrin filaments and the lipid bilayer, and study the effects of the attractive forces on the band-3 diffusion. We find that application of the attractive forces slows down the band-3 diffusion in the normal and defective RBC membrane. Especially, large attractive forces can prevent the band-3 hop diffusion in the normal RBC membrane and in the membrane with protein defects in the vertical interactions. But, in the cases of the membrane with protein defects in the horizontal interactions, the band-3 diffusion coefficients become independent of the attractive forces when attractive forces are large. This is because the band-3 particles travel to different cytoskeleton compartments through the broken spectrin filaments. Moreover, we show that the band-3 diffusion coefficients measured at the large attractive forces generally agree with the percolation analysis in (116). By comparing the band-3 diffusion coefficients from our simulation with the experimental measured band-3 diffusion coefficients in HS and HE (112, 115, 147, 197), we estimated the scale of the effective attractive force between the spectrin filaments and lipid bilayer is at least 20 times smaller than the binding forces at the two major binding sites at the immobile band-3 proteins and the glycophorin C. The approach used here for the simulation of band-3 diffusion in the defective RBC membrane provides a basis for the future study of the mechanism of the membrane vesiculation in HS and HE.

We model single HbS fibers using a quadruple chain of particles and the hexagonally shaped cross-section of the HbS fiber is coarse grained into four particles. The motions of all the particles are governed by the Langevin equation. In order to simulate the thermal behaviors of HbS fibers, five different interaction potentials are applied between the particles, namely a FENE potential, a truncated L-J potential, a bending potential, a torsional potential and a L-J potential. By employing these five potentials, the proposed model is able to derive the experimentally measured bending rigidity, and the torsional rigidity of a single HbS fiber. Then, the model is used in the study of the zippering process of initially parallel free fibers and of frustrated fibers. Finally, the behavior of zippered fibers under compression is explored. The results show that while low interaction energy between free fibers is enough to cause zippering, frustrated fibers or fibers under compression require much higher interaction energy to remain zippered. This means that fiber frustration and compression can result to partial unzipping of bundles of polymerized HbS fibers. Continuous polymerization of the unzippered fibers via heterogeneous nucleation and additional unzippering under compression can explain the formation of HbS fiber networks and consequently the wide variety of shapes of deoxygenated sickle cells.

We apply a CGMD model with high computational efficiency to simulate single HbS fibers as one chain of particles. The model is able to derive the experimentally measured bending and torsional rigidities of single HbS fibers. In addition, it is employed to investigate the effect of the Van der Waals force and the depletion force on the zippering of two HbS fibers to improve our understanding on the HbS fiber aggregation in SCD. The present model has the potential to significantly facilitate the studies of HbS polymerization, fiber aggregation and gel formation. The present model could be combined with a hybrid method of dynamic coarse-graining, to simulate HbS polymerization. Since the HbS polymerization process is governed by monomer addition as opposed to oligomer addition (319), the polymerization has to be studied at the atomistic level. Therefore, the tip of a HbS fiber, where polymerization occurs, could be simulated at the atomistic level while the already polymerized HbS fiber could be simulated by the proposed coarse-grained model to reduce the overall computational cost. A similar technique has been introduced in phospholipid bilayer models (320, 321).

# Chapter 9.

315. Li H & Lykotrafitis G (2011) A coarse-grain molecular dynamics model for sickle hemoglobin fibers. *Journal of the mechanical behavior of biomedical materials* 4(2):162-173.
316. Coffey WT, Kalmykov YP, & Titov SV (2002) Langevin equation method for the rotational Brownian motion and orientational relaxation in liquids *Journal of Physics A: Mathematical and General* 35:6789-6803.
317. Ross P, D. & Minton A, P. (1977) Hard Quasispherical Model for The Viscosity of Hemoglobin Solution. *BioChemical and Biophysical Research Communications* 76(4):971-976.
318. Gittes F, Mickey B, Nettleton J, & Howard J (1993) Flexural rigidity of microtubules and actin-filaments measured from thermal fluctuation in shape. *Journal of Cell Biology* 120(4):923-934.
319. Aprelev A, Liu Z, & Ferrone Frank A (2011) The Growth of Sickle Hemoglobin Polymers. *Biophysical Journal* 101(4):885-891.
320. Izvekov S & Voth GA (2005) A Multiscale Coarse-Graining Method for Biomolecular Systems. *Journal of Physical Chemistry B* 109(7):2469-2473.
321. Shi Q, iZvekov S, & Voth GA (2006) Mixed Atomistic and Coarse-Grained Molecular Dynamics: Simulation of a Membrane-Bound Ion Channel. *Journal of Physical Chemistry B* 110(31):15045–15048.
197